\documentclass[a4paper,11pt]{article}

\usepackage{graphicx}
\usepackage{color}
\usepackage{appendix}
\usepackage{psfrag}
\usepackage{amsmath,amssymb}

\RequirePackage[sort&compress,square,comma,numbers]{natbib}
\RequirePackage{hyperref}%
\RequirePackage{hypernat}

\allowdisplaybreaks[1]
\newcommand{\bea}{\begin{eqnarray}}
\newcommand{\beas}{\begin{eqnarray*}}
\newcommand{\nn}{\nonumber}
\newcommand{\eea}{\end{eqnarray}}
\newcommand{\eeas}{\end{eqnarray*}}
\newcommand{\bd}{\begin{displaymath}}
\newcommand{\ed}{\end{displaymath}}
\newcommand{\be}{\begin{equation}}
\newcommand{\ee}{\end{equation}}
\newcommand{\bi}{\begin{itemize}}
\newcommand{\ei}{\end{itemize}}

\newcommand{\as}{\alpha_s}

\newcommand{\RE}{{\rm Re}}
\newcommand{\IM}{{\rm Im}}

\newcommand{\hc}{\mbox{h.c.}}

\newcommand{\gev}{\mbox{GeV}}

\newcommand{\Br}{{\rm Br}}

\newcommand{\rC}{r_{\textrm{C}}}
\newcommand{\rT}{r_{\textrm{T}}}
\newcommand{\rEW}{r_{\textrm{EW}}}
\newcommand{\rEWC}{r_{\textrm{EW}}^{\textrm{C}}}
\newcommand{\rEWA}{r_{\textrm{EW}}^{\textrm{A}}}
\newcommand{\rtEW}{\tilde{r}_{\textrm{EW}}}
\newcommand{\rtEWC}{\tilde{r}_{\textrm{EW}}^{\textrm{C}}}
\newcommand{\rtEWA}{\tilde{r}_{\textrm{EW}}^{\textrm{A}}}
\newcommand{\rtEWi}{\tilde{r}_{\textrm{EW},\,i}}
\newcommand{\rtEWCi}{\tilde{r}_{\textrm{EW},\,i}^{\textrm{C}}}
\newcommand{\rtEWAi}{\tilde{r}_{\textrm{EW},\,i}^{\textrm{A}}}

\newcommand{\captionfonts}{\small}
\makeatletter  
\long\def\@makecaption#1#2{%
  \vskip\abovecaptionskip
  \sbox\@tempboxa{{\captionfonts #1: #2}}%
  \ifdim \wd\@tempboxa >\hsize
    {\captionfonts #1: #2\par}
  \else
    \hbox to\hsize{\hfil\box\@tempboxa\hfil}%
  \fi
  \vskip\belowcaptionskip}
\makeatother   

\begin{document}

\begin{titlepage}

\begin{flushright}
\normalsize
SFB/CPP-10-122 \\
TTP10-48\\
MZ-TH/10-44\\
November 2010
\end{flushright}

\begin{center}
\Large\bf\boldmath
$\bar B_s\rightarrow \phi \rho^0$ and $\bar B_s\rightarrow \phi \pi^0$
as a handle on\\ isospin-violating New Physics
\unboldmath
\end{center}

\vspace*{0.8cm}
\begin{center}

{\sc Lars~Hofer}$^{a,b}$,
{\sc Dominik~Scherer}$^{a}$
and {\sc Leonardo~Vernazza}$^{c}$\footnote{Alexander-von-Humboldt Fellow}\\
\vspace{0.7cm}
{\sl $^a$
Institut f\"ur Theoretische Teilchenphysik, \\
Karlsruhe Institute of Technology, D--76128 Karlsruhe, Germany}\\[0.4cm]
{\sl $^b$
Institut f\"ur Theoretische Physik und Astrophysik, \\
Universit\"at W\"urzburg, D--97074 W\"urzburg, Germany}\\[0.4cm]
{\sl $^c$
Institut f\"ur Physik (THEP),\\ Johannes Gutenberg-Universit\"at, D--55099 Mainz, Germany}\\[0.4cm]
\end{center}

\vspace*{0.8cm}
\begin{abstract}

The $2.5\sigma$ discrepancy between theory and experiment observed in the difference
\linebreak\mbox{$A_{\rm CP}(B^-\to\pi^0 K^-)-A_{\rm CP}(\bar{B}^0\to\pi^+ K^-)$} can
be explained by a new electroweak penguin amplitude. Motivated by this result, we
analyse the purely isospin-violating decays $\bar B_s\to\phi\rho^0$ and
$\bar B_s\to\phi\pi^0$, which are dominated by electroweak penguins, and show that
in presence of a new electroweak penguin amplitude their branching ratio can be
enhanced by up to an order of magnitude, without violating any constraints from
other hadronic $B$ decays. This makes them very interesting modes for LHCb and future
$B$ factories. We perform both a model-independent analysis and a
study within realistic New Physics models such as a modified-$Z^0$-penguin scenario,
a model with an additional $Z^{\prime}$ boson and the MSSM. In the latter cases the
new amplitude can be correlated with other flavour phenomena, such as semileptonic
$B$ decays and $B_s$-$\bar{B}_s$ mixing, which impose stringent constraints on the
enhancement of the two $B_s$ decays. In particular we find that, contrary to claims
in the literature, electroweak penguins in the MSSM can reduce  the discrepancy in
the $B\to\pi K$ modes only marginally. As byproducts we update the SM predictions
to $\Br(\bar B_s\to\phi\pi^0)=1.6^{+1.1}_{-0.3}\cdot 10^{-7}$ and
$\Br(\bar B_s\to\phi\rho^0)=4.4^{+2.7}_{-0.7}\cdot 10^{-7}$ and perform a
state-of-the-art analysis of $B\rightarrow \pi K$ amplitudes in QCD factorisation.
\vspace*{0.8cm}
\end{abstract}
\vfil

\end{titlepage}

\newpage

\section{Introduction}\label{intro}

At present flavour physics has entered a new exciting era. The new experiment LHCb
and the planned super-B-factories will bring the precision of Standard Model (SM)
tests and the scope of searches for New Physics (NP) to unseen heights. Particularly
important thereby are flavour-changing neutral current (FCNC) decays, which in the
SM are highly-suppressed electroweak loop processes. In this work we present a
phenomenological analysis of two hadronic FCNC decays, namely
$\bar B_s\rightarrow \phi \rho^0$ and $\bar B_s\rightarrow
\phi \pi^0$. We argue that within the next years these decays will become very
interesting objects for experimental analyses of the electroweak penguin sector.
Up to now, this sector has been tested in hadronic decays only in $B\to\pi K$
modes, and the discrepancies found between the SM prediction and experimental
measurements is the main motivation for our work.\medskip

The four $B\to\pi K$ decay channels, first observed by the CLEO experiment in
the late 1990s \cite{Godang:1997we,CroninHennessy:2000kg}, have become by now
a classic in flavour physics thanks to the precise measurements by BABAR and
BELLE. This is also reflected in the large number of theoretical studies of
these decays in the SM and various extensions of it. Charged and neutral $B$
mesons can decay to a $\pi K$ final state due to a weak process at the
partonic level, $b\rightarrow s \bar q q$ with $q=u,d$.
This process is dominated by an FCNC loop governed  by the CKM factor
$V_{ts}^*V_{tb}$ and receives, in the $q=u$ case, also a small tree-level
contribution involving the smaller CKM factor $V_{us}^*V_{ub}$. The
$B\to\pi K$ branching fractions are therefore small, of order $\mathcal
O(10^{-6})$, and sensitive to new FCNCs arising in extensions of the SM. For
this reason they are, together with the corresponding CP asymmetries, important
observables for tests of the SM flavour structure and for NP searches.\medskip

With the data of the $B$ factories having become more and more precise, some
discrepancies between $B\to\pi K$ measurements and SM predictions have occurred,
provoking speculations on a ``$B\to\pi K$ puzzle''. To date, the measurements of
the branching fractions have fluctuated towards the SM predictions, the latter
still suffering from large hadronic uncertainties, and only the CP asymmetries
show an unexpected behavior \cite{Gronau:1998ep,Baek:2007yy,Fleischer:2007mq}
manifesting itself in the quantity
\be\label{a1}
\Delta A_{\rm CP} \equiv A_{\rm CP}(B^-\to\pi^0 K^-)
-A_{\rm CP}(\bar{B}^0\to\pi^+ K^-).
\ee
For this observable we find in the framework of QCD factorisation (QCDF)
\be\label{a2}
\Delta A_{\rm CP}\stackrel{\text{SM}}{=} 1.9^{+5.8}_{-4.8}\,\%
\ee
as the SM prediction, which differs significantly from the experimental value
\be\label{a3}
\Delta A_{\rm CP}\stackrel{\text{exp.}}{=} (14.8 \pm 2.8), \%.
\ee
\cite{Barberio:2008fa}. Adopting a frequentist approach where we consider a theoretical ``error bar'' as
a range of values definitely containing the true theory result but without
assigning any statistical meaning to it \cite{Hocker:2001xe}, this amounts to
a $2.5\sigma$ discrepancy.\medskip

A point which has received much attention in the literature (see e.g.
\cite{Buras:2003yc} and references therein) is the fact that the formerly
observed discrepancies as well as the currently existing anomaly in $\Delta
A_{\rm CP}$ suggest a violation of the strong isospin symmetry beyond the amount
expected in the SM. This has often been interpreted as a hint for enhanced
electroweak penguins (EW penguins)
\cite{Buras:2003dj,Yoshikawa:2003hb,Mishima:2004um}. We will give a brief
overview and discuss the current status of this topic in section \ref{BpiK}.
Whether the $2.5\sigma$ discrepancy in $\Delta A_{\rm CP}$ is a hint for NP in
EW penguins or a non-perturbative hadronic effect or simply a statistical
fluctuation is controversial. The point that we want to make is that, in order
to assess this question, it is highly desirable to obtain further information
from other hadronic decays which are sensitive to EW penguin contributions. For
this reason we study the \textit{purely} isospin-violating decays $\bar
B_s\rightarrow \phi \rho^0$ and $\bar B_s\rightarrow \phi \pi^0$, which are
dominated by EW penguins, extending and updating our analysis presented in ref.~\cite{Hofer:2009ct}. If NP in this sector exists at a level where it can
explain the $\Delta A_{\rm CP}$ puzzle, it could be clearly visible in these
purely isospin-violating decays. The upcoming new generation of flavour
experiments will have the opportunity to detect these modes for the first time
and to measure their branching fractions. The aim of our work is to provide a
detailed analysis from the theory side, both in the SM and beyond.\medskip

Since the decays $\bar B_s\rightarrow \phi \rho^0$ and $\bar B_s\rightarrow \phi
\pi^0$ are not related to other decay modes via flavour symmetries, the
non-perturbative part of their decay amplitudes has to be determined from first
principles. This can be achieved using the framework of QCDF
\cite{Beneke:1999br,Beneke:2000ry,Beneke:2001ev,Beneke:2003zv,Beneke:2006hg}.
This method amounts to a calculation of the hadronic matrix elements up to
corrections of order $\Lambda_\text{QCD}/m_B$, where
$\Lambda_\text{QCD}\sim\mathcal O(200 \,\text{MeV})$ is a typical
non-perturbative energy scale of strong interactions. We will use this method
throughout the paper in all analyses of $B$ decays to light mesons.\medskip

The plan of the paper is as follows: In Chapter \ref{qual}, we discuss the issue
of isospin-violation in $B\to\pi K$ decays and the phenomenology of $\bar
B_s\rightarrow \phi \rho^0$ and $\bar B_s\rightarrow \phi \pi^0$. As a byproduct
we provide simple formulas which allow for an easy calculation of various
observables concerning these decay modes, taking into account NP effects in EW
penguins. Chapter \ref{modind} contains a detailed quantitative analysis of
$\bar B_s\rightarrow \phi \rho^0$ and $\bar B_s\rightarrow \phi \pi^0$ in
different scenarios of a model-independent parameterisation of NP in EW
penguins. This analysis is performed in light of our present knowledge on EW
penguins from other $B$ decays, in particular  $B\to\pi K$. It is complemented
in Chapter \ref{moddep} with studies of particular extensions of the SM which
feature enhanced EW penguins. We conclude in Chapter \ref{concl}. We keep the
main body of the paper free of technicalities and refer the reader interested
in technical details to the appendices.\bigskip

\boldmath
\section{Isospin-violation in hadronic \texorpdfstring{$B$}{B} decays}\label{qual}

\subsection{The \texorpdfstring{$B\to\pi K$}{B to Pi,K} modes}\label{BpiK}
\unboldmath

The $B\to \pi K$ decays are dominated by the isospin-conserving QCD penguin
amplitude. Nevertheless, they receive small contributions from the tree and the
EW penguin amplitude, which are isospin-violating. Combining
measurements of the four different decay modes $B^-\to \pi^-\bar{K}^0$, $B^-\to
\pi^0K^-$, $\bar{B}^0\to \pi^+K^-$ and $\bar{B}^0\to \pi^0\bar{K}^0$, it is
possible to construct observables in which the leading contribution from the QCD
penguin drops out, so that they are sensitive to isospin violation.\medskip

The mesons participating in $B\to \pi K$ decays transform under isospin
rotations as
\be\label{g1}
   (\bar{B}^0,-B^-)_{1/2}\,,\hspace{1cm}
   (\bar{K}^0,-K^-)_{1/2}\,,\hspace{1cm}
   (\pi^+,-\pi^0,-\pi^-)_1\,.
\ee
Furthermore we can assign isospin to the operators appearing in the
effective Hamiltonian
\be\label{g1a}
{\cal H}_{\rm eff} = \frac{G_F}{\sqrt{2}}
\sum_{p=u,c} \lambda_{p}^{(s)} \left(C_1
Q_1^p+C_2Q_2^p+\sum_{i=3}^{10} C_i Q_i
+C_{7\gamma}Q_{7\gamma}+C_{8g}Q_{8g}\right)+\hc,
\ee
which mediates the $B\to \pi K$ transitions. Here $\lambda_{p}^{(s)}=V_{pb}V_{ps}^*$
represents a product of elements of the quark mixing (CKM) matrix, $Q_{1,2}^{p}$ are
the so-called current-current operators, $Q_{3,\ldots,6}$ are QCD penguin operators,
$Q_{7\gamma}$ and $Q_{8g}$ represent the electromagnetic and chromomagnetic operators and
\begin{align}
 Q_7 = (\bar s_{\alpha} b_{\alpha})_{V-A}  \textstyle{ \sum_q\frac{3}{2}}e_q (\bar q_{\beta} q_{\beta})_{V+A}, \qquad & & Q_8 &= (\bar s_{\alpha} b_{\beta})_{V-A}  \textstyle{ \sum_q\frac{3}{2}}e_q (\bar q_{\beta} q_{\alpha})_{V+A},\\
 Q_9 = (\bar s_{\alpha} b_{\alpha})_{V-A}  \textstyle{ \sum_q\frac{3}{2}}e_q (\bar q_{\beta} q_{\beta})_{V-A}, \qquad & & Q_{10} &= (\bar s_{\alpha} b_{\beta})_{V-A}  \textstyle{ \sum_q\frac{3}{2}}e_q (\bar q_{\beta} q_{\alpha})_{V-A},
\end{align}
are the EW penguin operators ($\alpha,\beta$ denote colours). The latter are of great
importance for our work. We define the operators as in \cite{Beneke:2001ev} so
that $C_1(M_W)=1$ at leading order. Containing $\bar{u}u$- and
$\bar{d}d$-bilinears, the operators $Q_1$,...,$Q_{10}$ can be distributed among
\be\label{g2}
   \mathcal{H}_{\textrm{eff}}\,=\, \mathcal{H}_{\textrm{eff}}^{\Delta I=1}\,+\,
                                  \mathcal{H}_{\textrm{eff}}^{\Delta I=0}
\ee
according to the decomposition $1/2\otimes 1/2=1\oplus 0$ \cite{Neubert:1997wb}.
Since the QCD penguin operators $Q_{3,\ldots,6}$ involve the isosinglet combination
$(\bar{u}u+\bar{d}d)$, they contribute solely to
$\mathcal{H}_{\textrm{eff}}^{\Delta I=0}$ whereas the other operators give
contributions to both parts of $\mathcal{H}_{\textrm{eff}}$. The $B\to \pi K$
decays thus follow the isospin pattern
\be\label{g3}
   1/2\,\stackrel{\Delta I=1,0}{\longrightarrow}\,1\otimes 1/2\,=\,3/2\oplus
1/2,
\ee
implying that all four decay amplitudes can be decomposed into three independent
isospin amplitudes, $\mathcal{A}^{\Delta I=0}_{1/2}$, $\mathcal{A}^{\Delta
I=1}_{1/2}$ and $\mathcal{A}^{\Delta I=1}_{3/2}$ with the lower index denoting
the total isospin of the final state.

One finds that $B\to \pi K$ is dominated by the QCD penguin contribution and
thus $|\mathcal{A}^{\Delta I=0}_{1/2}|\gg |\mathcal{A}^{\Delta
I=1}_{3/2,\,1/2}|$. To a first approximation, all the decay modes can be
described by the amplitude $\mathcal{A}^{\Delta I=0}_{1/2}$ only, dictating the
relative size of the branching fractions to be $1:2:1:2$ (in the same
order as in tab.~\ref{tab1}).\medskip

\begin{figure}
   \includegraphics[width=1.0\textwidth]{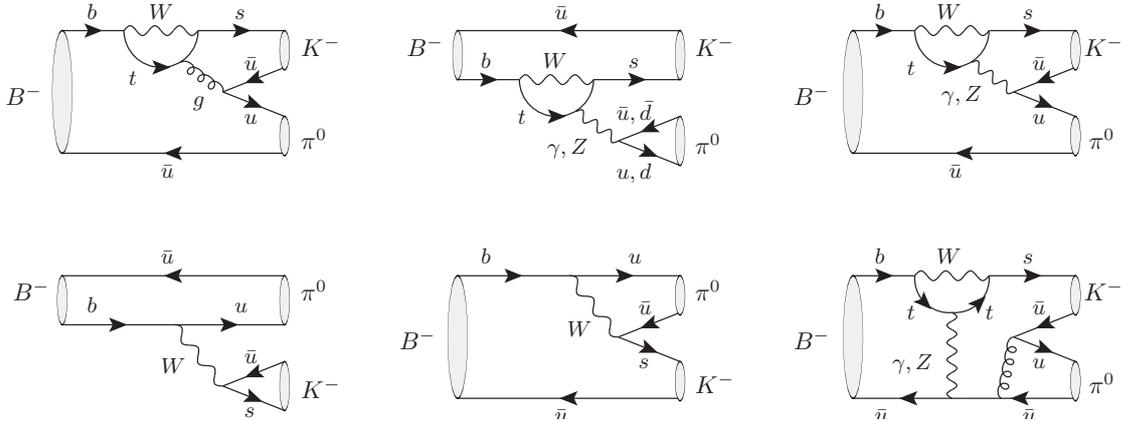}
   \caption{\label{fig:BKPi} Diagrams representing the topological
parameterisation in eq.~(\ref{g5}) for $B^-\to \pi^0 K^-$. First line from left to
right: QCD penguin ($P$), colour-allowed EW penguin ($\rEW$),
colour-suppressed EW penguin ($\rEWC$). Second line from left to right:
colour-allowed tree ($\rT$), colour-suppressed tree ($\rC$), EW penguin
annihilation ($\rEWA$).}
 \end{figure}

The isospin-invariant amplitudes receive contributions from various SM quark
diagrams. It is only at the level of these diagrams that the pattern of CP
violation can be correctly implemented, i.e.\ that the amplitudes
$\mathcal{A}^{\Delta I=0}_{1/2}$, $\mathcal{A}^{\Delta I=1}_{3/2,\,1/2}$
can be related to their CP-conjugated counterparts
$\overline{\mathcal{A}}^{\Delta I=0}_{1/2}$, $\overline{\mathcal{A}}^{\Delta
I=1}_{3/2,\,1/2}$. This suggests an alternative parameterisation of the
amplitudes in terms of the topologies of the underlying quark-level transitions
\cite{Zeppenfeld:1980ex,Gronau:1994rj}:
\bea\label{g5}
{\cal A}(B^-\to  \pi^- \bar{K}^0)&\simeq& P
\left(1 - \frac{1}{3}r^C_{\rm EW}+\frac{2}{3}r^{\rm A}_{\rm EW} \right)
,\nonumber\\
\sqrt{2}\,{\cal A}(B^-\to \pi^0 K^-)&\simeq& P
\left(1+r_{\rm EW} +\frac{2}{3}r^C_{\rm EW}+\frac{2}{3}r^{\rm A}_{\rm EW}
-(r_{\rm T}+r_{\rm C})e^{-i \gamma}\right),\nonumber\\
{\cal A}(\bar{B}^0\to \pi^+ K^-)&\simeq& P
\left(1 +\frac{2}{3}r^C_{\rm EW}-\frac{1}{3}r^{\rm A}_{\rm EW} - r_{\rm T} e^{-i
\gamma}\right),\nonumber\\
\sqrt{2}\,{\cal A}(\bar{B}^0\to \pi^0 \bar{K}^0)&\simeq&-\,P
\left(1 -r_{\rm EW} - \frac{1}{3}r^C_{\rm EW}-\frac{1}{3}r^{\rm A}_{\rm EW} +
r_{\rm C}e^{-i \gamma}\right).
\eea
This topological parameterisation is illustrated by the corresponding Feynman
diagrams for $B^-\to \pi^0 K^-$ in fig.~\ref{fig:BKPi}. In eq.~(\ref{g5}) we
have factored out the dominant QCD penguin amplitude $P$ and neglected penguin
amplitudes suppressed by $|V_{us}^*V_{ub}|/|V_{cs}^*V_{cb}|$. The dependence on
the weak CKM phase $\gamma$ has been made explicit, while strong phases are
contained in the ratios $r_i$ which fulfill $|r_i|<1$. These quantities denote
corrections from different types of Feynman diagrams: $\rT$ and $\rC$ stem from
colour-allowed and colour-suppressed tree diagrams, $\rEW$ and $\rEWC$ from
colour-allowed and colour-suppressed EW penguins, respectively.
Annihilation via QCD penguin diagrams is absorbed into $P$ whereas weak
annihilation via EW penguin diagrams is parameterised by $\rEWA$ and
colour-suppressed tree annihilation is neglected. With our QCDF setup explained
in Appendix~\ref{hadr} and the expressions for the ratios given in Appendix~\ref{IsoRatios} we obtain
\bea\label{g6}\nn
    \rT&=&0.17^{+0.07}_{-0.06}\,+\,0.03^{+0.03}_{-0.10}\,i\,, \\ \nn
    \rC&=&0.07^{+0.04}_{-0.06}\,+\,(-0.01)^{+0.03}_{-0.05}\,i\,, \\ \nn
    \rEW&=&\,0.13^{+0.05}_{-0.05}\,+\,0.02^{+0.02}_{-0.07}\,i\,, \\ \nn
    \rEWC&=&0.04_{-0.03}^{+0.02}\,+\,(-0.01)_{-0.03}^{+0.02}\,i\,, \\
    \rEWA&=&0.007_{-0.010}^{+0.002}\,+\,(-0.004)_{-0.003}^{+0.011}\,i\,.
\eea
The result displays the typical features of QCDF predictions, namely small
strong phases and large uncertainties of colour-suppressed topologies. The
smallness of the $r_i$ reflects the domination of the isospin-conserving QCD
penguin and justifies the expansion of physical observables in the $r_i$. Among
the isospin-violating contributions the colour-allowed tree gives the largest
corrections followed by the EW penguin which dominates over the
colour-suppressed tree. The colour-suppressed EW penguin ratio $\rEWC$ and
especially the EW penguin annihilation ratio $\rEWA$ are quite small and
consequently they have been omitted in most analyses of $B\to \pi K$ decays. In
particular, the possibility of having NP in the EW penguin annihilation
amplitude $\rEWA$ has to our knowledge not been considered so far. However,
we want to point out that such an approximation is not valid in the analysis
of CP asymmetries: non-vanishing direct CP asymmetries are caused by the
interference of parts of the decay amplitude with different weak and strong
phases. Consequently direct CP asymmetries in $B\to \pi K$ cannot be generated
by the QCD penguin amplitude alone and are automatically sensitive to subleading
contributions, encoded in the imaginary parts of the $r_i$ coefficients. These,
in turn, are generated in QCDF either perturbatively at $\mathcal{O}(\alpha_s)$ or
non-perturbatively at $\mathcal{O}(\Lambda_{\textrm{QCD}}/m_B)$. At
$\mathcal{O}(\alpha_s)$ the colour-suppression of $\rEWC$ is not present anymore
and the $\Lambda_{\textrm{QCD}}/m_B$\,-\,suppressed $\rEWA$ can compete as well.
Therefore we keep $\rEWC$ and $\rEWA$ in our calculation and we will see
in later chapters that we can indeed have a large NP contribution in these
amplitudes.\medskip
\begin{table}[t]
  \centering
  \begin{tabular}{|l|r|r|}
    \hline
    Observable & Theory & Experiment \\
    \hline
    Br$(\bar{B}^0\to\pi^0 \bar{K}^0) \cdot 10^{6}$            & $5.8^{+5.7}_{-3.6}$ &
$9.5^{+0.5}_{-0.5}$ \\
    Br$(\bar{B}^0\to\pi^+ K^-)\cdot 10^{6}$            & $14.0^{+12.1}_{-7.8}$   &
$19.4^{+0.6}_{-0.6}$ \\
    Br$(B^-\to\pi^0 K^-)\cdot 10^{6}$                  & $9.6^{+7.3}_{-4.9}$ &
$12.9^{+0.6}_{-0.6}$ \\
    Br$(B^-\to\pi^- \bar{K}^0)\cdot 10^{6}$                  & $15.7^{+13.7}_{-8.9}$  &
$23.1^{+1.0}_{-1.0}$ \\
    $R_c^B(\pi K)$                                 & $1.22^{+0.17}_{-0.15}$ &
$1.12^{+0.07}_{-0.07}$ \\
    $R_n^B(\pi K)$                                 & $1.22^{+0.18}_{-0.16}$ &
$1.02^{+0.06}_{-0.06}$ \\
    $R_c^K(\pi K)$                               & $1.27^{+0.12}_{-0.11}$ &
$1.24^{+0.07}_{-0.07}$ \\
    $R_n^K(\pi K)$                               & $1.27^{+0.13}_{-0.15}$ &
$1.13^{+0.08}_{-0.07}$ \\
    $R_c^{\pi}(\pi K)$                           & $1.04^{+0.10}_{-0.08}$ &
$1.11^{+0.06}_{-0.06}$ \\
    $R_n^{\pi}(\pi K)$                           & $1.55^{+0.38}_{-0.31}$ &
$1.26^{+0.09}_{-0.09}$ \\
    $R(\pi K)$                                   & $1.02^{+0.02}_{-0.02}$ &
$1.05^{+0.05}_{-0.05}$ \\
    $A_{\rm CP}(\bar{B}^0\to\pi^0 \bar{K}^0)$    & $-0.003^{+0.057}_{-0.108}$ &
$-0.01^{+0.10}_{-0.10}$ \\
    $A_{\rm CP}(\bar{B}^0\to\pi^+ K^-)$    & $-0.047^{+0.187}_{-0.047}$ &
$-0.098^{+0.012}_{-0.011}$ \\
    $A_{\rm CP}(B^-\to\pi^0 K^-)$          & $ -0.028^{+0.221}_{-0.059}$ &
 $0.050^{+0.025}_{-0.025}$ \\
    $A_{\rm CP}(B^-\to\pi^- \bar{K}^0)$          & $ 0.003^{+0.012}_{-0.003}$ &
$0.009^{+0.025}_{-0.025}$ \\
    $\Delta A_{\rm CP}=\Delta A_{\rm CP}^-$      & $0.019^{+0.058}_{-0.048}$ &
$0.148^{+0.027}_{-0.028}$ \\
    $\Delta A_{\rm CP}^0$                        & $0.006^{+0.118}_{-0.057}$ &
$0.019^{+0.103}_{-0.103}$ \\
    $S_{\rm CP}(\bar{B}^0\to\pi^0 \bar{K}^0)$    & $0.80^{+0.06}_{-0.08}$ &
$0.57^{+0.17}_{-0.17}$ \\
    \hline
  \end{tabular}
  \caption{Theoretical vs.\ experimental results for $\bar{B}\to \pi \bar{K}$ decays.
The experimental data is taken from \cite{Barberio:2008fa}. The original results can be found in
\cite{Asner:1995hc,Chen:2000hv,Bornheim:2003bv,Chang:2004um,Aubert:2006fha,Aubert:2006gm,Abe:2006xs,Abe:2006qx,Morello:2006pv,Aubert:2007hh,Aubert:2008bj,:2008zza,Aubert:2008sb,babar:2008se,Fujikawa:2008pk}.}\label{tab1}
\end{table}

One can easily see from eqs.~(\ref{g5},\ref{g6}) that the two amplitudes
involved in $\Delta A_\text{CP}$ differ only by the subdominant contributions
$\rC$, $\rEW$ and $\rEWA$, all of which are isospin-violating.
Turning to the CP asymmetries, one finds in the SM
\bea\label{g7}\nn
A_{\rm CP}(B^-\to\pi^0 K^-)&\simeq&-2\,\IM \left(r_{\rm T}
+r_{\rm C}\right) \sin\gamma, \\
A_{\rm CP}(\bar{B}^0\to\pi^+ K^-)&\simeq&-2\,\IM (r_{\rm T}) \sin\gamma,
\eea
with terms quadratic in the $r_i$ being neglected. Thus the only possible
explanation for a large $\Delta A_{\rm CP}$ in the SM seems to be a large
imaginary part of $\rC$, i.e.\ a large absolute value and large strong phase of
the colour-suppressed tree amplitude, generated by some hadronic
effects at the low scale $\Lambda_\text{QCD}$ which can hardly be calculated
perturbatively. However, QCDF predicts only a small
$\textrm{Im}(\rC)$, insufficient to explain the data, even when all the theory
uncertainties are included. Therefore one is tempted to conclude that the
discrepancy in $\Delta A_{\rm CP}$ is not due to our lack of understanding of
strong interactions but due to isospin-violating NP.\medskip

For this reason, $\Delta A_{\rm CP}$ has been studied in various NP models in
recent publications
\cite{Fleischer:2007mq,Buras:2005cv,Gronau:2006xu,Gronau:2006ha,Fleischer:2007hj,Kim:2007kx,Feldmann:2008fb,Fleischer:2008wb,Baek:2009pa,Baek:2009hv}.
The main ingredient of these analyses is usually an enhancement of the EW penguin
topologies by effects of virtual heavy particles. Such contributions can be
included into the amplitudes (\ref{g5}) by the replacements
\be\label{g8}
r_\text{EW}  \rightarrow r_\text{EW} + \tilde{r}_\text{EW}
e^{-i\delta},\hspace{1.0cm}
r_\text{EW}^{\textrm{C}}  \rightarrow r_\text{EW}^{\rm C} +
\tilde{r}_\text{EW}^{\rm C} e^{-i\delta}, \hspace{1.0cm}
r_\text{EW}^{\textrm{A}}  \rightarrow r_\text{EW}^{\rm A} +
\tilde{r}_\text{EW}^{\rm A} e^{-i\delta},
\ee
where $\delta$ is a new weak phase and $\tilde{r}_\text{EW}^{(i)}$
are complex numbers including a strong phase. The CP asymmetries
then become
\bea\label{g9}\nn
A_{\rm CP}(B^-\to\pi^0 K^-)&\simeq&-2\,\IM \left(r_{\rm T}
+r_{\rm C}\right) \sin\gamma + 2\,\IM\left(\tilde{r}_\text{EW}
+\frac{2}{3}\tilde{r}_\text{EW}^\textrm{C}
+\frac{2}{3}\tilde{r}_\text{EW}^\textrm{A}\right)\sin\delta, \\
A_{\rm CP}(\bar{B}^0\to\pi^+ K^-)&\simeq&-2\,\IM (r_{\rm T})\sin\gamma
+2\,\IM \left(\frac{2}{3}\tilde{r}_\textrm{EW}^\textrm{C}
-\frac{1}{3}\tilde{r}_\textrm{EW}^\textrm{A} \right)\sin\delta,
\eea
such that
\be\label{g10}
\Delta A_{\rm CP}\,\simeq\,-2\,\IM \left(r_{\rm C}\right) \sin\gamma +
2\,\IM\left(\tilde{r}_\text{EW} +
\tilde{r}_\text{EW}^\textrm{A}\right)\sin\delta
\ee
can turn out to be much larger than in the SM. The observed discrepancy can be
solved by a $\tilde{r}_\text{EW}$ or a $\tilde{r}_\text{EW}^A$ term comparable
in size to the corresponding SM term $r_\text{EW}$.\medskip

Apart from $\Delta A_{\rm{CP}}$ one can also construct other observables from
the $B\to\pi K$ data which are sensitive to isospin violation, for example
certain ratios of branching fractions. Even though tensions with experimental
data in these observables raised the formulation of a ''$B\to\pi K$ puzzle'' in
the first place
\cite{Buras:2003yc,Buras:2003dj,Mishima:2004um,Buras:2004ub,Buras:2004th,Kim:2005jp,Nandi:2004dx},
in the meantime these quantities are in reasonable agreement with
the SM predictions. However, they serve as important constraints for NP in EW
penguins and we define and discuss them in Appendix~\ref{IsoRatios}. Note in
particular that the quantity $\Delta A_\text{CP}^0$ defined there, which is
the difference of the two remaining CP asymmetries not appearing in
$\Delta A_\textrm{CP}$, probes the same combination of $\IM(\rtEW)$ and
$\IM(\rtEWA)$ as $\Delta A_\text{CP}$. Unfortunately, data on
$A_{\rm CP}(B^-\to \pi^-\bar{K}^0)$ and especially on
$A_{\rm CP}(\bar{B}^0\to  \pi^0\bar{K}^0)$ are not good enough yet to gain
any information from these observables. Experimental results and SM predictions
for the $B\to\pi K$ observables are given in tab.~\ref{tab1}.\medskip

The main problem which makes it difficult to single out a possible NP
contribution in $B\to\pi K$ decays is evident from (\ref{g5}): the colour-allowed
EW penguin contributions and colour-suppressed tree contributions enter the
amplitudes in (\ref{g5}) exclusively in the combination
\begin{equation}\label{eq:ParaCombi}
    \rEW\,-\,\rC\,e^{-i\gamma}.
\end{equation}
 This implies that colour-allowed EW penguins and colour-suppressed trees are
inextricably linked with each other, reflecting the fact that the topological
parameterisation contains some redundancy. Physical effects found in any
experiment cannot unambiguously be attributed to one or the other partner of
this topology pair. A new EW penguin contribution $\tilde{r}_{\textrm{EW}}
e^{-i\delta}$ can be probed only in one of the four physical combinations
\begin{eqnarray}\label{eq:rtEWrCLink}
   &&\RE(\rtEW)\cos\delta\,-\,\RE(\rC)\cos\gamma\,+\,\RE(\rEW)\,,\nonumber\\
   &&\IM(\rtEW)\sin\delta\,-\,\IM(\rC)\sin\gamma\,,\nonumber\\
   &&\IM(\rtEW)\cos\delta\,-\,\IM(\rC)\cos\gamma\,+\,\IM(\rEW)\,,\nonumber\\
   &&\RE(\rtEW)\sin\delta\,-\,\RE(\rC)\sin\gamma\,.
\end{eqnarray}
Therefore, probing $\tilde{r}_\text{EW}e^{-i\delta}$ is challenged by the large
hadronic uncertainties in the QCDF prediction for $\rC$, which can mimic or hide
such a NP signal. One possible way to constrain $r_C$ is the approximate $SU(3)$ flavour symmetry which relates it to a corresponding $B\to\pi\pi$ topology. Using this symmetry it has been found that current data on $CP$ violation in $\bar B^0\to\pi^0 K_S$ is also in disagreement with the SM, independently of $\Delta A_{CP}$, and can be explained by adding $\tilde r_\text{EW}$ to the amplitude \cite{Fleischer:2008wb}.\medskip

The perspective of our work is the following: In order to find out whether the
$\Delta A_{\rm CP}$ discrepancy really is a manifestation of isospin-violating
physics beyond the SM, one should also study other observables on which such a
kind of NP could have a large impact and see whether similar effects appear in
measurements of these observables. Our proposal in this work
is to test the hypothesis of isospin-violating NP by looking at processes which
are highly sensitive to it, namely \textit{purely isospin-violating $B_s$
decays}.\bigskip

\boldmath
\subsection{Purely isospin-violating \texorpdfstring{$B_s$}{Bs} decays}\label{isospinviolating}
\unboldmath

EW penguin contributions to hadronic $B$ decays are usually overshadowed by the
larger QCD penguins. This problem can be avoided if one succeeds in probing
exclusively the $\Delta I=1$ part of the effective Hamiltonian which is
orthogonal to the QCD penguin operators. To achieve this for $B\to \pi K$, we
had to single out the $\Delta I=1$ part of the transition in eq.~(\ref{g3}) by
combining different isospin-related decay modes, for example by considering the
observable $\Delta A_{\textrm{CP}}$. Our proposal now is to consider decays to
which QCD penguins do not contribute at all, i.e.\ pure $\Delta I=1$ decays,
where no such procedure is needed.\medskip

There are no two-body decays of the $B^0$ or $B^{\pm}$ meson with this property.
In these cases the final state would have to be a pure $|3/2,\pm 1/2\rangle$
isospin state which cannot be constructed out of two mesons. The $B_s$ meson, on
the other hand, is an isosinglet and it can decay as
\be\label{g11}
   0\,\stackrel{\Delta I=1}{\longrightarrow}\,0\otimes 1\,=\,1\,.
\ee
The final state must consist of an isospin triplet, i.e.\ $\pi^0$ or $\rho^0$,
and an isosinglet, i.e.\ a meson with the  flavour structure $s\bar{s}$. In order
to avoid complications stemming from \mbox{$\eta-\eta'$-mixing}, we restrict ourselves
to the vector-meson $\phi$ which is to a good approximation a pure $s\bar{s}$
state. This leaves us with the two $\Delta I=1$ channels
\begin{center}
\fbox{ $\bar{B}_s \to \phi \rho^0 \qquad \text{and} \qquad \bar B_s \to \phi
\pi^0$.}
\end{center}\medskip

So far only an upper limit $\Br(\bar{B}_s \to \phi \rho^0)\leq 6.17\cdot 10^{-4}$
exists \cite{Abe:1999ze} and no detailed theory analysis of
$\bar{B}_s\to\phi\rho^0,\phi\pi^0$ has been published. Only the SM
branching fractions and CP asymmetries have been calculated in general surveys
on $B$ decays to light mesons \cite{Beneke:2003zv,Beneke:2006hg}. In addition,
$\bar{B}_s \to \phi \pi^0$ has been suggested as a tool to measure $\gamma$ via the mixing-induced CP asymmetry \cite{Fleischer:1994rs}. Since
in the era of LHCb and super B-factories these two processes will become
interesting objects for tests of isospin-violation and potential NP we will in
the following study their phenomenology in full detail, in the SM and
beyond.\medskip

\begin{figure}[t]
  \includegraphics[width=4.8cm]{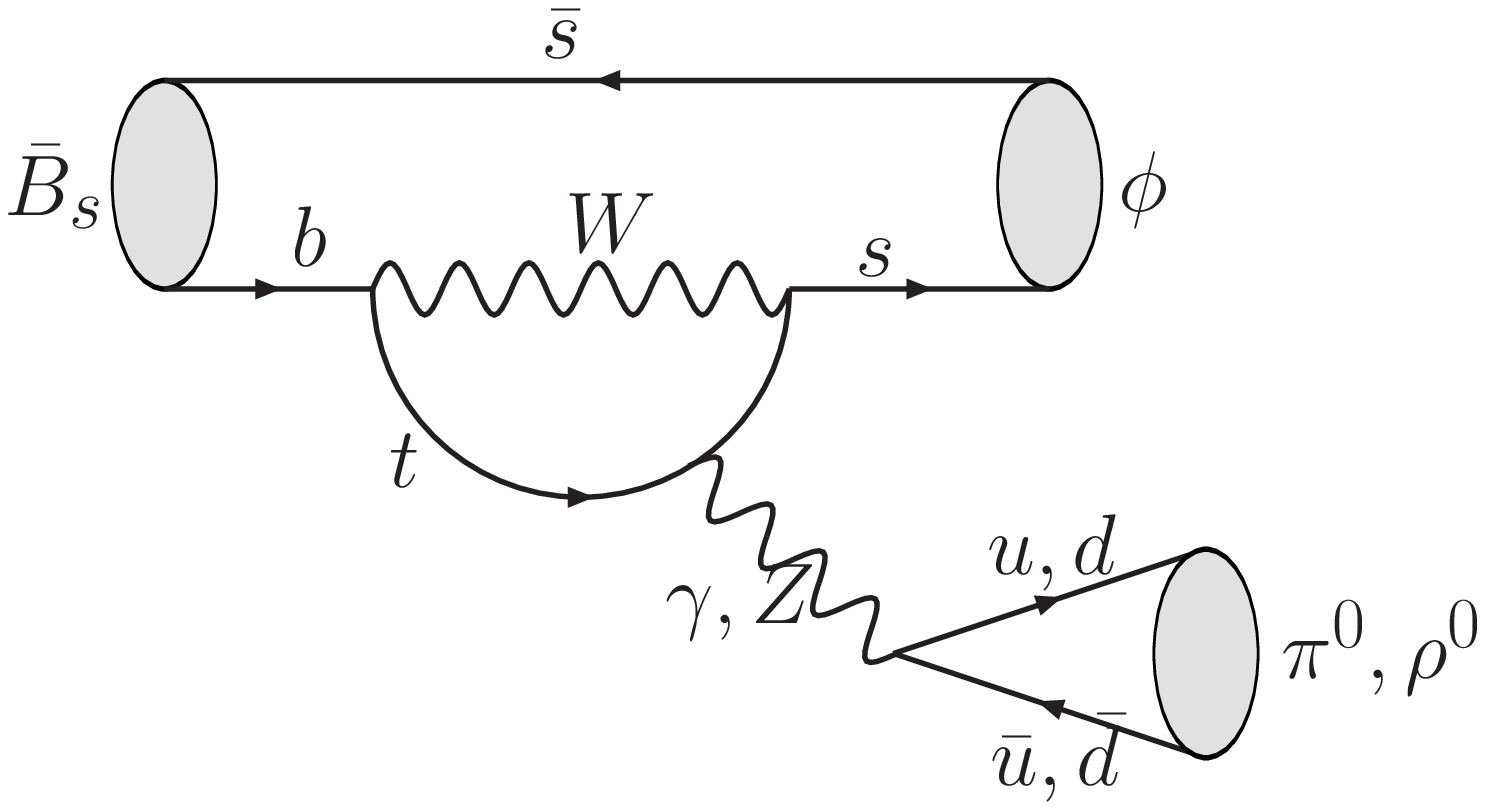} \quad
  \includegraphics[width=4.8cm]{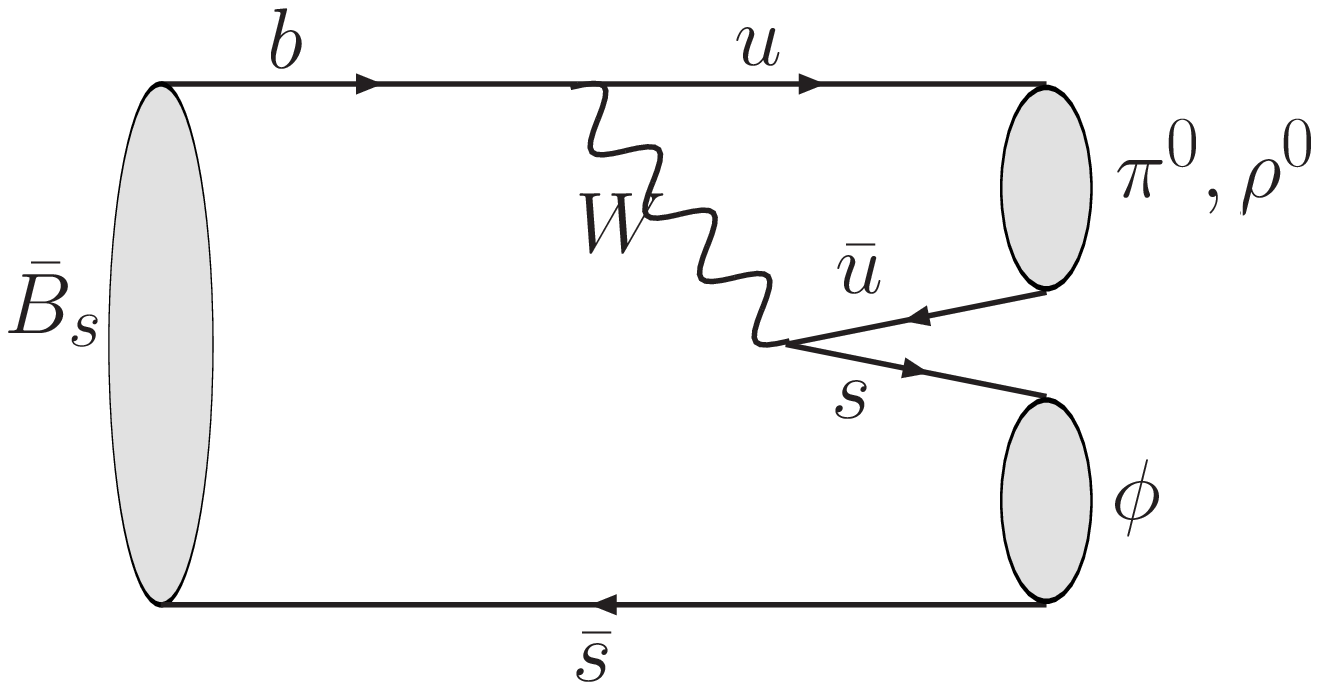} \quad
 \includegraphics[width=4.8cm]{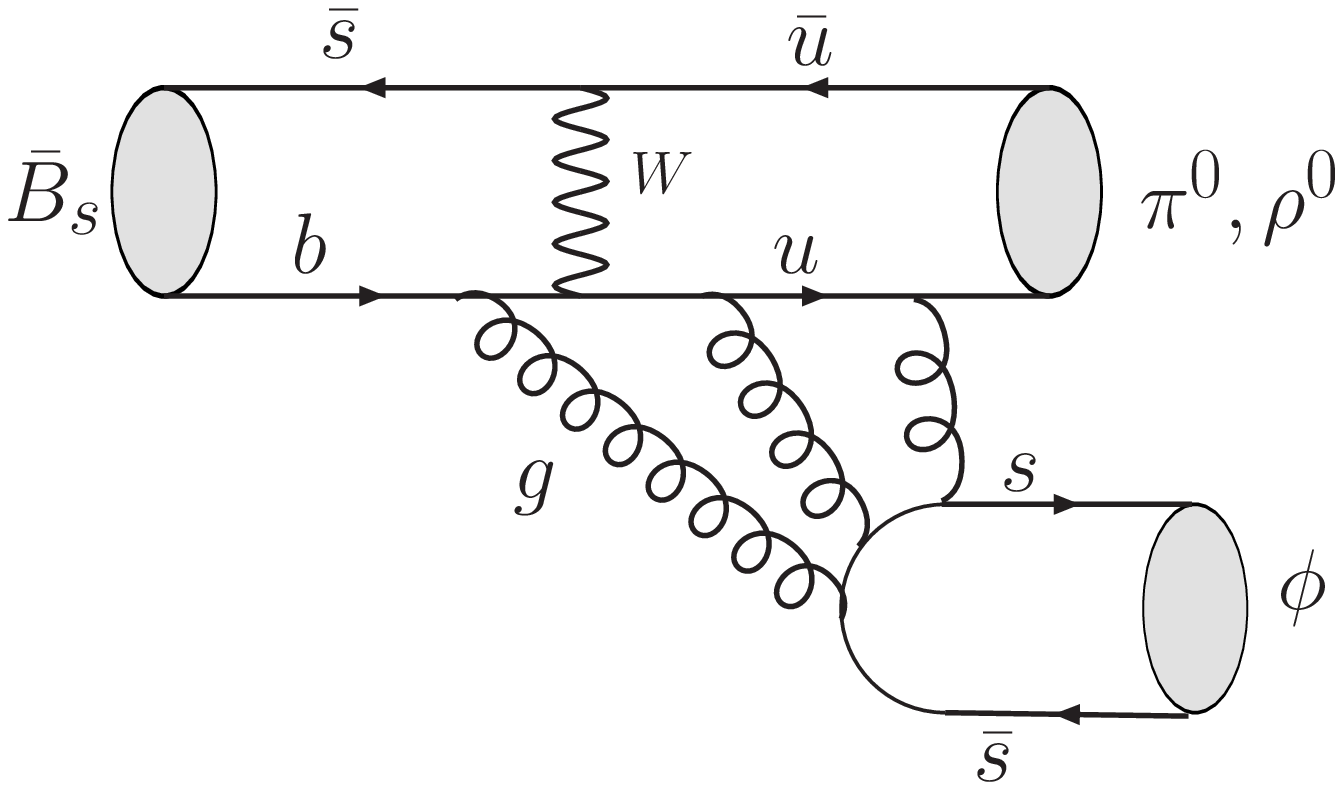}
  \caption{penguin, tree and annihilation topologies contributing to $\bar
B_s\rightarrow \phi\rho^0,\phi\pi^0$}
  \label{ampsphipi}
\end{figure}

\noindent In the SM only three basic topologies are present in these decays,
depicted in fig. \ref{ampsphipi}:
\begin{itemize}
 \item EW penguins
 \item CKM- and colour-suppressed tree diagrams
 \item Singlet-annihilation diagrams.
\end{itemize}\medskip

Since the flavour-structure of $\pi^0$ and $\rho^0$ excludes their production via
gluon-exchange, annihilation can only contribute if the $\phi$ meson (the flavour
singlet) in the final state is produced from gluons and the second meson comes
from weak (as depicted in fig. \ref{ampsphipi}) or electromagnetic interactions.
Since the $\phi$ is colour-neutral and it is odd under charge-conjugation,
at least three gluons are needed, so that the singlet-annihilation amplitude is
formally of higher-order in $\alpha_s$ and does not appear in QCDF
at the next-to-leading order $\mathcal O(\alpha_s^1)$ \cite{Beneke:2003zv}.
However, annihilation topologies in general do not factorise and cannot be
calculated perturbatively, because the exchanged gluons may be soft. This means
that we can, from a theoretical point of view, only rely on the suppression of
these contributions by $\Lambda/m_B$, where $\Lambda$ is a non-perturbative
scale, and by $1/N_c$. This leads to the expectation that both the tree and the
EW penguin amplitudes can receive corrections of $10\%-20\%$ from
singlet-annihilation. However, we can also argue from a phenomenological point
of view that $\phi$-production from three gluons is suppressed by the OZI rule
\cite{OZI,Okubo:1963fa,Iizuka:1966wu,Iizuka:1966fk} and should thus be only a
small effect, even though this rule is theoretically not well understood. In
short, our reasoning leads us to the conclusion that in order to test NP in
$\bar{B}_s\to\phi\rho^0,\phi\pi^0$, we have to look for new effects which are
much larger than this intrinsic uncertainty.\medskip

In all our calculations of $\bar{B}_s\to\phi\rho^0,\phi\pi^0$ we use the full
QCDF decay amplitudes, see refs.~\cite{Beneke:2003zv,Beneke:2006hg}. However, since
these are quite involved, we now quote simple approximative formulas
which can be used as building blocks for an easy calculation of various
observables such as branching fractions, CP asymmetries and polarisation
fractions. Neglecting singlet-annihilation we can parameterise the amplitudes in analogy to
eq.~(\ref{g5}) as
\be\label{g13}
\sqrt{2}\mathcal A(\bar B_s \to \phi M_2)\, =\,
P_{\rm EW}^{M_2} \left( 1 - r_{\textrm{C}}^{M_2} e^{- i \gamma} \right),
\ee
with $M_2$ representing a $\pi^0$, a longitudinal $\rho^0$ or a $\rho^0$ with
negative helicity. The positive helicity amplitude can be neglected in the SM
because of its $\Lambda_{\textrm{QCD}}^2/m_B^2$\,--\,suppression. We have
factored out the EW penguin amplitude $P_{\rm EW}^{M_2}$ anticipating its
dominance over the colour-suppressed tree represented by the tree-to-penguin
ratio $r_{\textrm{C}}^{M_2}$. A new contribution to the $B\to \pi K$ amplitudes
of the form (\ref{g8})  would also enter the $\bar B_s\to\phi\pi^0,\phi\rho^0$ amplitude
(\ref{g13}) modifying it as
\begin{equation}\label{eq:BsPhiPiNP}
   \sqrt{2}\,A(\bar{B}_s\to\phi
M_2)\,=\,P_{\textrm{EW}}^{M_2}\,\left(1\,-\,r^{M_2}_{\textrm{C}}\,e^{-i\gamma}\,
+\,\tilde{r}_{\textrm{EW}}^{M_2}\,e^{-i\delta}\right)\,
\end{equation}
where $\tilde{r}_{\textrm{EW}}^{M_2}$ contains a strong phase and $\delta$ is
the weak phase introduced in (\ref{g8}).
If we assume the new contribution to be of the order of the SM EW penguin, as
required by a solution of the ``$\Delta A_{\textrm{CP}}$-puzzle'', we have
$|\tilde{r}_{\textrm{EW}}^{M_2}|\sim\mathcal{O}(1)$ and expect a large
enhancement of the $\bar B_s\to\phi\pi^0,\phi\rho^0$ branching fractions, up to an order
of magnitude. In order to obtain the same effect within the SM one would have to
assume an even larger enhancement of the soft non-perturbative physics entering
the colour-suppressed tree topology in $r_{\textrm{C}}^{M_2}$.\medskip

\noindent
Choosing a phase convention such that $P^{M_2}_{\textrm{EW}}$ is real, we find
\begin{equation}\label{eq:NumPEW}
   P_{\textrm{EW}}^{\pi}\,=\,6.45^{+1.87}_{-0.54}\cdot 10^{-9},\hspace{1cm}
   P_{\textrm{EW}}^{\rho,0}\,=\,9.95^{+2.83}_{-0.79}\cdot 10^{-9},\hspace{1cm}
   P_{\textrm{EW}}^{\rho,-}\,=\,4.27^{+1.34}_{-0.81}\cdot 10^{-9},
\end{equation}
for the isotriplet meson being $\pi^0$, longitudinal $\rho^0$ and  $\rho^0$ with
negative helicity, respectively. We further have
\begin{eqnarray}\label{eq:NumRc}
   r^{\pi}_{\textrm{C}}&=& 0.41_{-0.41}^{+0.37}\,-\,0.13_{-0.30}^{+0.30}\,i
\,,\nonumber\\[0.2cm]
   r^{\rho,0}_{\textrm{C}}&=&
0.39_{-0.39}^{+0.35}\,-\,0.13_{-0.29}^{+0.28}\,i\,, \nonumber\\[0.2cm]
   r^{\rho,-}_{\textrm{C}} &=&
0.21_{-0.46}^{+0.49}\,+\,0.15_{-0.45}^{+0.45}\,i\,.
\end{eqnarray}
Inserting these numbers into eq.~(\ref{g13}) we obtain a good approximation of
the SM amplitudes for the $\bar{B}_s$ decays. Replacing $\gamma \rightarrow
-\gamma$ in eq.~(\ref{g13}) yields the corresponding CP-conjugated amplitudes
($B_s$ decays). Subsequently one can use the formulas in Appendix~\ref{Observables}
to convert the amplitudes into physical observables. In section \ref{sec:ModIndPara}
we extend these prescriptions to physics beyond the
SM.\medskip

One should keep in mind that the numbers above are calculated using
state-of-the-art values for the non-perturbative input parameters, summarised in
Appendix~\ref{input}. They are based on lattice QCD, QCD sum rules and
experimental data. Since our knowledge on these parameters is hopefully going to
improve in the future it is desirable to have an additional parameterisation of
the decay amplitudes where the non-perturbative input can be changed. We find
the dominant sources of theory uncertainties to be (ordered by importance)
\begin{itemize}
\item the form factors $A_0^{B_s\rightarrow\phi}(0)$ and
$F_\pm^{B_s\rightarrow\phi}(0)$,
\item the CKM angle $\gamma$,
\item the non-factorisable spectator-scattering amplitudes, parameterised by the
complex number $X_H$ and the first inverse moment $\lambda_{B_s}$ of the
$B_s$-meson light-cone distribution amplitude.
\end{itemize}
The remaining uncertainties, stemming from decay constants, Gegenbauer moments,
quark masses and CKM parameters, are much less important so we do not need to
display them explicitly. Setting the less important theory parameters to their
default values we arrive at the following approximate expressions for the
quantities in eqs.~(\ref{eq:NumPEW},\ref {eq:NumRc}):
\begin{align}\nonumber
 P_{\textrm{EW}}^{\pi}& \,=\, & 17.0 \, A_0^{B_s\rightarrow\phi}(0)\cdot
10^{-9}, \qquad  r^{\pi}_{\textrm{C}} & \, = \, &  -0.12i  - 0.02  + \frac{0.01
\text{GeV} (1+X_H)}{A_0^{B_s\rightarrow\phi}(0) \lambda_{B_s}}  \\ \nonumber
 P_{\textrm{EW}}^{\rho,0}& \,=\, & 26.2\, A_0^{B_s\rightarrow\phi}(0)\cdot
10^{-9}, \qquad r^{\rho,0}_{\textrm{C}} & \, = \, &  -0.13i  - 0.02  +
\frac{0.01 \text{GeV} (1+X_H)}{A_0^{B_s\rightarrow\phi}(0) \lambda_{B_s}} \\
 P_{\textrm{EW}}^{\rho,-}& \,=\, &  6.6\, F_-^{B_s\rightarrow\phi}(0) \cdot
10^{-9}, \qquad r^{\rho,-}_{\textrm{C}} & \, = \, & 0.14i - 0.06 -
\frac{0.02\text{GeV} (1-X_H)}{ F_-^{B_s\rightarrow\phi}(0) \lambda_{B_s}}.
\label{eq:Approx}
\end{align}
The tree topologies $r^{M_2}_{\textrm{C}}$ suffer from the large
spectator-scattering uncertainties due to a strong cancellation between the
leading order and QCD vertex corrections. Again one can insert these formulas
into eq.~(\ref{g13}), this time with arbitrary values and uncertainties for the
form factors and spectator-scattering parameters, and use the definitions in
Appendix~\ref{Observables} to calculate physical observables. CP conjugation
again amounts to replacement $\gamma\rightarrow -\gamma$. \medskip

We conclude this section quoting our QCDF results for the SM values of the
$\bar B_s\to\phi\pi^0,\phi\rho^0$ observables. As for the CP-averaged branching fractions
we obtain
\begin{equation}\label{eq:BrAvBs}
    \Br(\bar{B}_s\to\phi\pi^0)\,=\,1.6^{+1.1}_{-0.3}\cdot 10^{-7},\hspace{1.5cm}
    \Br(\bar{B}_s\to\phi\rho^0)\,=\,4.4^{+2.7}_{-0.7}\cdot 10^{-7}.
\end{equation}
For comparison we also quote the approximate result according to
(\ref{eq:Approx}):
\begin{equation}\label{eq:BrAvBsappr}
    \Br(\bar{B}_s\to\phi\pi^0)\,=\,1.6^{+1.0}_{-0.3}\cdot 10^{-7},\hspace{1.5cm}
    \Br(\bar{B}_s\to\phi\rho^0)\,=\,4.4^{+2.4}_{-0.7}\cdot 10^{-7}.
\end{equation}
The smallness of the SM branching ratios compared to other hadronic $B$ decays
is due to the absence of QCD penguins and non-suppressed tree-level
contributions. The measurement of these branching fractions is thus challenging
and has not been achieved yet. However, we will show in later chapters that NP in
EW penguins has the chance to enhance the BRs by up to an order of magnitude,
such that this measurement is a very interesting project. We expect that LHCb
will be able to measure $\Br(\bar B_s \rightarrow \phi\rho^0)$ while the $\bar
B_s \rightarrow \phi\pi^0$ mode is more suitable for a super B-factory where a
full reconstruction can cure the notorious difficulties with the identification
of neutral pions. In case of a strong enhancement $\bar B_s \rightarrow \phi\rho^0$ should also be visible in the Tevatron data \cite{punziprivate}. The branching ratio $\Br(\bar{B}_s\to\phi\rho^0)$ is dominated by the longitudinal polarisation state as can be seen in
\begin{equation}\label{eq:BrAv0Bs}
   \Br(\bar{B}_s\to\phi_L\rho^0_L)\,=\,3.7^{+2.5}_{-0.7}\cdot
10^{-7}\,
\end{equation}
and the longitudinal polarisation fraction
\begin{equation}
   f_{\textrm{L}}\,=\,0.84^{+0.08}_{-0.11}\,.
\end{equation}

As stated above, one of the main sources of uncertainty in the QCDF predictions
is the form factor $A_0^{B_s\to\phi}$. It can in principle be eliminated by
considering the ratios
\begin{equation}
      \frac{\Br(\bar B_s \rightarrow \phi\rho^0)}{\Br(\bar B_s \rightarrow
\phi\pi^0)}
       = 2.83^{+0.35}_{-0.23}, \qquad \qquad
       \frac{\Br(\bar B_s \rightarrow \phi_L\rho^0_L)}{\Br(\bar B_s \rightarrow
\phi\pi^0)} = 2.38^{+0.10}_{-0.08}\,.
\end{equation}
NP could still be visible in these ratios because in many scenarios it enters
$\bar B_s \rightarrow \phi\rho^0$ and $\bar B_s \rightarrow \phi\pi^0$ in
different ways. The cancellation of $A_0^{B_s\to\phi}$ also occurs in the ratios
\bea\label{g21}\nn
\frac{\Br(\bar B_s \rightarrow \phi\pi^0)}{\Br(\bar B_s \rightarrow
\phi\phi)}&=&0.007^{+0.008}_{-0.004}, \\
\frac{\Br(\bar B_s \rightarrow \phi\rho^0)}{\Br(\bar B_s \rightarrow
\phi\phi)}&=&0.020^{+0.023}_{-0.010}, \quad
\frac{\Br(\bar B_s \rightarrow \phi_L\rho^0_L)}{\Br(\bar B_s \rightarrow
\phi\phi)}=0.017^{+0.019}_{-0.009}.
\eea
There however this gain is compensated by additional uncertainties arising
from the QCD-penguin-dominated decay $\bar B_s\rightarrow \phi\phi$. The
experimental benefit in these last ratios is that at LHCb absolute branching
ratios cannot be measured because the absolute number of $B_s$ mesons is unknown.
Finally, we find the direct CP asymmetries to be very uncertain:
\be\label{g18}
A_{\rm CP}^\text{dir} (\bar B_s \rightarrow \phi\rho^0) = 0.19^{+0.53}_{-0.61}, \qquad\qquad
A_{\rm CP}^\text{dir} (\bar B_s \rightarrow \phi\pi^0) = 0.27^{+0.50}_{-0.62}.
\ee
Due to the smallness of the branching ratios, these CP asymmetries are also
difficult to access experimentally, therefore we will not consider them any further.\bigskip

\section{Model-independent analysis}\label{modind}

In the previous chapter we proposed to test the hypothesis of NP in the EW
penguin sector, as suggested by the discrepancy in the $B\to \pi K$ observable
$\Delta A_{\textrm{CP}}$, by a measurement of the decays
$\bar B_s\to\phi\pi^0,\phi\rho^0$. In this chapter we support our proposal by a
quantitative analysis pursuing the following strategy: We parameterise NP in EW
penguins in a model-independent way by adding corresponding terms to the Wilson
coefficients $C_7^{(\prime)},...,C_{10}^{(\prime)}$. By performing a
$\chi^2$-fit we determine the NP parameters in such a way that they describe
well the $B\to \pi K$ data. In particular they should allow for a solution of
the $\Delta A_{\textrm{CP}}$ discrepancy. Further hadronic decays like
$B\to \rho K,\pi K^*,\rho K^*$ are used to impose additional constraints at the
$2\,\sigma$ level. With respect to the resulting fit we study the decays
$\bar B_s\to\phi\pi^0,\phi\rho^0$ and quantify a potential enhancement of their branching
fractions. Note that such an exhaustive analysis, correlating different hadronic
decay modes with sensitivity to isospin violation, is only possible if hadronic
matrix elements are calculated from first principles like in the framework of
QCDF. A method based on flavour symmetries, as it has been used in most studies
of $B\to \pi K$ decays so far, could not achieve this. In particular, the decays
$\bar B_s\to\phi\pi^0,\phi\rho^0$, which are our main interest, are not related to any
other decay via $SU(3)_{F}$ so their branching fractions cannot be predicted in
this way.\bigskip

\subsection{Modified EW penguin coefficients}\label{sec:ModIndPara}

In the SM the Wilson coefficients $C_7,...,C_{10}$ obey
the hierarchy $|C_9|\gg |C_7|\gg |C_8|,|C_{10}|$ at the electroweak scale. This
is because $C_9$ receives $1/\sin^2\theta_W$-enhanced contributions from $Z$-penguin
and box diagrams in contrast to $C_7$, while $C_{8,10}$ are generated for the first
time at two-loop level due to their colour structure. For our model-independent
analysis we consider arbitrary NP contributions to the coefficients $C_7$ and
$C_9$ as well as to their mirror counterparts $C_7^{\prime}$ and $C_9^{\prime}$.
Normalizing the new coefficients to the SM value $C_9^{\rm LO}$ defined in
eq.~(\ref{locoeffs}) in the appendix, we have
\be\label{g23}
C_{7,9}^{(\prime)\rm NP}(M_W) = C_9^{\rm LO}(M_W)\,q_{7,9}^{(\prime)},
\qquad \qquad q_{7,9}^{(\prime)}  = |q_{7,9}^{(\prime)}|e^{i
\phi_{7,9}^{(\prime)}},
\ee
where $\phi_{7,9}^{(\prime)}$ are new weak phases. The coefficient $C_9^{\rm
LO}$ contains the parts of $C_9^{\textrm{SM}}$ enhanced by $m_t^2/M_W^2$ and
$1/\sin^2\theta_W$, as explained in Appendix~\ref{heff}. There we also describe
the scheme which we use for the renormalisation-group evolution. Applying it to
the NP coefficients leads to the low-scale values displayed in tab.~\ref{tab:WiCoNP}.
They can be compared to the dominant SM coefficient $C^\text{SM}_9(m_b)/\alpha=-1.203$.
\medskip

\begin{table}
 \centering
 \begin{tabular}{|c|r|r|}
\hline
        & $C^{\rm NP}_i(m_b)/\alpha$ & $C_i^{\rm{NP}\,\prime}(m_b)/\alpha$
\\  \hline
 $C_7$    &  $-0.966\,q_7 + 0.009\,q_9$ &   $ -0.966\,q_7^{\prime} +
0.009\,q_9^{\prime}$ \\
 $C_8$    &   $- 0.387\,q_7 + 0.002\,q_9$ &   $- 0.387\,q_7^{\prime} +
0.002\,q_9^{\prime}$ \\
 $C_9$    & $0.010\,q_7 - 1.167\,q_9$ & $0.010\,q_7^{\prime} -
1.167\,q_9^{\prime}$ \\
 $C_{10}$ & $- 0.001\,q_7 + 0.268\,q_9$ & $- 0.001\,q_7^{\prime} +
0.268\,q_9^{\prime}$ \\ \hline
 \end{tabular}
\caption{NLO short-distance coefficients of the EW penguin operators at the scale
$m_b$. Modifications to other short-distance coefficients are negligible. }
\label{tab:WiCoNP}
\end{table}

In our analysis we will study several different scenarios. First, we consider
the cases where only one of the coefficients $q_7$, $q_9$, $q_7^{\prime}$,
$q_9^{\prime}$ is different from zero. This means we assume the dominance of an
individual NP operator as it has also been done for example in
ref.~\cite{Feldmann:2008fb}. Second, we consider the possibilities of having
$q_7=q_9$, $q_7^{\prime}=q_9^{\prime}$, $q_7=q_9^{\prime}$ and
$q_7^{\prime}=q_9$. Finally, we study parity-symmetric new contributions
corresponding to the three cases $q_7=q_7^{\prime}$, $q_9=q_9^{\prime}$ and
$q_7=q_7^{\prime}=q_9=q_9^{\prime}$.
Each of these scenarios can be described by means of two real parameters, the
absolute value $|q|$ and phase $\phi$ of the NP contribution under
consideration. This reduced number of free parameters allows us to perform a fit
to $B\to \pi K$ data and to draw meaningful conclusions on the
$\bar B_s\to\phi\pi^0,\phi\rho^0$ decays. The study of this large set of well-motivated
simplified scenarios is assumed to represent all relevant features of the
general framework with unrelated $q_7$, $q_9$, $q_7^{\prime}$,
$q_9^{\prime}$.\medskip

Our main motivation for adding NP to the coefficients
$C_7^{(\prime)},C_{9}^{(\prime)}$ was the claim that the $\Delta
A_{\textrm{CP}}$ discrepancy can be solved in this way, namely by generating the
terms $\rtEW$, $\rtEWC$, $\rtEWA$ introduced in eq.~(\ref{g8}).
Introducing individual terms for each of the four relevant Wilson coefficients, we obtain
\begin{eqnarray}
\sum\limits_{i=7,9,7',9'}\rtEWi\, e^{-i\delta_i}&=&(q_7-q_7^{\prime})\,\left[\,(-0.12)^{+0.04}_{-0.05}\,+\,(-0.02)^{
+0.07}_{-0.02}\,i\,\right]\,+\nonumber\\[0.1cm]
&&(q_9-q_9^{\prime})\,\left[\,0.12^{+0.05}_{-0.04}\,+\,0.02^{+0.02}_{-0.07}\,i\,\right]\,,\nonumber\\[0.2cm]
\sum\limits_{i=7,9,7',9'}\rtEWCi\, e^{-i\delta_i}&=&(q_7-q_7^{\prime})\,\left[\,0.10^{+0.03}_{-0.02}\,+\,0.01^{+0.01}_{
-0.06}\,i\,\right]\,+\nonumber\\[0.1cm]
&&(q_9-q_9^{\prime})\,\left[\,0.04^{+0.02}_{-0.03}\,+\,(-0.005)^{+0.016}_{
-0.026}\,i\,\right]\,,\nonumber
\end{eqnarray}
\begin{eqnarray}\label{NPRew}
\sum\limits_{i=7,9,7',9'}\rtEWAi\, e^{-i\delta_i}&=&(q_7-q_7^{\prime})\,\left[\,0.03^{+0.04}_{-0.07}\,+\,(-0.06)^{+0.12
}_{-0.01}\,i\,\right]\,+\nonumber\\[0.1cm]
&&(q_9-q_9^{\prime})\,\left[\,0.007^{+0.003}_{-0.010}\,+\,(-0.006)^{+0.012}_
{-0.003}\,i\,\right]\,.
\end{eqnarray}
Let us briefly discuss the main characteristics of these coefficients:
\begin{itemize}
 \item  First of all, note that parity-symmetric models obviously do not
contribute to $B\to \pi K$ at all. This general feature of $B$ decays into
two pseudoscalar mesons ($PP$ decays) follows from eq.~(\ref{b6}). Therefore
such a scenario cannot solve the $\Delta A_{\textrm{CP}}$ discrepancy.
\item The contributions $\tilde{r}_{\textrm{EW},\,7^{(\prime)}}$
and $\tilde{r}_{\textrm{EW},\,9^{(\prime)}}$
tend to cancel each other. Hence in the scenarios with $q_7=q_9$ and
$q_7^{\prime}=q_9^{\prime}$ only a negligible new colour-allowed EW penguin
contribution is generated.
\item Whereas $\RE(\tilde{r}_{\textrm{EW},\,9^{(\prime)}}^{\textrm{C}})$
features the typical colour-suppression with respect to
$\RE(\tilde{r}_{\textrm{EW},\,9^{(\prime)}})$,
this pattern is not obeyed by the $q_7^{(\prime)}$ terms. This is due to a
conspirative interplay of the large mixing of $C_7^{(\prime)}$ into $C_8^{(\prime)}$
(compare tab.~\ref{tab:WiCoNP}), constructive interference of the new $C_7^{(\prime)}$
and $C_8^{(\prime)}$ contributions in the QCDF coefficient $a_8^{(\prime)}$ and a chiral
enhancement factor $r_{\chi}^{\pi,K}\approx 1.5$ multiplying $a_8^{(\prime)}$ in
eq.~(\ref{b5}) for the topological amplitude. None of these three effects is present
in the $q_9^{(\prime)}$ case.
 \item The annihilation coefficient $\tilde{r}_{\textrm{EW},\,7^{(\prime)}}^{\textrm{A}}$
develops a large imaginary part. In scenarios with non-vanishing $q_7^{(\prime)}$
this term gives the dominant contribution to $\Delta A_{\textrm{CP}}$.
\end{itemize}

\begin{figure}
    \begin{minipage}{0.3\linewidth}
      \psfragscanon
      \psfrag{Imq7DACP}[tc][Bc][1][0]{\small{$\IM(q_7)$}}
      \psfrag{Req7DACP}[cc][cc][1][0]{\small{$\RE(q_7)$}}
      \includegraphics[width=\linewidth]{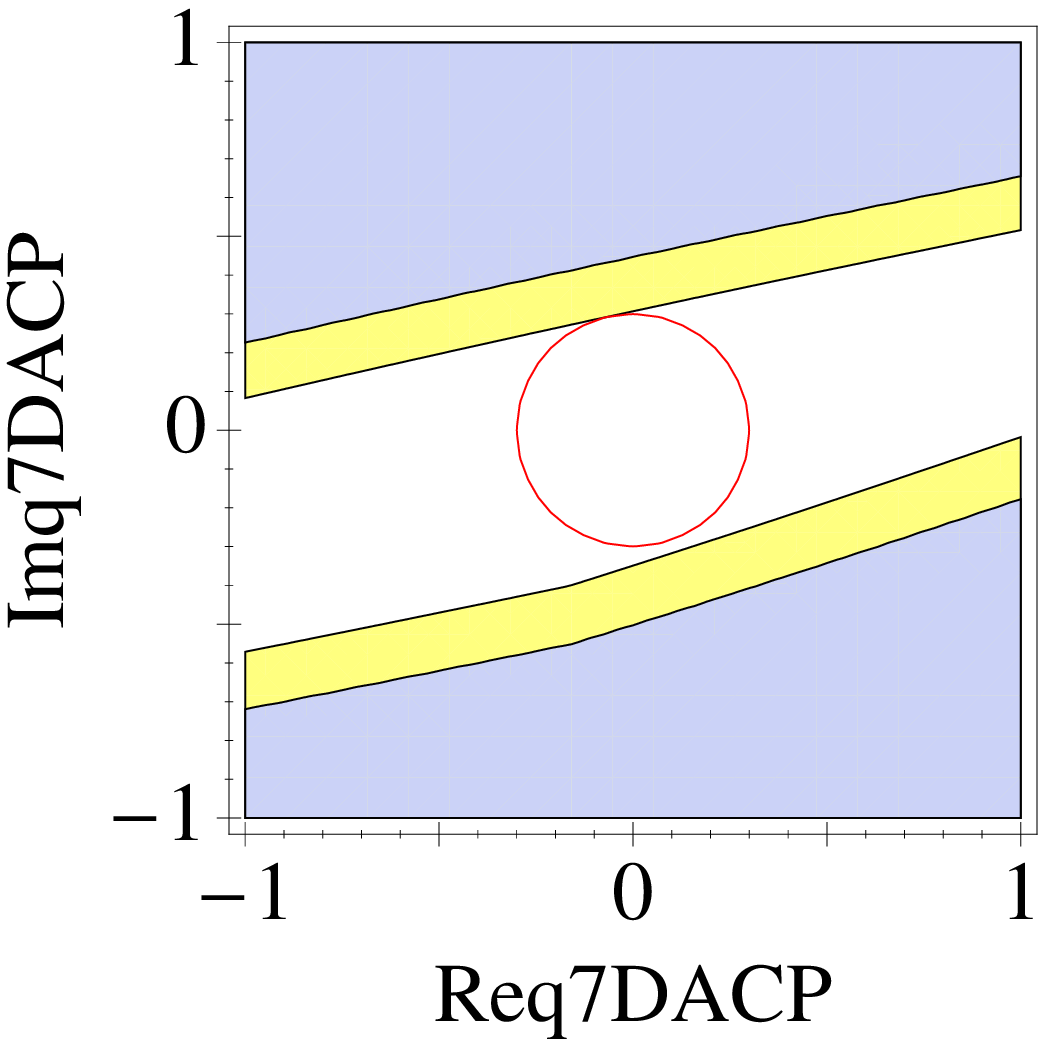}
   \end{minipage}\hspace{0.04\linewidth}
   \begin{minipage}{0.3\linewidth}
      \psfragscanon
      \psfrag{Imq9DACP}[tc][Bc][1][0]{\small{$\IM(q_9)$}}
      \psfrag{Req9DACP}[cc][cc][1][0]{\small{$\RE(q_9)$}}
      \includegraphics[width=\linewidth]{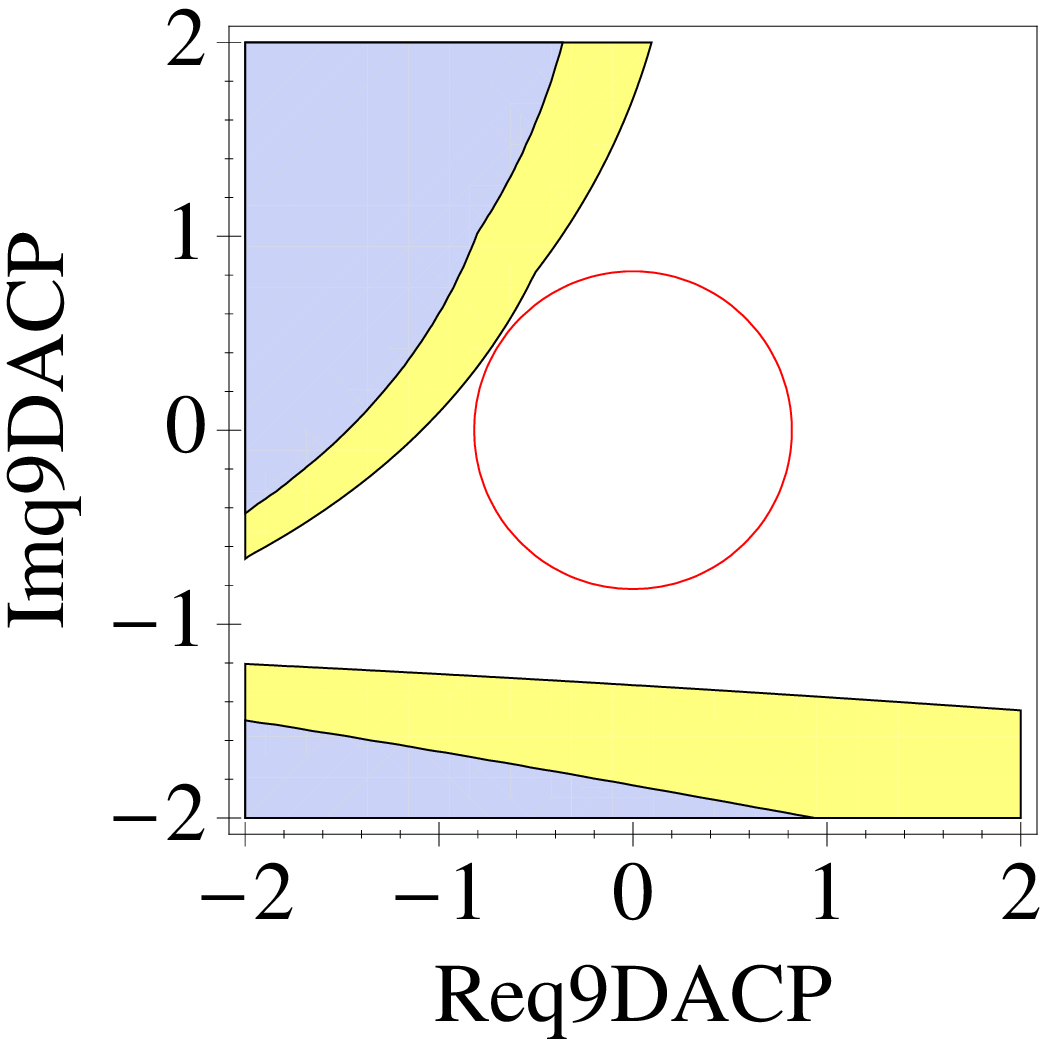}
   \end{minipage}\hspace{0.04\linewidth}
   \begin{minipage}{0.3\linewidth}
      \psfragscanon
      \psfrag{Imq79DACP}[tc][Bc][1][0]{\small{$\IM(q_{7,9})$}}
      \psfrag{Req79DACP}[cc][cc][1][0]{\small{$\RE(q_{7,9})$}}
      \includegraphics[width=\linewidth]{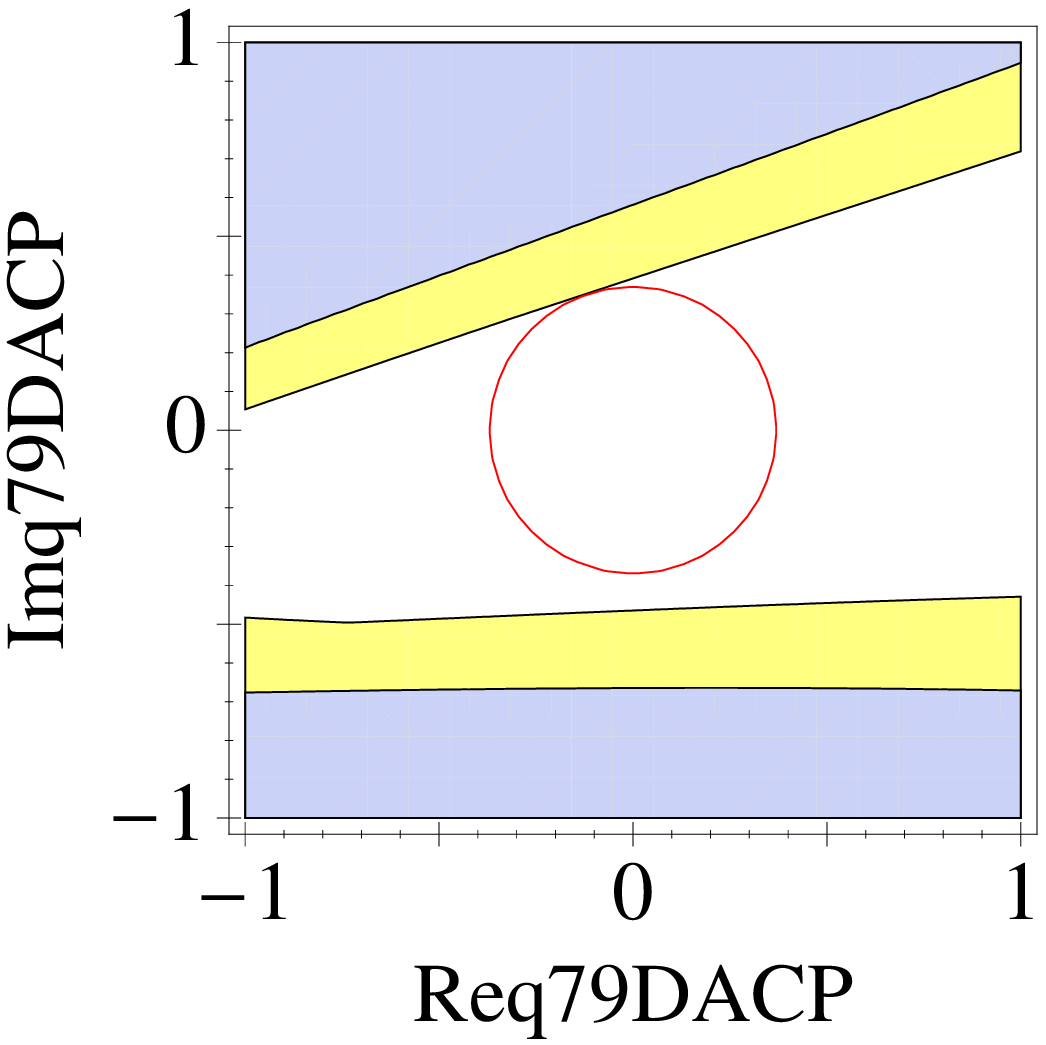}
   \end{minipage}
\caption{NP contribution needed to solve the $\Delta A_{\textrm{CP}}$
discrepancy in the three scenarios (from left to right) with single $q_7$,
single $q_9$ and equal $q_7=q_9$ contribution. Yellow region: Theory error band
and experimental $1\,\sigma$ region overlap. Blue region: Theory error band and
experimental central value overlap. Red circle: Minimal magnitude of the NP
contribution needed to reduce the $\Delta A_{\textrm{CP}}$ discrepancy below the
$1\,\sigma$ level. \label{fig:DeltaACP}}
\end{figure}

From eq.~(\ref{g10}) we see that the $\Delta A_{\textrm{CP}}$ discrepancy can be
solved either through $\rtEW$ or through $\rtEWA$. Except for the parity-symmetric
models, all the scenarios mentioned above can achieve such a solution.
In fig.~\ref{fig:DeltaACP} this is illustrated for the cases with a single $q_7$
or $q_9$ and for the $q_7=q_9$ scenario. Graphs for the respective mirror
scenarios are obtained by a $180^{\circ}$ rotation. The yellow region contains
those points of the $(\RE(q_i),\IM(q_i))$\,-\,plane for which the theory error
band overlaps with the experimental $1\,\sigma$ region, whereas the blue
region represents those points for which also the experimental central value lies
within the theory error interval. The red circle illustrates the minimal
$|q|$\,-\,value needed to reduce the $\Delta A_{\textrm{CP}}$ tension below the
$1\,\sigma$ level. For the three scenarios in fig.~\ref{fig:DeltaACP} we read
off $|q_7|\gtrsim 0.3$, $|q_9|\gtrsim 0.8$ and $|q_7|=|q_9|\gtrsim 0.4$. The
fact that in the $q_7=q_9$ case only a small NP contribution is needed, in spite
of the absence of $\rtEW$, demonstrates the importance of the annihilation term
$\rtEWA$. Finally, we like to stress that the solution of the $\Delta
A_{\textrm{CP}}$ discrepancy via a minimal $|q|$\,-\,value requires the
adjustment of the phase $\phi$ to a certain value. Realistic scenarios avoiding
such a fine-tuning have larger $|q|$\,-\,values,
typically $|q|\sim 1$.\medskip

Our main goal is to study the impact of such a NP scenario on the decays $\bar B_s\to
\phi\pi^0,\phi\rho^0$. The NP contributions to
$C_7^{(\prime)},...,C_{10}^{(\prime)}$ generate the
$\tilde{r}_{\textrm{EW}}^{M_2}$\,-\,terms introduced in
eq.~(\ref{eq:BsPhiPiNP}). Introducing again individual terms for contributions
from the various Wilson coefficients, they read for the four different amplitudes
\begin{eqnarray}\label{eq:rtEWPhiNP}
   \sum\limits_{i=7,9,7',9'}\tilde{r}_{\textrm{EW},\,i}^{\pi}\, e^{-i\delta_i}
& = & - 0.9 \,\left(q_7\,+\,q_7^{\prime}\,-\,q_9\,-\,q_9^{\prime}\right),  \nonumber \\
   \sum\limits_{i=7,9,7',9'}\tilde{r}_{\textrm{EW},\,i}^{\rho,0}\, e^{-i\delta_i}
& = & \phantom{-} 0.9  \,\left(q_7\,-\,q_7^{\prime}\,+\,q_9\,-\,q_9^{\prime}\right),  \nonumber \\
   \sum\limits_{i=7,9,7',9'}\tilde{r}_{\textrm{EW},\,i}^{\rho,-}\, e^{-i\delta_i}
& = & - 0.6  \,\left(q_7\,+\,q_9\right),  \nonumber \\
   \sum\limits_{i=7,9,7',9'}\tilde{r}_{\textrm{EW},\,i}^{\rho,+}\, e^{-i\delta_i}
& = & \phantom{-}0.6  \,\left(q_7^{\prime}\,+\,q_9^{\prime}\right)
\times P_{\textrm{EW}}^{\rho,-}/P_{\textrm{EW}}^{\rho,+}\,,
\end{eqnarray}
where we have neglected $q_{7,9}$-contributions to
$\tilde{r}_{\textrm{EW}}^{\rho,+}$ and $q_{7,9}^{\prime}$-contributions to
$\tilde{r}_{\textrm{EW}}^{\rho,-}$ according to their
$\Lambda_{\textrm{QCD}}^2/m_B^2$ suppression. The SM EW penguin amplitude
$P_{\textrm{EW}}^{\rho,+}$ drops out of the total expression
(\ref{eq:BsPhiPiNP}) of the amplitude, $P_{\textrm{EW}}^{\rho,-}$ is given in
eq.~(\ref{eq:NumPEW}). The parameters $\tilde{r}_{\textrm{EW}}^{M_2}$ develop
only very small strong phases and uncertainties not indicated in
(\ref{eq:rtEWPhiNP}). This is because they are ratios of equal topologies such
that uncertainties and strong phases approximately cancel.\medskip

We have stated the expressions in eq.~(\ref{eq:rtEWPhiNP}) for two reasons: Firstly we
want to show the main consequences of non-vanishing $q_{7,9}^{(\prime)}$ for the
$\bar B_s\to\phi\pi^0,\phi\rho^0$ decays. We see that for $|q_i|=\mathcal{O}(1)$ indeed
new contributions with the magnitude of the leading SM EW penguin are generated.
While parity-symmetric NP was invisible in $B\to \pi K$, it could be
detected in $\bar B_s\to\phi\pi^0$ and in principle also in $\bar B_s\to\phi\rho^0$ due to the
different interference patterns of $\tilde{r}_{\textrm{EW}}^{\rho,-}$ and
$\tilde{r}_{\textrm{EW}}^{\rho,+}$ with the corresponding SM contributions.
Furthermore, left- and right-handed NP could be distinguished by a polarisation
measurement of $\bar B_s\to\phi\rho^0$. This general feature of decays to vector-vector
final states has been pointed out by Kagan \cite{Kagan:2004ia}. Note that the
question of left- vs.\ right-handed NP cannot be answered from $B\to \pi K$
alone since, as we have seen, the two scenarios  differ only by a rotation
in the NP parameter space.\medskip

The second benefit of eq.~(\ref{eq:rtEWPhiNP}) is that it allows for a
simple calculation of  $\bar{B}_s\to\phi\pi,\phi\rho$ observables to a very good
accuracy. In this way it permits a study of these decays without the extensive
implementation of QCDF. One simply evaluates the amplitude
eq.~(\ref{eq:BsPhiPiNP}) inserting eqs.~(\ref{eq:NumPEW},\ref{eq:NumRc}) - or
alternatively eq.~(\ref{eq:Approx}) - and the NP part from eq.~(\ref{eq:rtEWPhiNP}).
The CP-conjugated amplitude is obtained by flipping the sign of $\gamma$ and replacing
the $q_i$ by their complex conjugates. Subsequently one can use the formulas in
Appendix~\ref{Observables} to calculate observables. \bigskip

\boldmath
\subsection{Fit to \texorpdfstring{$B\to\pi K$}{B to Pi,K} data and constraints from other
decays}\label{BpiKconstr}
\unboldmath

The four $B\to\pi K$ channels are the most precisely measured hadronic
$b\rightarrow s$ decays. For this reason, we use experimental information from
these channels as input for our quantitative NP analysis by performing a fit of
$q_{7}^{(\prime)}$ and $q_{9}^{(\prime)}$ to $B\to\pi K$ data. This fit will be
an indication for values of the $q_{i}^{(\prime)}$ that are realistic to expect
and consequently will enable us to find an expected size of enhancement of the
branching ratios of $\bar B_s\to\phi\rho^0$ and $\bar B_s\to\phi\pi^0$. In the treatment of
theoretical and experimental uncertainties in the fit we follow the
\textit{R}fit scheme \cite{Hocker:2001xe}. More details on this issue are given
in Appendix~\ref{fit}.\medskip

Since $B\to\pi K$ decays are greatly dominated by QCD penguins and since they
suffer from large theoretical uncertainties, it is obvious that NP effects
residing in EW penguins are difficult to find in branching ratios and direct CP
asymmetries. It is more useful to consider instead  particular combinations of
these basic observables which highlight the isospin-violating contributions and
can be predicted with a better precision. For example, it is difficult to see
a need for isospin-violating NP by looking at the two CP asymmetries
entering in $\Delta A_{\rm CP}$ (see tab.~\ref{tab1}) since both of them have
more than $100\%$ theoretical uncertainty, reflecting the difficulty of
predicting strong phases in QCDF. In the difference $\Delta A_{\rm
CP}$, the theory uncertainties cancel to a large extent such that the
discrepancy with experimental data becomes clearer. Moreover, in realistic
models a new contribution in the EW penguin sector usually comes in combination
with NP of comparable size in the QCD penguins since the new contribution in
general matches onto a linear combination of the QCD and EW penguin operators.
By considering mainly isospin-violating observables, one reduces the sensitivity
to additional effects from new contributions to $C_3,...,C_6$ which we did not
include in our parameterisation (\ref{g23}).\medskip

For our $B\to\pi K$ fit, we use suitable ratios of branching fractions
and differences of CP asymmetries as well as the mixing-induced CP asymmetry in
$\bar B^0\to\pi^0 \bar K^0$ as input. The definitions of these quantities are
summarised in Appendix~\ref{IsoRatios}. Many of these quantities have also been
considered in the past in the context of flavour symmetry analyses of $B\to\pi
K$ decays. A summary of theoretical predictions vs.\ experimental results for all
these observables as well as for the $B\to\pi K$ branching fractions and CP
asymmetries is provided in tab.~\ref{tab1}. At present none of these quantities
deviates from the SM prediction by more than $~ 1 \sigma$ apart from $\Delta
A_{\rm CP}$. We thus expect $\Delta A_{\rm CP}$ (and to a lower degree also
$S_{\textrm{CP}}$) to pull the fit towards non-zero values of
the $q_i^{(\prime)}$ whereas the other observables will favour values close to
the origin of the complex plane.\medskip

In addition to the fit we consider constraints on the NP parameters arising
from a large number of hadronic B decays. To this end we compare the theoretical
prediction of an observable as a function of the $q_{i}^{(\prime)}$ to its measured
value and extract a $2\sigma$-constraint as follows:
\be\label{g231b}
{\rm Point[q_{7,9}^{(\prime)}\,\,space]} =\left\{\begin{array}{cl}
           {\rm allowed}
           & \mbox{  if  } \left\{\begin{array}{l}
                              (x_{\rm theo}+\sigma_{\rm theo,\,sup})>(x_{\rm
exp}-2 \sigma_{\rm exp,\,inf}) \\
                          \textrm{ and }(x_{\rm theo}-\sigma_{\rm
theo,\,inf})<(x_{\rm exp}+2 \sigma_{\rm exp,\,sup}),
                           \end{array}\right. \\
           {\rm excluded} & \mbox{  otherwise}.\\
         \end{array} \right.
\ee
Here $(x_{\rm theo})^{+\sigma_{\rm theo,\,sup}}_{-\sigma_{\rm theo,\,inf}}$
represents the theoretical prediction for the respective physical observable.
The uncertainty does not imply a particular probability distribution but the
true value is supposed to lie within the error interval. For the experimental
value $x_{\rm exp}\pm \sigma_{\rm exp}$ a Gaussian error is assumed.\medskip

This procedure is applied to data from $B\to\pi K$ as well as to data
from the $B\to \rho K$, $B\to \pi K^*$ and $B\to \rho K^{(*)}$
decay channels, which are simply the pseudoscalar-vector ($PV$) and vector-vector
($VV$) modes corresponding to $B\to\pi K$. The $PV$ and $VV$ modes
turn out to be more sensitive to isospin-violating flavour topologies than their
$PP$ counterparts because the leading QCD penguin amplitude is smaller.
Experimental information on these decays, however, is not (yet) as precise as the
available data for the $\pi K$ modes. Therefore we do not include the $PV$, $VV$
modes into the fit but prefer to consider them as constraints at the $2\sigma$
level only. Nonetheless, the constraints from $B\to \rho K$ and
$B\to \pi K^{\star}$ give some information complementary to the one from
$B\to\pi K$ because they test different chirality structures than
$B\to \pi K$ and are therefore sensitive to other linear combinations of
the $q_{7,9}^{(\prime)}$. Moreover, we apply eq.~(\ref{g231b}) also to data from
$B\to K^{(*)} \phi$, $\bar B_s\to \phi\phi$ and $\bar B_s\to \bar{K}K$ decays even
if they only carry a small sensitivity to EW penguins.

\subsection{Results of the model-independent analysis}\label{modindres}

We now discuss the results of the analysis outlined in the previous section.
The aim is to make predictions for the $B_s$ decays in combination with
the regions of the $q_{7,9}^{(\prime)}$ parameter space which are preferred,
or not yet excluded, by experimental data from $B\to\pi K$ and related
decays.\medskip

\begin{figure}[t]
\begin{center}
  \includegraphics[width=0.99\textwidth]{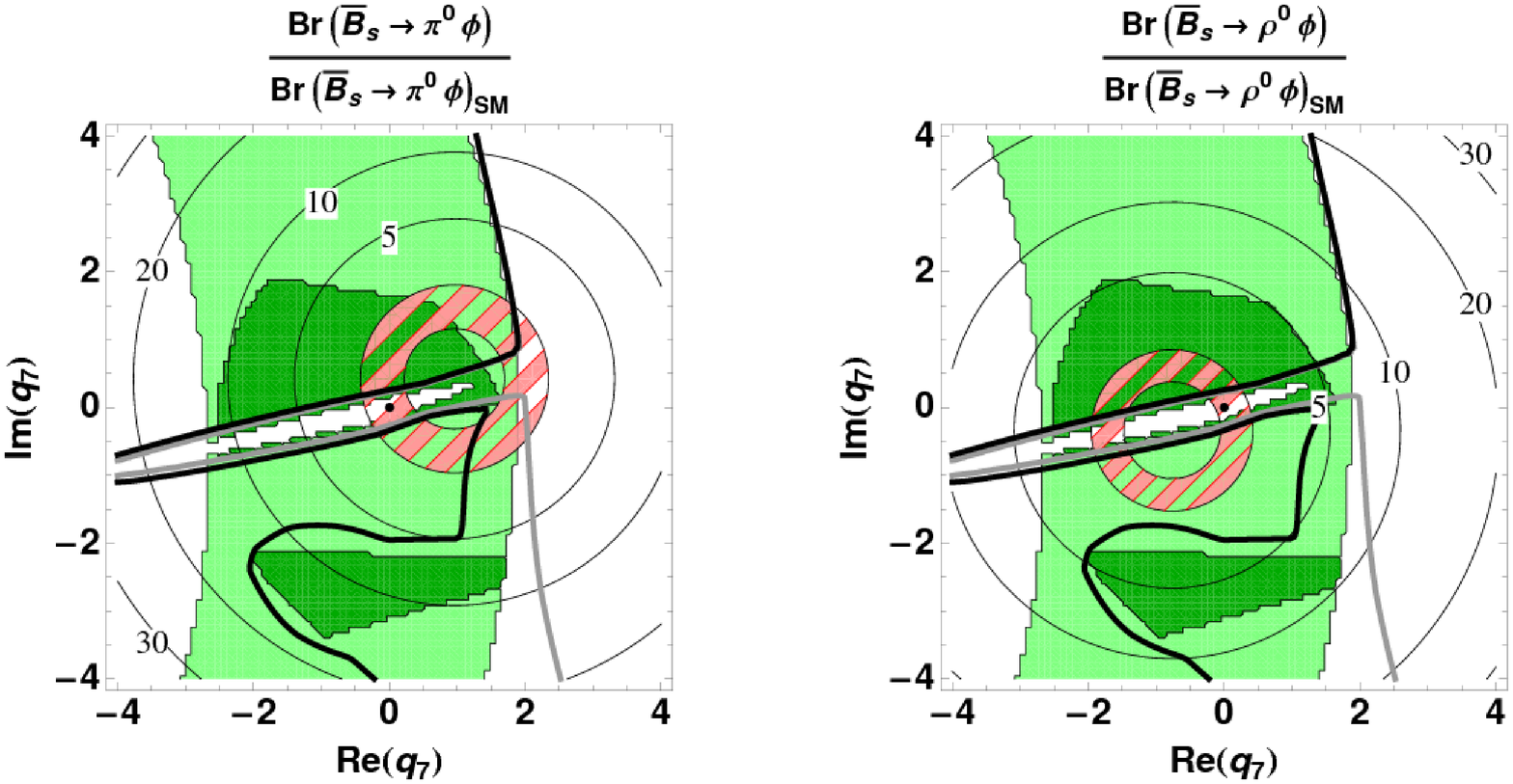}
  \includegraphics[width=0.99\textwidth]{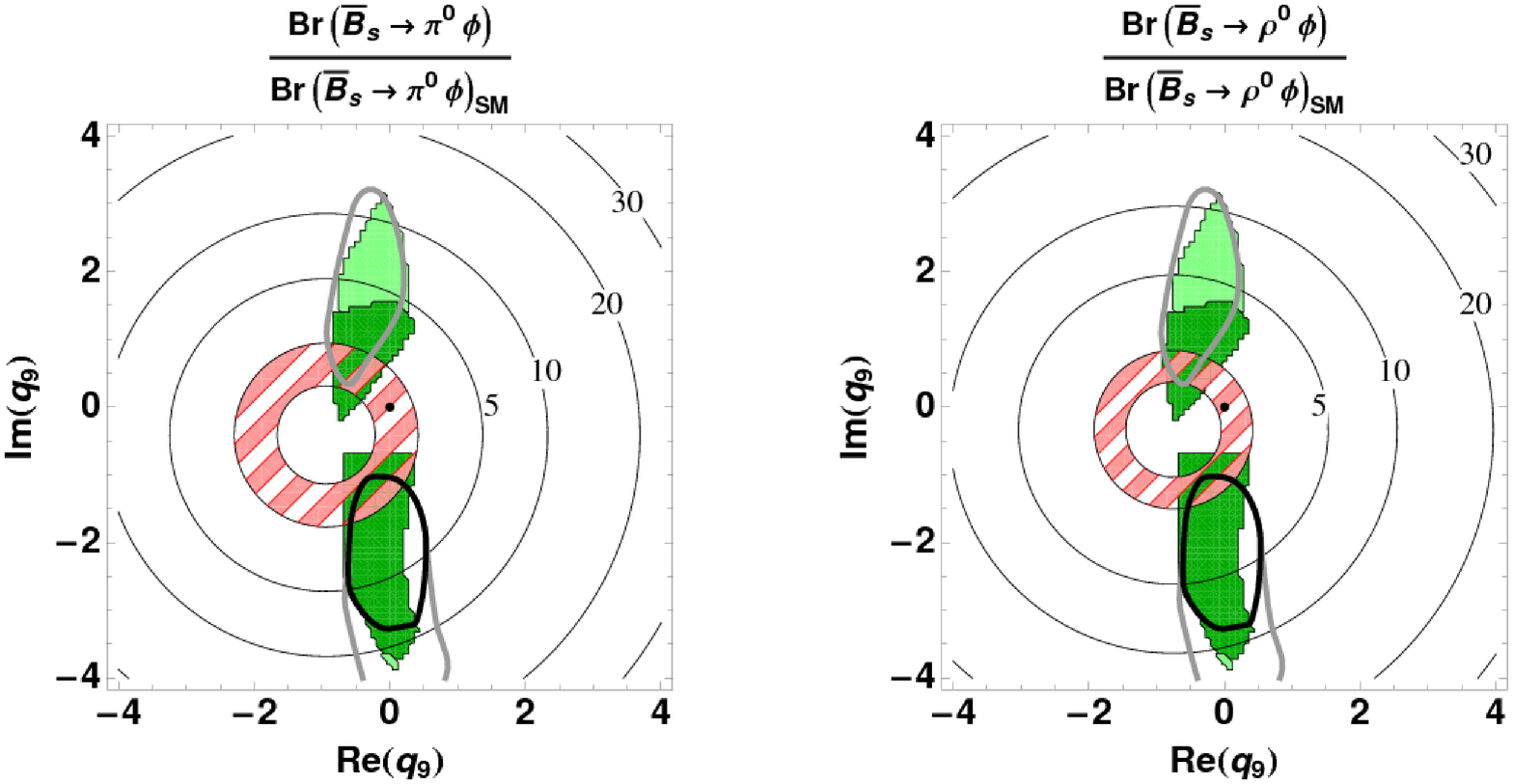}
\end{center}
  \caption{Enhancement factors of the $\bar{B_s}\rightarrow~\phi\rho^0,\phi\pi^0$
  branching ratios with respect to their SM values. The black dot represents the
  SM result while the red striped region shows the theoretical uncertainty
  in the SM. The dark green area is the region allowed by the $2\,\sigma$
  constraints from $\bar B \to \pi K^{(*)},\rho K^{(*)},\phi K^{(*)}$
  and $\bar B_s\to\phi\phi,\bar K K$ decays; for comparison, the light green area represents
  the area allowed by constraints from isospin-sensitive observables only, considering
  only $\bar B \to \pi K,\pi K^{(*)},\rho K$ decays. The solid black line
  represents the $1\sigma$ CL of the fit with $S_{CP}(\bar B^0 \to \pi \bar{K}^0)$,
  while the solid grey line represents the $1\sigma$ CL of the fit without it.
  Here the scenarios $q_7\neq 0$ (upper row) and $q_9\neq 0$ (lower row) are displayed. For the $q_7$-scenario the $1\sigma$ region of the fit is the region to the left of the black (grey) curve.}
  \label{figq7q9}
\end{figure}

\begin{figure}[t]
\begin{center}
  \includegraphics[width=0.8\textwidth]{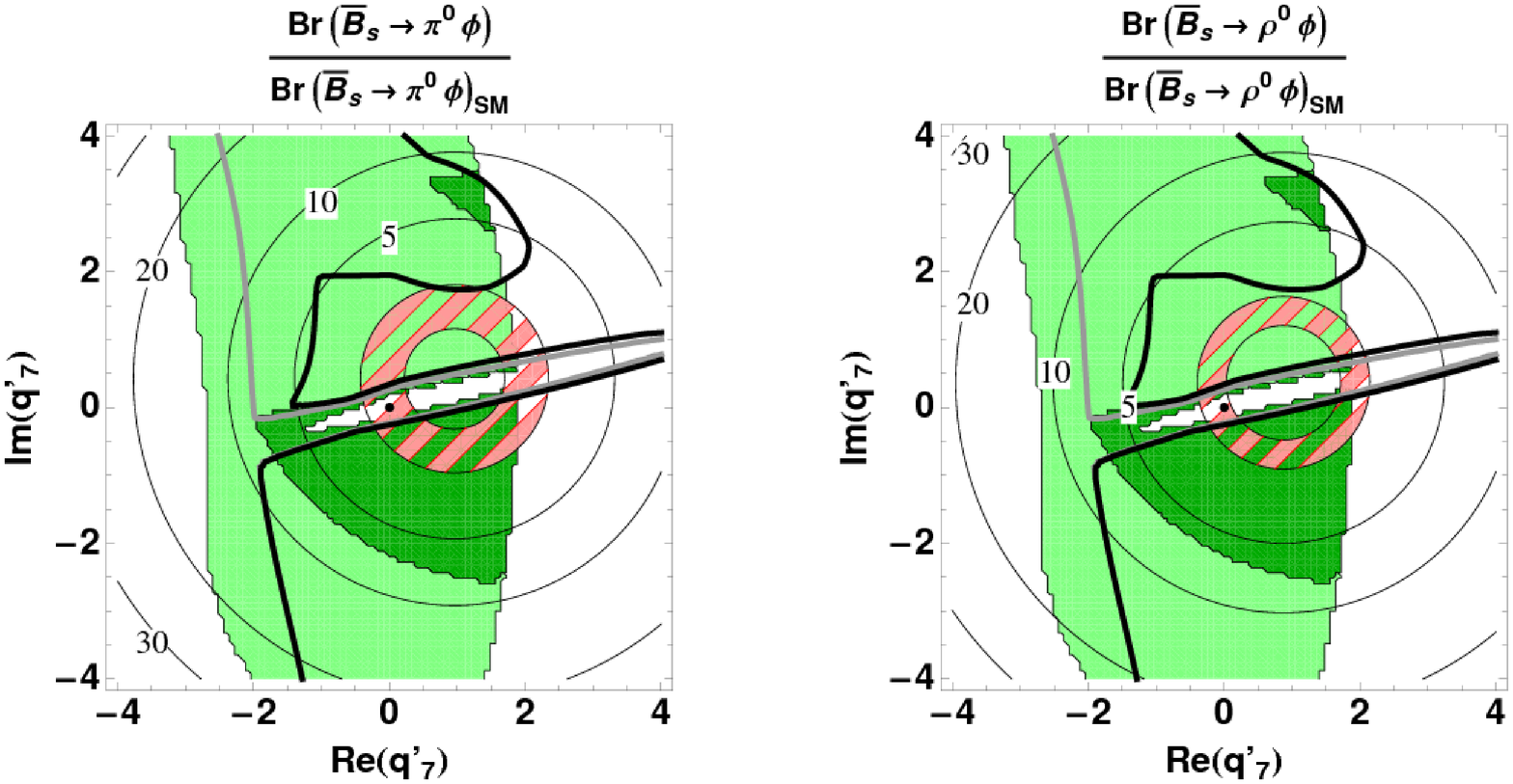}
  \includegraphics[width=0.8\textwidth]{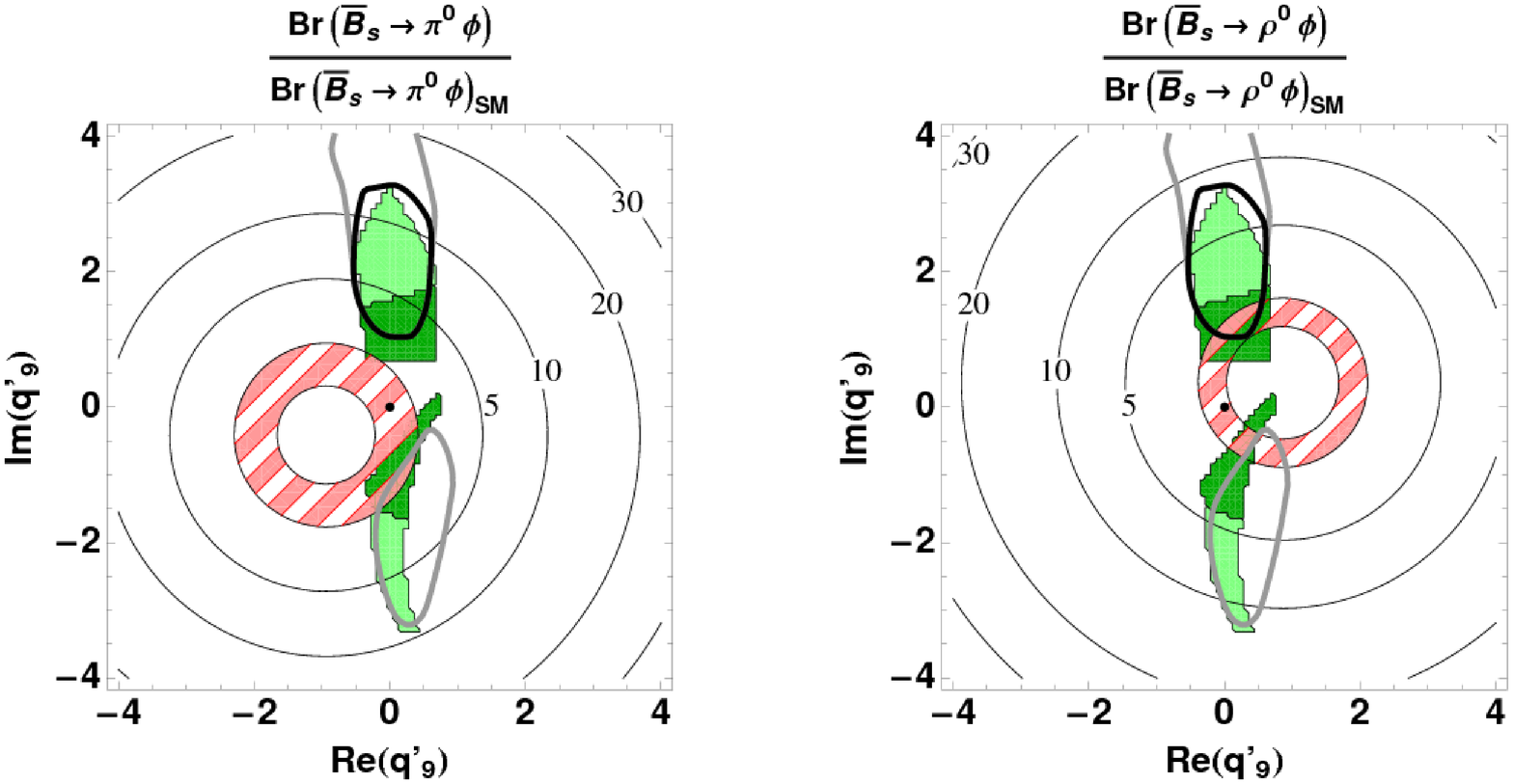}
\includegraphics[width=0.4\textwidth]{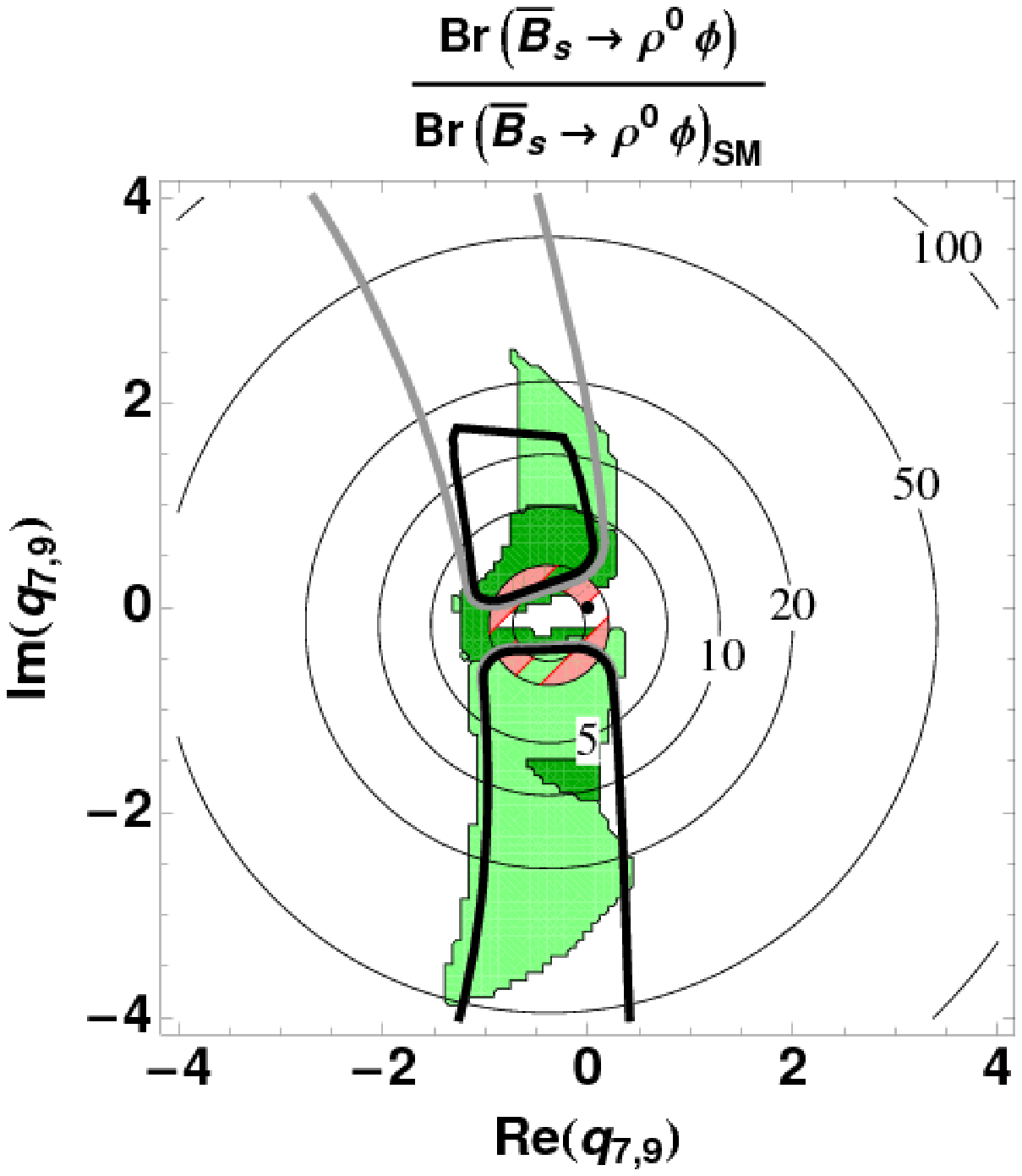}
\includegraphics[width=0.4\textwidth]{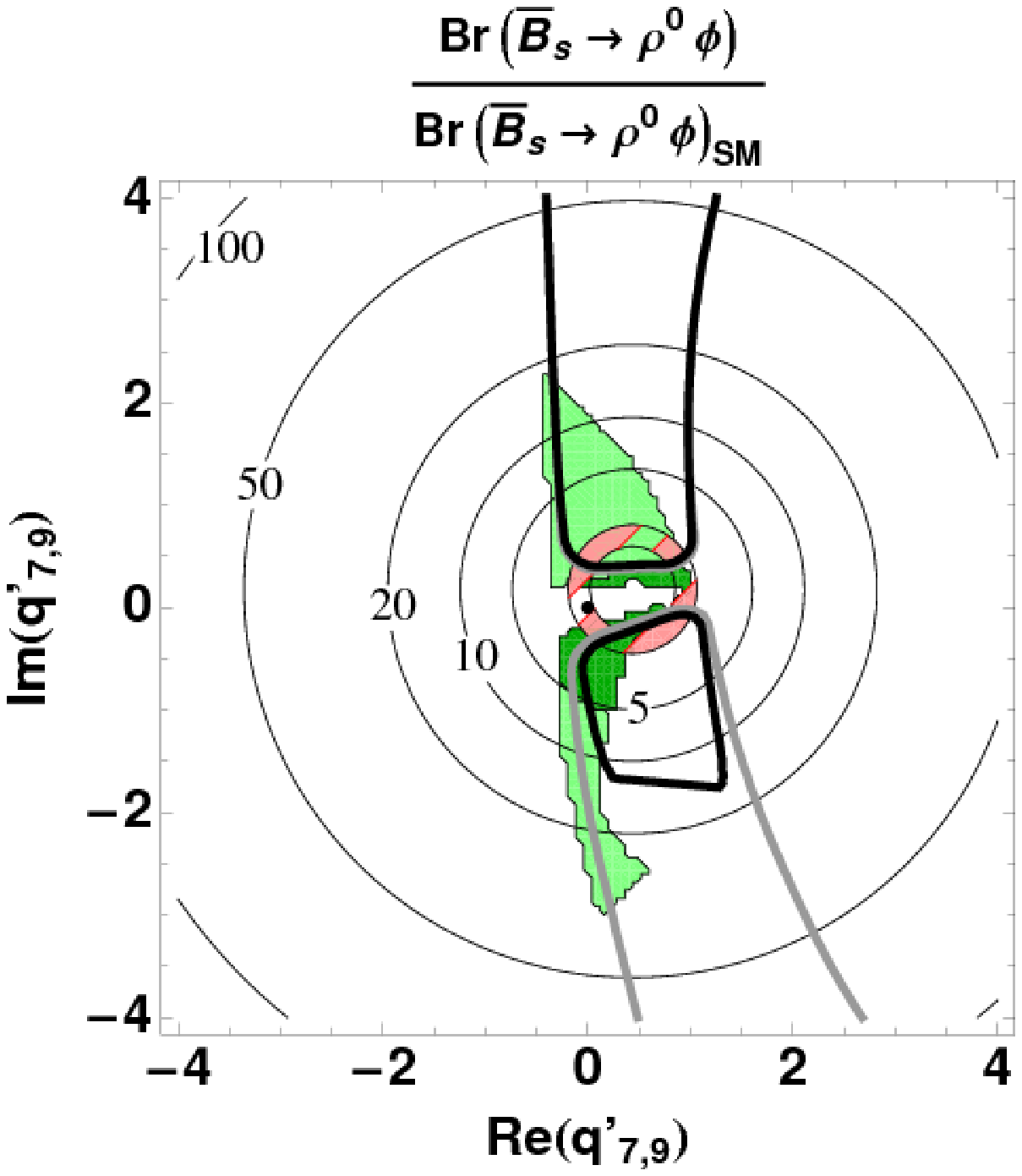}
\end{center}
  \caption{Enhancement factors of the $\bar{B_s}\rightarrow~\phi\rho^0,\phi\pi^0$
  branching ratios with respect to their SM values. The meaning of the contours and regions
  is the same as in fig.~\ref{figq7q9}. Here the scenarios $q_7^{\prime}\neq0$
  (upper row) and $q_9^{\prime}\neq 0$ (center row) and $q_7=q_9\neq 0$ and
$q_7^{\prime}=q_9^{\prime}\neq 0$ (lower row) are displayed. For the $q_7^{\prime}$-scenario the $1\sigma$ region of the fit is the region to the right of the black (grey) curve.}
  \label{figq7q9P}
\end{figure}

\begin{table}[ht]
 \centering
 \begin{tabular}{|c|r|r|r|} \hline
   Scenario     & $\frac{\Br(\bar B_s \rightarrow \phi\pi^0)}{\Br^{\rm SM}(\bar
B_s \rightarrow \phi\pi^0)}$
                & $\frac{\Br(\bar B_s \rightarrow \phi_L\rho^0_L)}{\Br^{\rm
SM}(\bar B_s \rightarrow \phi_L\rho^0_L)}$
                & $\frac{\Br(\bar B_s \rightarrow \phi\rho^0)}{\Br^{\rm SM}(\bar
B_s \rightarrow \phi\rho^0)}$ \\ \hline
 $q_7\neq 0$                               & 11.0            (18.7) &  6.0  \phantom{0}(9.9) &  5.3  \phantom{0}(8.4) \\
 $q_9\neq 0$                               &  8.8            (16.4) &  9.3            (17.0) &  8.7            (15.1) \\
 $q_7=q_9\neq 0$                           &  1.0  \phantom{0}(1.7) & 11.5            (21.1) & 10.8            (18.7) \\
 $q_7^{\prime}\neq 0$                      &  8.3            (15.6) &  8.8            (16.4) &  8.4            (14.7) \\
 $q_9^{\prime}\neq 0$                      &  6.2  \phantom{0}(9.8) &  2.8  \phantom{0}(5.6) &  2.7  \phantom{0}(5.0) \\
 $q_7^{\prime} = q_9^{\prime}\neq 0$       &  1.0  \phantom{0}(1.7) &  5.7  \phantom{0}(8.9) &  5.2  \phantom{0}(7.9) \\
 $q_7=q_9^{\prime}\neq 0$                  &  1.0  \phantom{0}(1.7) &  1.0  \phantom{0}(1.7) &  1.6  \phantom{0}(2.3) \\
 $q_7^{\prime}=q_9\neq 0$                  &  1.0  \phantom{0}(1.7) &  1.0  \phantom{0}(1.7) &  1.1  \phantom{0}(1.8) \\ \hline
 $q_7=q_7^{\prime}\neq 0$                  & 29.5            (48.1) &  1.0  \phantom{0}(1.7) &  2.1  \phantom{0}(3.0) \\
 $q_9=q_9^{\prime}\neq 0$                  & 11.1            (20.5) &  1.0  \phantom{0}(1.7) &  1.5  \phantom{0}(2.2) \\
 $q_7=q_7^{\prime}=q_9=q_9^{\prime}\neq 0$ &  1.0  \phantom{0}(1.8) &  1.0  \phantom{0}(1.7) &  2.3  \phantom{0}(3.4) \\ \hline
 \end{tabular}
\caption{Maximal possible enhancement of branching ratios compatible
with the constraints from $B \to \pi K,\rho K,\pi K^*,\rho K^*,\phi K,\phi K^*$
and $\bar B_s\to\phi\phi,\bar{K}K$ decays at the 2$\sigma$\,-\,level as well as the $1\sigma$
confidence level from $B \to \pi K$ decays (method a)). We use the default SM value and the
default (the maximal theoretical) total value of each branching ratio for an
optimally chosen $q_i$ value. Numbers in the second part of the table are obtained ignoring $\Delta A_{\rm CP}$.
\label{tab3}}
\end{table}

\begin{table}[t]
 \centering
 \begin{tabular}{|c|r|r|r|} \hline
   Scenario     & $\frac{\Br(\bar B_s \rightarrow \phi\pi^0)}{\Br^{\rm SM}(\bar
B_s \rightarrow \phi\pi^0)}$
                & $\frac{\Br(\bar B_s \rightarrow \phi_L\rho^0_L)}{\Br^{\rm
SM}(\bar B_s \rightarrow \phi_L\rho^0_L)}$
                & $\frac{\Br(\bar B_s \rightarrow \phi\rho^0)}{\Br^{\rm SM}(\bar
B_s \rightarrow \phi\rho^0)}$ \\ \hline
 $q_7\neq 0$                               & 77.4           (134.1) & 72.5           (117.6) & 66.9           (104.7) \\
 $q_9\neq 0$                               & 12.0 \phantom{0}(21.9) & 12.6 \phantom{0}(22.8) & 11.8 \phantom{0}(20.3) \\
 $q_7=q_9\neq 0$                           &  1.0 \phantom{00}(1.7) & 52.9 \phantom{0}(90.9) & 49.4 \phantom{0}(81.0) \\
 $q_7^{\prime}\neq 0$                      & 56.6            (99.2) & 59.5           (103.2) & 54.0            (90.5) \\
 $q_9^{\prime}\neq 0$                      & 13.0 \phantom{0}(20.6) & 13.0 \phantom{0}(20.5) & 11.7 \phantom{0}(18.1) \\
 $q_7^{\prime} = q_9^{\prime}\neq 0$       &  1.0 \phantom{00}(1.8) & 36.3 \phantom{0}(58.2) & 32.8 \phantom{0}(51.2) \\
 $q_7=q_9^{\prime}\neq 0$                  &  1.0 \phantom{00}(1.7) &  1.0 \phantom{00}(1.7) &  2.5 \phantom{00}(3.8) \\
 $q_7^{\prime}=q_9\neq 0$                  &  1.0 \phantom{00}(0.0) &  1.1 \phantom{00}(1.8) &  2.5 \phantom{00}(3.8) \\\hline
 $q_7=q_7^{\prime}\neq 0$                  & 76.0           (131.9) &  1.0 \phantom{00}(1.7) &  3.8 \phantom{00}(5.5) \\
 $q_9=q_9^{\prime}\neq 0$                  & 13.0 \phantom{0}(20.6) &  1.0 \phantom{00}(1.7) &  1.5 \phantom{00}(2.2) \\
 $q_7=q_7^{\prime}=q_9=q_9^{\prime}\neq 0$ &  1.0 \phantom{00}(1.8) &  1.0 \phantom{00}(1.7) &  4.0 \phantom{00}(5.9) \\ \hline
 \end{tabular}
\caption{Maximal possible enhancement of branching ratios compatible
with the constraints from isospin sensitive observables in
$B \to \pi K,\rho K,\pi K^*$ decays at the 2$\sigma$\,-\,level as well
as the 1-$\sigma$ confidence level from $B \to \pi K$ decays, without
including $S_{CP}(\bar B^0 \to \pi \bar{K}^0)$ (method b)). We use the default
SM value and the default (the maximal theoretical) total value of
each branching ratio for an optimally chosen $q_i$ value. Numbers in the second part of the table are obtained ignoring $\Delta A_{\rm CP}$.}
\label{tab3b}
\end{table}

In figs.~\ref{figq7q9} and \ref{figq7q9P} we present graphs
showing the enhancement $\Br^{\textrm{SM}+\textrm{NP}}/\Br^{\textrm{SM}}$
of the $\bar{B_s}\rightarrow~\phi\rho^0,\phi\pi^0$ branching ratios
as a function of the parameters $q_i^{(\prime)}$ in some representative
scenarios. The displayed numbers are obtained with our default hadronic
input. In order to be distinguishable from the SM, a particular scenario
must at least provide a value for
$\Br^{\textrm{SM}+\textrm{NP}}/\Br^{\textrm{SM}}$ which exceeds a potential
enhancement factor faked by hadronic uncertainties in the SM prediction.
Parameter points for which the enhancement factor lies within the theory
error band of the SM prediction are represented by the red-striped ring.
The SM itself corresponds, of course, to the origin of the plots and is
highlighted by a black dot.\medskip

In our sample models we introduced NP exclusively in the EW penguin
operators. In realistic models, however, a new contribution in the EW
penguin sector usually comes in combination with NP of comparable size
in the QCD penguins since the new contribution in general matches onto
a linear combination of the QCD and EW penguin operators. In order to
account for this fact, we use the experimental data in two different ways:
\begin{itemize}
 \item[a)] We present a fit using isospin-sensitive quantities in the
$B \to \pi K$ decays, such as ratios of branching fractions and the
differences of CP asymmetries as discussed in Appendix~\ref{IsoRatios}
and~\ref{fit}, plus the time dependent CP asymmetry
$S_{CP}(\bar B^0 \to \pi \bar{K}^0)$. In \mbox{figs.~\ref{figq7q9},\ref{figq7q9P}}
we individuate the 1$\sigma$ region by a solid black line.
At the same time we consider the constraints one obtains from all
non-leptonic $B \to \pi K,\rho K,\pi K^*,\rho K^*,\phi K,\phi K^*$
and $\bar B_s\to\phi\phi,\bar{K}K$ decays at the 2$\sigma$\,-\,level and mark the
allowed region by a (dark-)green area.
\item[b)] We exclude the time dependent CP asymmetry
$S_{CP}(\bar B^0 \to \pi \bar{K}^0)$ from the fit and we restrict
the constraints to the subset of observables which are particularly
sensitive to isospin violation, see Appendix~\ref{IsoRatios}. This
procedure enlarges the 1$\sigma$ confidence-level of the $B\to\pi K$ fit
(indicated by the grey line) as well as the region of $2\sigma$-allowed
parameter points by the areas depicted by lighter colours in the figures.
\end{itemize}
Whereas the results from a) are valid only if NP is strictly limited to
the electroweak penguin operators, the results from b) are expected to
remain approximately valid also in presence of NP in QCD penguins, since
such a kind of NP has only a minor impact on the quantities considered
in b).\medskip

We find that the $B\to\pi K$ and related decays set quite
strong constraints on the parameter space, especially in scenarios
where $q_9\neq 0$ or $q_9^{\prime}\neq 0$. This basically rules out
the possibility of having $|q_i|\gtrsim 5$, i.e.\ NP corrections
cannot be much larger than the EW penguins of the SM. The fact that
the SM point is always excluded at the $2\,\sigma$ level is a direct
consequence of the $\Delta A_{CP}$ data. According to the sign pattern
in eq.~(\ref{NPRew}), the $B\to \pi K$ fits of the primed and unprimed
scenarios in \mbox{figs.~\ref{figq7q9},\ref{figq7q9P}} are related
to each other through rotation by $180^{\circ}$. The fit works best in
the $q_9^{(\prime)}$ scenario where (using method a)) the best
fit point is given by
\begin{equation}\label{eq:BestFit}
    |\hat{q}_9^{(\prime)}|\,=\,1.9\hspace{2cm}\hat{\varphi}_{9}^{(\prime)}\,=\,-100^{\circ}\,(+180^{\circ}).
\end{equation}
This parameter point yields a full agreement of all the $B\to\pi K$
observables with the experimental mean values (for $S_{\textrm{CP}}$ the nearly
exact value $S_{\textrm{CP}}=0.55$ is obtained).
In the $q_7^{(\prime)}=q_9^{(\prime)}$ case a plateau of $\chi^2=0$ points
arises due to the large theoretical errors. It turns out that the
$B\to \pi K$ observables are not very sensitive to the $q_7^{(\prime)}$-only scenarios and so the fit does not work well here. Hence
within the $q_7^{(\prime)}$-only setting one can mainly rely on the
$2\,\sigma$ constraints. The total sets of constraints of the primed and
unprimed scenarios are not anymore related to each other in a simple way
since they involve $PP$ decays subject to a $180^{\circ}$ rotation together
with $PV$ decays which are unaffected by a
$q_{7,9}\leftrightarrow q_{7,9}^{\prime}$ exchange. It turns out that the
constraints are stronger in the $q_9^{\prime}$-only and in the
$q_7^{\prime}=q_9^{\prime}$ scenarios than in their unprimed counterparts
and that the best fit regions are cut away in these cases.\medskip

From \mbox{figs.~\ref{figq7q9},\ref{figq7q9P}}
the enhancement $\Br^{\textrm{SM}+\textrm{NP}}/\Br^{\textrm{SM}}$ of the
$B_s$ branching fractions can be read off with respect to the different
constraint- and fit-regions.
A large enhancement of the $\bar{B_s}\rightarrow~\phi\rho^0,\phi\pi^0$
branching ratio can be expected in many scenarios, especially in those
involving $q_7^{(\prime)}\neq 0$.  The fact that large parts of the
allowed regions do not overlap with the SM uncertainty regions is encouraging.
It means that, if such NP is realised in nature, it could be possible to
probe it easily. In tabs.~\ref{tab3} and \ref{tab3b} we quote the maximal
enhancement factors that can be obtained considering all points in parameter
space which lie within the $1\,\sigma$ region of the $B\to \pi K$ fit and
fulfill the additional $2\,\sigma$ constraints. The numbers in tab.~\ref{tab3}
refer to input a) while the numbers in tab.~\ref{tab3b} refer to input b). The
first number in each cell represents $\Br_{\rm med}^{\textrm{SM}+\textrm{NP}}
/\Br^{\rm SM}_{\rm med}$ while the number in brackets
represents $\Br_{\rm max}^{\textrm{SM}+\textrm{NP}}
/\Br^{\rm SM}_{\rm med}$, both evaluated for
the $q_i$ value which gives the largest enhancement. Here ``max'' and
``med'' refer to the upper limit of the theoretical uncertainty range
and to our default value, respectively, according to our input given in
Appendix~\ref{input}. Exploiting the theory error in favour of an enhancement,
the number in brackets gives the absolutely maximal enhancement possible for
each scenario whereas the first number gives a typical enhancement factor, but
still for the most enhancing parameter point. Concerning the parity-symmetric
scenarios one should have in mind that like the SM they violate
$\Delta A_{\textrm{CP}}$ at the $>2\,\sigma$ level since they have no impact
on $B\to \pi K$ decays. The corresponding enhancement factors shown in
tabs.~\ref{tab3} and \ref{tab3b} are obtained ignoring $\Delta A_{\textrm{CP}}$
but taking into account all other constraints.\medskip

In most scenarios an enhancement of more than an order of magnitude is possible.
Exceptions are $\bar{B}_s\to\phi\pi$ for $q_7^{(\prime)}=q_9^{(\prime)}$ and
$\bar{B}_s\to\phi_{L}\rho_{L}$ for parity-symmetric models and have their
origin in the pattern of eq.~(\ref{eq:rtEWPhiNP}).
Furthermore, effects in the $q_9^{\prime}$ and the
$q_7^{\prime}=q_9^{\prime}$ scenarios are limited by the small allowed region
resulting from the $B\to \pi K$ fit. Largest effects occur as expected in the
scenarios which are least constrained by $B\to \pi K$, i.e. the single
$q_7^{(\prime)}$ and the parity-symmetric models. Especially in these cases
a $\bar{B}_s\to\phi\pi$ measurement would complement $B\to \pi K$ data and,
while the parity-symmetric models lack the motivation via the
$\Delta A_{\textrm{CP}}$ discrepancy, the $q_7^{\prime}$ setting resolves
it with ease (see Fig.~\ref{fig:DeltaACP}). Moreover, we like to stress that
$B\to \pi K$ data alone cannot distinguish among opposite-parity scenarios
because such scenarios generate equal results for the $B\to \pi K$ observables
(for $180^{\circ}$-rotated parameter points). Therefore an analysis of
$B\to \pi K$ should for example be supported by the analysis of a $PV$ decay,
suggesting $\bar B_s\to\phi\pi^0$ as an ideal candidate. \medskip

We have seen that NP in the EW penguin coefficients allows for an enhancement
of $\Br(\bar{B}_s\to\phi\pi,\phi\rho)$ of more than an order of magnitude.
According to the simple topological structure of these decays, the observation
of such an effect would be a clear and unambiguous signal for such a scenario.
It is interesting to raise also the reversed question, i.e. whether the absence
of such an effect would rule out a NP solution of the $\Delta A_{\textrm{CP}}$
discrepancy, at least for a specific scenario. This is, however, not compulsory.
In nearly all the considered settings there are points within the $1\,\sigma$
region of the $B\to \pi K$ fit which do not generate an enhancement of
$\Br(\bar{B}_s\to\phi\pi,\phi\rho)$. The only exception is the $q_9^{\prime}$-only case: Here an enhancement factor of at least 2.1 would occur
in $\bar{B}_s\to\phi\pi$. This time we have exploited the theoretical error
in disfavour of an enhancement (for the default value the factor is 2.7). Finally we provide in tab.~\ref{tab4} the $\bar{B}_s\to\phi\pi,\phi\rho$ branching
ratios for the best fit point in the $q_9$-only scenario.\medskip
\begin{table}
  \centering
  \begin{tabular}{|l|r|r|}
    \hline
    Observable & $|q_9|=1.9$,         \\
               & $\phi_9 = -100^{\circ}$, \\
\hline
    $\Br(\bar B_s \rightarrow \phi\pi^0)\cdot 10^{6}$     &
$0.35^{+0.41}_{-0.19}$ \\
    $\Br_L(\bar B_s \rightarrow \phi\rho^0)\cdot 10^{6}$  &
$0.90^{+0.98}_{-0.46}$ \\
    $\Br(\bar B_s \rightarrow \phi\rho^0)\cdot 10^{6}$    &
$1.13^{+0.95}_{-0.38}$ \\ \hline
  \end{tabular}
  \caption{Values of various observables at our best fit point in the scenario $q_{9}\neq 0$.}\label{tab4}
\end{table}
\clearpage

\section{Analysis of viable New-Physics models}\label{moddep}

In view of the results in chapter \ref{modind}, the question arises which
concrete models for NP can provide a large new EW penguin amplitude
without being excluded by present data. In this section we consider a number of
well-motivated NP models. The main difference with respect to the
model-independent analysis is the possibility of adding constraints from other
flavour processes beyond the hadronic $B$ modes, e.g.\ the semileptonic decay
$\bar{B}\to X_se^+e^-$, the radiative decay $\bar B\to X_s \gamma$ and
$B_s$-$\bar{B}_s$ mixing. These processes usually yield tight constraints on new
flavour structures and it has to be investigated if the effects in $B\to \pi K$
and $\bar B_s\to\phi\rho^0,\phi\pi^0$ survive these constraints.\bigskip

\boldmath
\subsection{Constraints from semileptonic decays and \texorpdfstring{$B_s-\bar{B}_s$}{Bs} mixing}\label{SemilepDeltaMs}
\unboldmath

Before turning to the models we summarise here how we implement constraints from
semileptonic $B$ decays and $B_s$-$\bar{B}_s$ mixing. The inclusive semileptonic decay $\bar{B}\to X_se^+e^-$ is generated by
electroweak interactions and therefore correlated to the hadronic EW penguins in
many models. We describe it by the effective Hamiltonian (\ref{g1a}), adding the
operators
\be\label{c1}
Q_{9V}\,=\,(\bar{s}_{\alpha}b_{\alpha})_{V-A}\,(\bar{l}l)_V\hspace{1cm}{\rm{and}}\hspace{1cm}
Q_{10A}\,=\,(\bar{s}_{\alpha}b_{\alpha})_{V-A}\,(\bar{l}l)_A
\ee
and the corresponding mirror copies $Q_{9V}^{\prime}$, $Q_{10A}^{\prime}$.
The SM expressions for the Wilson coefficients
can be found e.g.\ in refs. \cite{Buras:1994dj,Buchalla:1993bv}. Following
\cite{Buras:1994dj} and extending the formulae therein to include effects
of the mirror operators, we use the effective Hamiltonian to calculate the
ratio
\be\label{c3}
R_{e^+e^-}(q^2) \equiv \frac{\frac{d}{d q^2} \Gamma(b\to s\,e^+e^-)}{\Gamma(b\to
c\,e\bar\nu)},
\ee
where $q^2 = (p_{e^+}+p_{e^-})^2$ is the squared invariant mass of the lepton
pair. This ratio has the advantage that its theoretical uncertainty is
considerably reduced with respect to the simple branching fraction. We integrate
over a continuum region below the $\psi$ resonances to find the integrated ratio
\be\label{c4}
R_{e^+e^-}|_{[1,6]}\equiv\int_{1\rm GeV^2}^{6 \rm GeV^2} R_{e^+e^-}(q^2)d
q^2,
\ee
which we can finally compare to the experimental result
\cite{Glenn:1997gh,Aubert:2004it,Abe:2004sg,Iwasaki:2005sy}
\be\label{c5}
\Br_{e^+e^-} |_{[1,6]} = (1.60\pm 0.51) \cdot 10^{-6},
\ee
also normalized to the semileptonic decay. We require $R_{e^+e^-}|_{[1,6]}$ to be compatible with experimental data according to (\ref{g231b}).\medskip

Besides the inclusive $\bar{B}\to X_se^+e^-$, also the exclusive mode
$\bar{B}\to K^*l^+l^-$ has been found to be a
useful constraint for NP \cite{Bobeth:2008ij,Egede:2008uy,Altmannshofer:2008dz}.
Here we focus only on the forward-backward asymmetry
$A_{\rm FB}$ of this process \cite{Bobeth:2008ij}, which gives a constraint
complementary to that of $R_{e^+e^-}|_{[1,6]}$. In the light of present
experimental data we require the sign of $A_{\rm FB}(q^2)$ integrated over
$q^2>14$ GeV$^2$ to be negative.\medskip

For completeness we note that we use a renormalisation-group evolution analogous
to the one of the EW penguin operators, treating the parts of $C_{9V}$ and
$C_{10A}$ enhanced by $x_{tW}=m_t^2/M_W^2$ and/or $1/\sin^2\theta_W$ as leading
order. This results in the following SM initial conditions at the scale
$\mu\sim\mathcal{O}(M_W)$:
\bea\label{c2}\nn
C^{(0)}_{9V}&=&\frac{\alpha}{2\pi}\,\left(\frac{Y_0(x_{tW})}{\sin^2\theta_W}
-\frac{x_{tW}}{2}\right),\\ \nn
C^{(1)}_{9V}&=&\frac{\alpha}{2\pi}\,\left(-4Z_0(x_{tW})+\frac{x_{tW}}{2}+\frac{4
}{9}\right)\,
+\,\frac{\alpha}{2\pi}\,\frac{\alpha_s}{4\pi}\,\left(\frac{Y_1(x_{tW})}{
\sin^2\theta_W}-4x_{tW}\,\left(\frac{4}{3}-\frac{\pi^2}{6}\right)\right),\\
  C^{(0)}_{10A}&=&-\frac{\alpha}{2\pi}\,\frac{Y_0(x_{tW})}{\sin^2\theta_W},
\hspace{2cm}
C^{(1)}_{10A}=-\frac{\alpha}{2\pi}\,\frac{\alpha_s}{4\pi}\,\frac{Y_1(x_{tW})}{
\sin^2\theta_W}.
\eea
The functions $Y_{0,1}$ and $Z_0$ can be found e.g.\ in
\cite{Buchalla:1995vs}.\medskip

Finally, we consider constraints coming from
$B_s$-$\bar{B}_s$ mixing, which is described by the
effective weak Hamiltonian
\begin{equation}\label{c6}
{\cal H}_{\rm eff}^{(2)} \,=\, \frac{G_F^2 M_W^2}{4\pi^2}\,
(\lambda_t^{(s)})^2\,\sum_i\, C_i\, Q_i\,,
\end{equation}
with the operators \cite{Buras:2000if}
\begin{eqnarray}
   Q^{\textrm{VLL}}&=&(\bar{s}_{\alpha}\gamma^{\mu}P_L b_{\alpha})\,
                (\bar{s}_{\beta}\gamma_{\mu}P_L b_{\beta}), \nonumber\\[0.1cm]
   Q_1^{\textrm{SLL}}&=&(\bar{s}_{\alpha}P_L b_{\alpha})\,(\bar{s}_{\beta}P_L b_{\beta}),\hspace{1.8cm}
   Q_2^{\textrm{SLL}}\,=\,(\bar{s}_{\alpha}\sigma^{\mu\nu}P_L b_{\alpha})\,(\bar{s}_{\beta}\sigma_{\mu\nu}P_L b_{\beta}), \hspace{1cm}\nonumber\\[0.1cm]
   Q_1^{\textrm{LR}}&=&(\bar{s}_{\alpha}\gamma^{\mu}P_L b_{\alpha})\,(\bar{s}_{\beta}\gamma_{\mu} P_R b_{\beta}),\hspace{1cm}
   Q_2^{\textrm{LR}}\,\,\,=\,(\bar{s}_{\alpha}P_L b_{\alpha})\,(\bar{s}_{\beta} P_R b_{\beta})
\end{eqnarray}
and the mirror copies $Q^{\textrm{VRR}}$, $Q_1^{\textrm{SRR}}$ and $Q_2^{\textrm{SRR}}$.
In the SM only $C^{\textrm{VLL}}\neq 0$, while in extensions
of the SM all operators can receive contributions. The matrix element relevant
for $B_s$-$\bar{B}_s$ mixing,
\begin{equation}\label{c7}
M_{12}^{B_s} = \frac{1}{2m_{B_s}}\langle B_s^0 |{\cal H}_{\rm eff}^{(2)}|\bar B_s^0\rangle,
\end{equation}
is evaluated using lattice results from ref.~\cite{Lubicz:2008am}.
Besides the $B_s$-$\bar{B}_s$ mass difference
\begin{equation}\label{c8}
\Delta M_s = 2 |M_{12}^{B_s}| \,\stackrel{\text{exp.}}{=} \, (17.77 \pm 0.12)\, {\rm ps}^{-1}\,,
\end{equation}
\cite{Abulencia:2006ze}, we use the quantity \cite{Lenz:2006hd}
\begin{equation}\label{c9}
\Delta_s \equiv \frac{M_{12}^{B_s}}{M_{12}^{B_s,\rm SM}} = |\Delta_s| e^{i \phi_s},
\end{equation}
as additional constraint. This observable has been analysed in ref.~\cite{Lenz:2010gu}
in different generic NP scenarios and evidence for a NP contribution with a large new weak phase has been found. A fit of $\Delta_s$ and the analogous quantity $\Delta_d$
to data shows a $3.6\,\sigma$ discrepancy for the SM value $\Delta_s=1$.
In our study of the $Z^\prime$ models we
take those points of the NP parameter space as excluded which give a $\Delta_s$
outside the $2\,\sigma$ region drawn in fig.~9 of ref.~\cite{Lenz:2010gu}.\bigskip

\boldmath
\subsection{The modified-\texorpdfstring{$Z^0$}{Z0}-penguin scenario}\label{Z0P}
\unboldmath

The simplest class of models with large new contributions to EW penguins
comprises models with a modified $Z\bar{s}b$ coupling. Such
a FCNC coupling can either be generated by integrating out new heavy particles,
e.g.\ in supersymmetric models or fourth-generation models, or it can exist at
tree-level in more exotic scenarios like models with non-sequential quarks,
see e.g.\ ref.~\cite{Langacker:1988ur}. Consequences for hadronic $B$ decays
have been considered for example in \cite{Grossman:1999av}, more detailed
analyses of the motivation and the effects in flavour physics have been
performed in \cite{Buras:2004ub,Buchalla:2000sk}.

\subsubsection{Effective Theory}\label{heffZ0}

Our parameterisation of the $Z\bar{s}b$ coupling follows ref.~\cite{Grossman:1999av}.
At the electroweak scale we have an effective theory with the Lagrangian
\be\label{c10}
{\cal L}^{eff}_{Z} = -\frac{g}{4\cos \theta_W}\sum_{I\neq J} \bar{d}_I
\left[\kappa_L^{IJ}\gamma^{\mu}(1-\gamma_5)
+\kappa_R^{IJ}\gamma^{\mu}(1+\gamma_5)\right]d_J Z_{\mu},
\ee
where $I,J$ are generation indices. Since the flavour-violating couplings
are expected to be small, the flavour-diagonal couplings of the $Z$ bosons
are to a first approximation the same as in the SM.
Matching tree-level diagrams with $Z$ exchange onto the $\Delta B=\Delta S=1$
effective Hamiltonian adds new contributions $\delta C_i$ to the SM Wilson
coefficients $C_i$ and generates coefficients $C_i^{\prime}$ of the
mirror operators. The resulting contributions at the electroweak scale read
\begin{eqnarray}\label{c11}\nn
\delta C_{3} &=& \phantom{-}\,\frac{1}{6}\,\frac{\kappa_L^{sb}}{\lambda_t^{(s)}}\,, \hspace{4.3cm}
C^{\prime}_5 \,=\, \phantom{-}\,\frac{1}{6}\,\frac{\kappa_R^{sb}}{\lambda_t^{(s)}}\,, \nonumber\\
\delta C_{7} &=& \phantom{-}\,\frac{2}{3}\,\frac{\kappa_L^{sb}}{\lambda_t^{(s)}}\,\sin^2\theta_W\,, \hspace{3cm}
C^{\prime}_7 \,=\,   -\,\frac{2}{3}\,\frac{\kappa_R^{sb}}{\lambda_t^{(s)}}\, \cos^2\theta_W\,,\nonumber\\
\delta C_{9} &=& -\,\frac{2}{3}\,\frac{\kappa_L^{sb}}{\lambda_t^{(s)}}\,\cos^2\theta_W\,, \hspace{3cm}
C^{\prime}_9 \,=\,   \phantom{-}\,\frac{2}{3}\,\frac{\kappa_R^{sb}}{\lambda_t^{(s)}}\, \sin^2\theta_W\,.
\end{eqnarray}
They reach the size of the dominant SM Wilson coefficient
$C_9(\mu_W) $ if
\begin{equation}\label{eq:kappaSM}
|\kappa^{sb}_{L,R}|\, \sim\,|\kappa_{\textrm{SM}}|\,\equiv \,
  \frac{\alpha}{\pi\sin^2\theta_W}\,\lambda_t^{(s)}\,C_0(x_{tW})\,\sim\,
  0.00035\,,
\end{equation}
 where $C_0(x)$ is a loop function,
see e.g. ref.~\cite{Buchalla:1995vs}. Such a scenario corresponds
to $q_i\sim\mathcal O(1)$ in our model-independent analysis, thus
we expect significant effects in hadronic $B$ decays for such values
of $\kappa^{sb}_{L,R} $.\medskip

From the Lagrangian (\ref{c10}) and the SM coupling of the $Z$ to leptons
we also obtain corrections to the short-distance coefficients of the
semileptonic operators (\ref{c1}), namely
\begin{eqnarray}\label{c12}
\delta C_{9V} &=& -\,\frac{\kappa_L^{sb}}{\lambda_t^{(s)}}\left(2\sin^2 \theta_W -\frac{1}{2}\right), \hspace{1cm}
C^{\prime}_{9V} \,=\, -\,\frac{\kappa_R^{sb}}{\lambda_t^{(s)}}\left(2\sin^2 \theta_W -\frac{1}{2}\right),\nonumber \\[0.1cm]
\delta C_{10A} &=& -\,\frac{\kappa_L^{sb}}{\lambda_t^{(s)}}\left(\frac{1}{2}\right), \hspace{1.7cm} \hspace{1.3cm}
C^{\prime}_{10A} \,=\, -\,\frac{\kappa_R^{sb}}{\lambda_t^{(s)}}\left(\frac{1}{2}\right).\hspace{0.5cm}
\end{eqnarray}
This enables us to study constraints on $\kappa^{sb}_{L,R} $ from
semileptonic $B$ decays as indicated in the previous section.\medskip

Diagrams with $Z$-exchange contribute also to $B_s$-$\bar{B}_s$ mixing
via the Wilson coefficients
\begin{eqnarray}\label{c13}
\delta C_1^{\rm VLL}&=& \frac{4 \pi^2}{\sqrt{2}\,G_F M_W^2}\,
         \left(\frac{\kappa_L^{sb}}{\lambda_t^{(s)}}\right)^2, \hspace{1cm}
C_1^{\rm VRR}\,=\, \frac{4 \pi^2}{\sqrt{2}\,G_F M_W^2}\,
         \left(\frac{\kappa_R^{sb}}{\lambda_t^{(s)}}\right)^2, \nonumber \\[0.1cm]
C_1^{\rm LR}&=& \frac{8 \pi^2}{\sqrt{2}\,G_F M_W^2}\,
         \frac{\kappa_L^{sb}}{\lambda_t^{(s)}}
         \frac{\kappa_R^{sb}}{\lambda_t^{(s)}}\,.
\end{eqnarray}
Explaining the discrepancy in $\Delta_s$ defined in eq.~(\ref{c9})
with the help of these new contributions would push the couplings
$\kappa_{L,R}^{sb}$ to large values. Note, however, that in most
realistic cases the couplings $\kappa_{L,R}^{sb}$ are loop-induced
with the consequence of eq.~(\ref{c13}) actually representing
two-loop effects. Usually such scenarios provide also one-loop
contributions from box diagrams which then are more likely to
account for the $\Delta_s$ discrepancy. Therefore we prefer
not to include $\Delta_s$ as a constraint into our analysis
and regard a potential relaxation of the $\Delta_s$ discrepancy
only as a bonus feature.\bigskip

\subsubsection{Results}\label{Z0res}

\begin{figure}[th]
\begin{center}
  \includegraphics[width=1.00\textwidth]{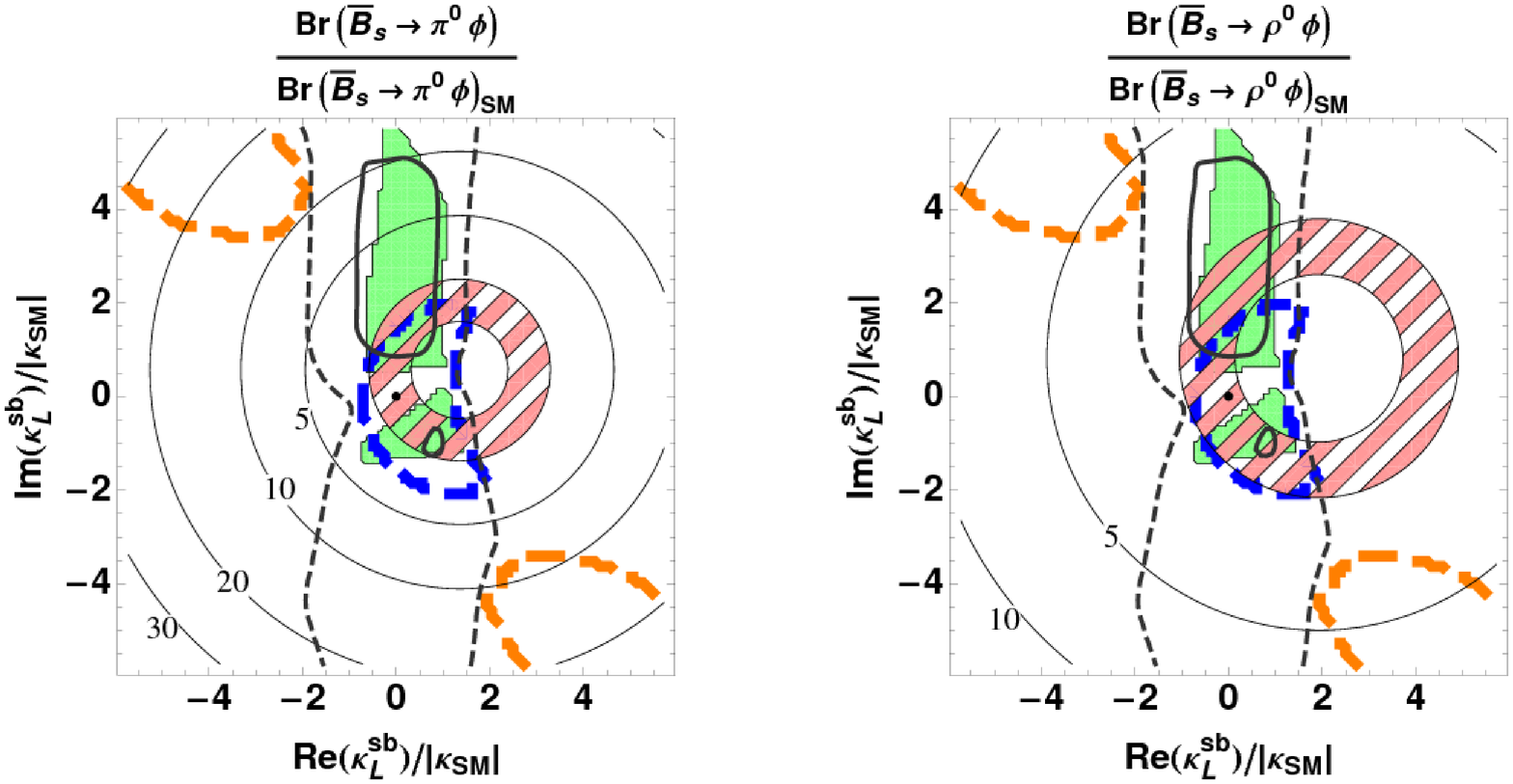}
  \includegraphics[width=1.00\textwidth]{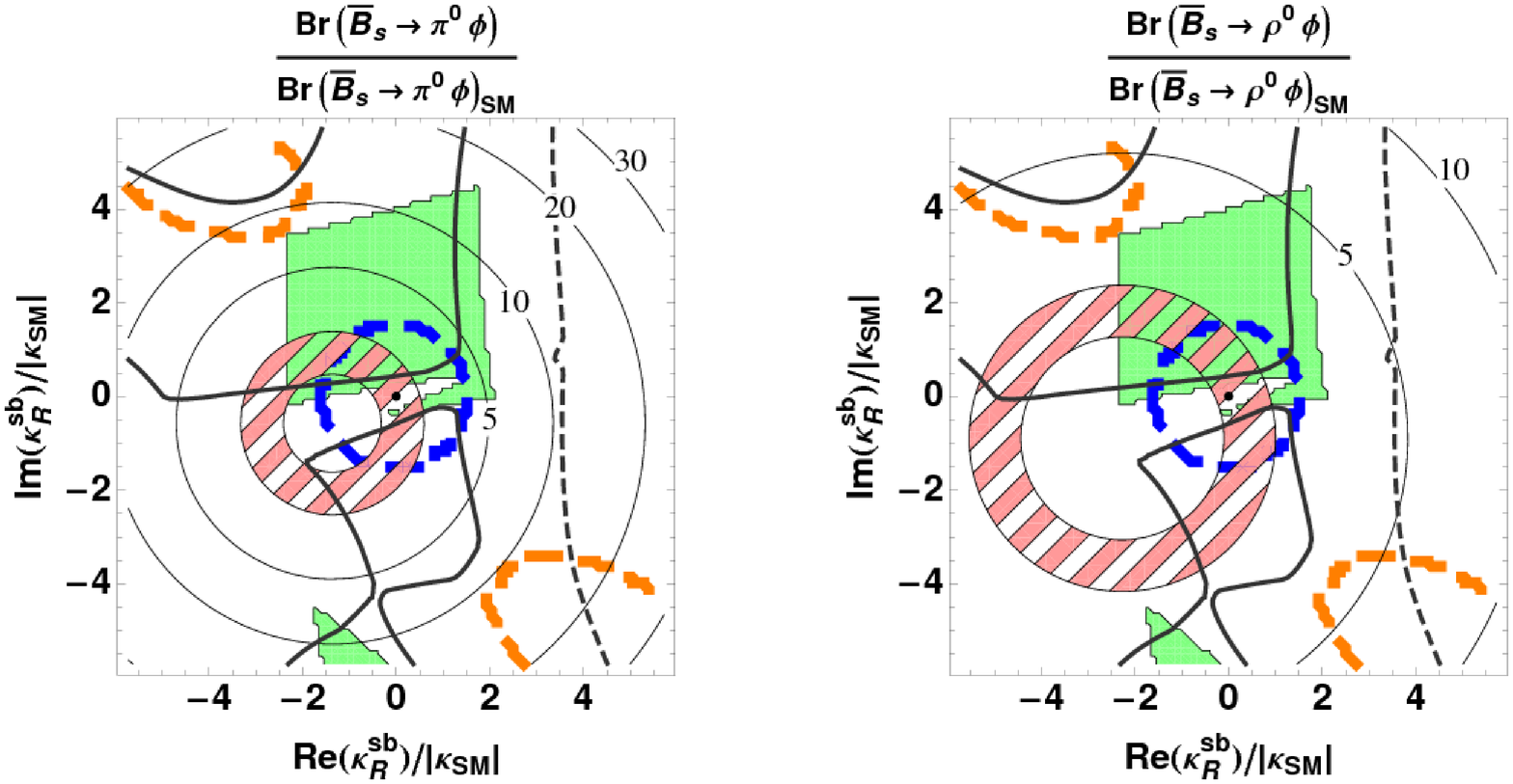}
\end{center}
  \caption{Enhancement factor for the $\bar{B_s}\rightarrow~\phi\rho^0,\phi\pi^0$
  branching ratios with respect to their SM values in the modified-$Z^0$-penguin
  scenario. The green area represents the region allowed by the $2\sigma$
  constraints from all the considered hadronic decays, while the area inside the
  dashed blue line represents the region allowed by the $2\sigma$ constraint
  from semi-leptonic decays. The areas inside the dashed orange line represent the
  parameter values for which the modified-$Z^0$-penguin would solve $\Delta_s$. See the text
  for further explanations.}
  \label{figZ0}
\end{figure}

In our study of the modified $Z$ coupling we consider the three special cases of non-vanishing $\kappa_L^{sb}$
only, $\kappa_R^{sb}$ only and $\kappa_L^{sb}=\kappa_R^{sb}$, similarly to the model-independent analysis. Since
$\cos^2\theta_W\gg\sin^2\theta_W$, the $\kappa_L^{sb}$ scenario shares its most
important features with the $q_9$ setup of the model-independent study and the
same holds for $\kappa_R^{sb}$ and $ q_7^\prime$. This expectation is confirmed
by the graphs in fig.~\ref{figZ0}, we only note
that we get a $180^\circ$ rotation due to the signs of $\delta C_9$ and $C^{\prime}_7$.
We have again marked the $1\sigma$\,-\,region of the $B\to\pi K$ fit by a black
line (as well as the additional $3\sigma$\,-\, black dotted line) and the region
allowed by the $2\,\sigma$ constraints from all hadronic decay
observables by a green area. The displayed regions refer to input a), as defined
in section \ref{modindres}, while we have refrained from showing the corresponding
regions for input b).\medskip

The main difference to the more general model-independent approach is that we now
face additional constraints from semileptonic decays and $B_s$-$\bar{B}_s$ mixing.
The allowed region for the former is given by the interior of the blue dashed curve,
the allowed region for the latter by the orange areas outside the zone preferred by the
$B\to \pi K$ fit. We see that the $\Delta_s$ anomaly of $B_s$-$\bar{B}_s$ mixing cannot
be resolved in a modified $Z$ scenario when fulfilling at the same time the semileptonic
constraints. This has already been noted in Ref.~\cite{Altmannshofer:2009ma}. Here we
recognise that also $B\to \pi K,\rho K,\pi K^*$ data are not compatible with a solution
of $\Delta_s$ in this way. In the previous section we remarked that it is plausible
to assign the explanation of $\Delta_s$ to other effects not directly related to the
modified $Z$ coupling. Pursuing this strategy, we are left with the semileptonic
decays which are compatible with the $1\,\sigma$ region of the $B\to \pi K$ fit
for all three cases but constrain the FCNC couplings $\kappa_{L,R}^{sb}$ to very
small values as can be seen from fig.~\ref{figZ0} where the coupling
$\kappa_{L,R}^{sb}$ is normalised to $|\kappa_{\textrm{SM}}|$ in
Eq.~(\ref{eq:kappaSM}).\medskip

As a consequence we expect no significant effects in $\bar B_s\to\phi\pi^0,\phi\rho^0$.
This expectation is confirmed by fig.~\ref{figZ0} and by the maximum enhancement
factors given in tab.~\ref{tab5}, which are determined in analogy to the ones in
tab.~\ref{tab3}. In the $\kappa_L^{sb}=\kappa_R^{sb}$ case no enhancement occurs
at all because of the pattern in eq.~(\ref{eq:rtEWPhiNP}): Equal contributions
to $C_7$ and $C_9^\prime$ and to $C_9$ and $C_7^\prime$ cancel pairwise. The
largest effect which one could gain in the other scenarios is a factor of
$\sim 4$ in the case where only $\kappa_R^{sb}\neq 0$. Therefore an enhancement of
$\bar B_s\to\phi\pi^0,\phi\rho^0$ due to a new modified $Z$ contribution  in practice becomes
indistinguishable from the potential enhancement caused by a
large non-factorisable SM effect. In fig.~\ref{figZ0} this is reflected
by the fact that the red-striped ring representing parameter points which
reproduce the SM result for the $B_s$ decays nearly fills the whole allowed
region of the parameter space.\medskip

Our results can be summarised as follows: The constraints from semileptonic
decays still allow for a solution of $\Delta A_{\textrm{CP}}$ via a modified
$Z$ coupling. This possibility would be excluded if an enhancement of
$\bar B_s\to\phi\pi^0$ or $\bar B_s\to\phi\rho^0$ by an order of magnitude was found.\bigskip

\begin{table}
 \centering
 \begin{tabular}{|c|r|r|r|} \hline
   Scenario     & $\frac{\Br(\bar B_s \rightarrow \phi\pi^0)}{\Br^{\rm SM}(\bar
B_s \rightarrow \phi\pi^0)}$
                & $\frac{\Br_L(\bar B_s \rightarrow \phi\rho^0)}{\Br_L^{\rm
SM}(\bar B_s \rightarrow \phi\rho^0)}$
                & $\frac{\Br(\bar B_s \rightarrow \phi\rho^0)}{\Br^{\rm SM}(\bar
B_s \rightarrow \phi\rho^0)}$ \\ \hline
 $\kappa_L^{sb}\neq 0$           &  10.3          (19.0) &   3.6\phantom{0}(7.0) &  3.4\phantom{0}(6.3) \\
 $\kappa_R^{sb}\neq 0$           &  48.3          (84.6) &  15.5          (28.2) & 14.2          (24.8) \\
 $\kappa_L^{sb}=\kappa_R^{sb} $ &   1.0\phantom{0}(1.7) &   1.0\phantom{0}(1.7) &  1.2\phantom{0}(1.8) \\ \hline
    \multicolumn{4}{|c|}{With additional semileptonic $B$ decay constraints} \\
\hline
 $\kappa_L^{sb}\neq 0$           &  1.6 (3.0) &  1.1 (2.2) &  1.1 (2.0) \\
 $\kappa_R^{sb}\neq 0$           &  4.0 (6.5) &  2.4 (3.9) &  2.2 (3.5) \\
 $\kappa_L^{sb}=\kappa_R^{sb} $ &  1.0 (1.7) &  1.0 (1.7) &  1.1 (1.7) \\ \hline
 \end{tabular}
\caption{Maximal possible enhancement of the $B_s$ branching ratios
in the modified-$Z^0$-penguin model. The upper part has been calculated
with the method a) of the model-independent analysis, the lower part
includes the $2\sigma$ constraints from semileptonic decays.}
\label{tab5}
\end{table}

\clearpage
\boldmath
\subsection{Models with an additional \texorpdfstring{$U(1)'$}{U(1)'} gauge symmetry}\label{Zprime}
\unboldmath

The presence of a heavy $Z^{\prime}$ boson associated with an additional
$U(1)'$ gauge symmetry is a well-motivated extension of the SM.
This additional symmetry has not been invented to solve a particular problem
of the SM, but rather occurs as a byproduct in many models
like e.g.\ Grand Unified Theories, various models of dynamical symmetry breaking
and Little-Higgs models. In many scenarios the $Z^{\prime}$ boson is expected
to have a mass at the TeV scale. It also appears in the form of a Kaluza-Klein
excitation of the SM $Z^0$ boson in theories with extra dimensions. An extensive
review about the physics of $Z^{\prime}$ gauge-bosons can be found in
\cite{Langacker:2008yv}. Here we are interested in implications
for flavour physics as discussed in
\cite{Grossman:1999av,Barger:2009eq,Chang:2009wt,Barger:2009qs}.

\subsubsection{Effective Theory}\label{heffZP}

We consider a model with an additional $Z'$ gauge-boson, neglecting
$Z$-$Z'$ mixing and assuming the absence of exotic fermions which
could mix with the SM fermions through non-universal $Z'$
couplings.  We write the general quark-antiquark-$Z^{\prime}$ coupling as
\cite{Grossman:1999av,Barger:2009qs}
\be\label{c20}
{\cal L}^{eff}_{Z^{\prime}} = -\frac{g_{U(1)'}}{2\sqrt{2}}\sum_{IJ} \bar{d}_I
\left[\zeta_L^{IJ}\gamma^{\mu}(1-\gamma_5)
+\zeta_R^{IJ}\gamma^{\mu}(1+\gamma_5)\right]d_J Z^{\prime}_{\mu}.
\ee
and similarly for the up-type quarks.
The couplings of interest are the flavour-changing $\zeta_{L,R}^{sb}$ as well
as the flavour-conserving charges $\zeta_{L,R}^{u}\equiv\zeta_{L,R}^{uu}$ and
$\zeta_{L,R}^{d}\equiv\zeta_{L,R}^{dd}$. Note that $SU(2)_L$ invariance implies
$\zeta_L^u=\zeta_L^d\equiv\zeta_L^q$ whereas no restrictions hold
in case of $\zeta_R^u$, $\zeta_R^d$. Following ref.~\cite{Grossman:1999av}
we introduce the parameter
\begin{equation}
    \xi \,\equiv\, \frac{g_{U(1)'}^2}{g^2}\,\frac{M_W^2}{M_{Z'}^2}
\end{equation}
with $g_{U(1)'}$ denoting the gauge coupling of the additional $U(1)'$ gauge
group and $M_{Z'}$ being the mass of the $Z'$-boson.
We then find the following additional contributions
to the short-distance coefficients at the electroweak scale:
\begin{eqnarray}\label{c21}
\delta C_{3}\, =\, -\,\frac{\zeta_L^{sb}}{\lambda_t^{(s)}}\,\zeta_L^{q}\,\xi\,, \hspace{3.8cm} &&
C^{\prime}_3 \,=\,- \,\frac{1}{3}\,\frac{\zeta_R^{sb}}{\lambda_t^{(s)}}\,
       \left(\zeta_R^{u}+2\zeta_R^{d}\right)\,\xi\,, \nonumber\\
\delta C_{5}\, =\,-\,\frac{1}{3}\, \frac{\zeta_L^{sb}}{\lambda_t^{(s)}}\,
                \left(\zeta_R^{u}+2\zeta_L^{d}\right)\,\xi\,,\hspace{1cm} \qquad &&
C^{\prime}_5\, =\,-\, \frac{\zeta_R^{sb}}{\lambda_t^{(s)}}\,\zeta_L^{q}\,\xi\,, \nonumber\\
\delta C_{7}\, =\,-\, \frac{2}{3}\,\frac{\zeta_L^{sb}}{\lambda_t^{(s)}}\,
                \left(\zeta_R^{u}-\zeta_R^{d}\right)\,\xi\,, \hspace{1.1cm}\qquad &&
C^{\prime}_7\, =\, \phantom{-}\,0\,, \nonumber\\
\delta C_{9}\, =\, \phantom{-}\,0\,, \hspace{5.1cm} &&
C^{\prime}_9\, =\, -\,\frac{2}{3}\,\frac{\zeta_R^{sb}}{\lambda_t^{(s)}}\,
                    \left(\zeta_R^{u}-\zeta_R^{d}\right)\,\xi\,.
\end{eqnarray}
Apart from $\xi$, FCNC transitions are controlled by the free parameters
$\zeta_{L,R}^{IJ}$. Depending on them, the flavour-changing
transitions contribute in general to both QCD and EW penguin operators,
as well as to their mirror copies. Here we follow the approach of refs.~\cite{Buras:2004ub,Barger:2009eq,Chang:2009wt,Barger:2009qs}
in which the main contribution is supposed to reside in the
EW penguins, i.e.\ $|\delta C_{3,5}(\mu_W)|\ll |\delta C_{7}(\mu_W)|$,
$|C_{3,5}^{\prime}(\mu_W)|\ll |C_{7}^{\prime}(\mu_W)|$. We implement
this assumption by setting $\zeta_R^{u}+2\zeta_R^{d} = \zeta_L^{q} = 0$.
The constant $\zeta_R^{u}-\zeta_R^{d}$ can then be absorbed into a
redefinition of $g_{U(1)'}$. After these simplifications we are left
with only two non-zero coefficients
\begin{equation}\label{c21b}
\delta C_{7}\, =\,-\, \frac{2}{3}\,\frac{\widetilde\zeta_L^{sb}}{\lambda_t^{(s)}}\,, \hspace{2.1cm}\qquad
C^{\prime}_9\, =\, -\,\frac{2}{3}\,\frac{\widetilde\zeta_R^{sb}}{\lambda_t^{(s)}}\,,
\end{equation}
where we have defined $\widetilde \zeta_{L,R}^{sb}\equiv \xi \zeta_{L,R}^{sb}$.\medskip

The coupling of the $Z'$ boson to quarks is not related to its
coupling to leptons. Therefore tight constraints from semileptonic
decays, as we encountered in the case of a modified $Z$ coupling,
can be avoided here by simply switching off the $Z'$ coupling to leptons.
Such ``leptophobic'' $Z'$ bosons can for example appear in models with
an $E_6$ gauge symmetry (see e.g. ref.~\cite{Rizzo:1998ut}). Since
leptophobic $Z'$ bosons avoid detection via traditional Drell-Yan
processes, their mass is much less constrained allowing for larger
values of the parameter $\xi$.\medskip

Besides constraints from hadronic $B$ decays we have to face
constraints from $B_s$-$\bar{B}_s$ mixing to which
tree-level $Z'$ exchange contributes. We find for the
$\Delta B=2$-Hamiltonian:
\begin{eqnarray}\label{c22}
\delta C_1^{\rm VLL}&=& \frac{4\pi^2\sqrt{2}}{G_F M_W^2}\,\left(\frac{\widetilde\zeta_L^{sb}}{\lambda_t^{(s)}}\right)^2\,\frac{1}{\xi}, \hspace{2cm}
C_1^{\rm VRR}\,=\, \frac{4\pi^2\sqrt{2}}{G_F M_W^2}\,\left(\frac{\widetilde\zeta_R^{sb}}{\lambda_t^{(s)}}\right)^2\,\frac{1}{\xi}, \nonumber\\[0.1cm]
C_1^{\rm LR}&=& \frac{8\pi^2\sqrt{2}}{G_F M_W^2}\,\left(\frac{\widetilde\zeta_L^{sb}}{\lambda_t^{(s)}}\right)\, \left(\frac{\widetilde\zeta_R^{sb}}{\lambda_t^{(s)}}\right)\,\frac{1}{\xi}\,.
\end{eqnarray}
In contrast to constraints from hadronic $B$ decays, the
$B_s$-$\bar{B}_s$ mixing constraint in the
$(\RE(\widetilde{\zeta}_I^{sb}),\IM(\widetilde{\zeta}_I^{sb}))$ \,-\,plane
depends on the parameter $\xi$ determined by the coupling constant
$g_{U(1)'}$ and the $Z'$ mass $M_{Z'}$. It gets stronger for smaller
$\xi$, i.e. for smaller $g_{U(1)'}$ and larger $Z'$ mass $M_{Z'}$.
This behaviour, which might seem counter-intuitive at first sight,
has its origin in the dependence of the hadronic decays on the parameter
combinations $\widetilde{\zeta}_{I}^{sb}=\xi\,\zeta_{I}^{sb}$. If one
chooses smaller $\xi$ values, one needs larger values of the FCNC
couplings $\zeta_{I}^{sb}$ in order to obtain the same effects in
the hadronic decays. Since the $B_s$-$\bar{B}_s$ mixing coefficients
in (\ref{c22}) depend quadratically on the $\zeta_{I}^{sb}$, this
procedure sharpens their constraints.\bigskip

\subsubsection{Results}\label{Zpres}

\begin{figure}[t]
\begin{center}
 \includegraphics[width=1.0\textwidth]{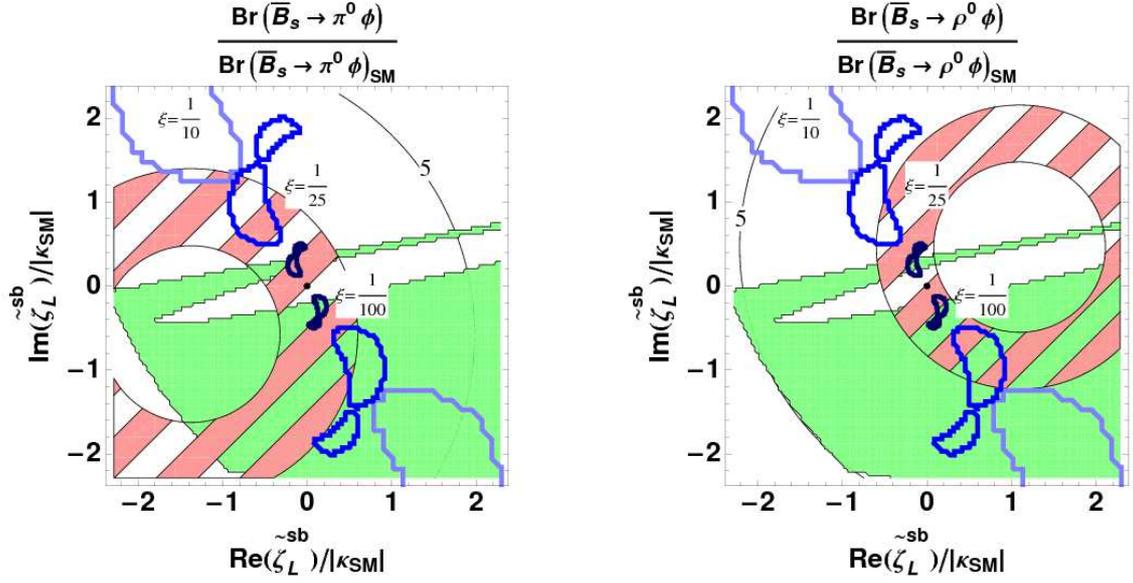}
\end{center}
 \caption{Enhancement factors of $\Br(\bar B_s \rightarrow \phi\pi^0)$ and $\Br(\bar B_s \rightarrow \phi\rho^0)$ for $\tilde \zeta_L^{sb}\neq 0$. The red-hatched ring corresponds to the SM uncertainty. The green area is allowed by the $2\sigma$ constraints from all
hadronic $B$ decays while the regions inside the blue lines are compatible with the constraint from $B_s$-$\bar{B}_s$
mixing. From the biggest to the smallest region they stand for the cases $\xi=1/10$, $\xi=1/25$ and $\xi=1/100$, respectively.}
 \label{figZPL}
\end{figure}

\begin{figure}[t]
\begin{center}
 \includegraphics[width=1.0\textwidth]{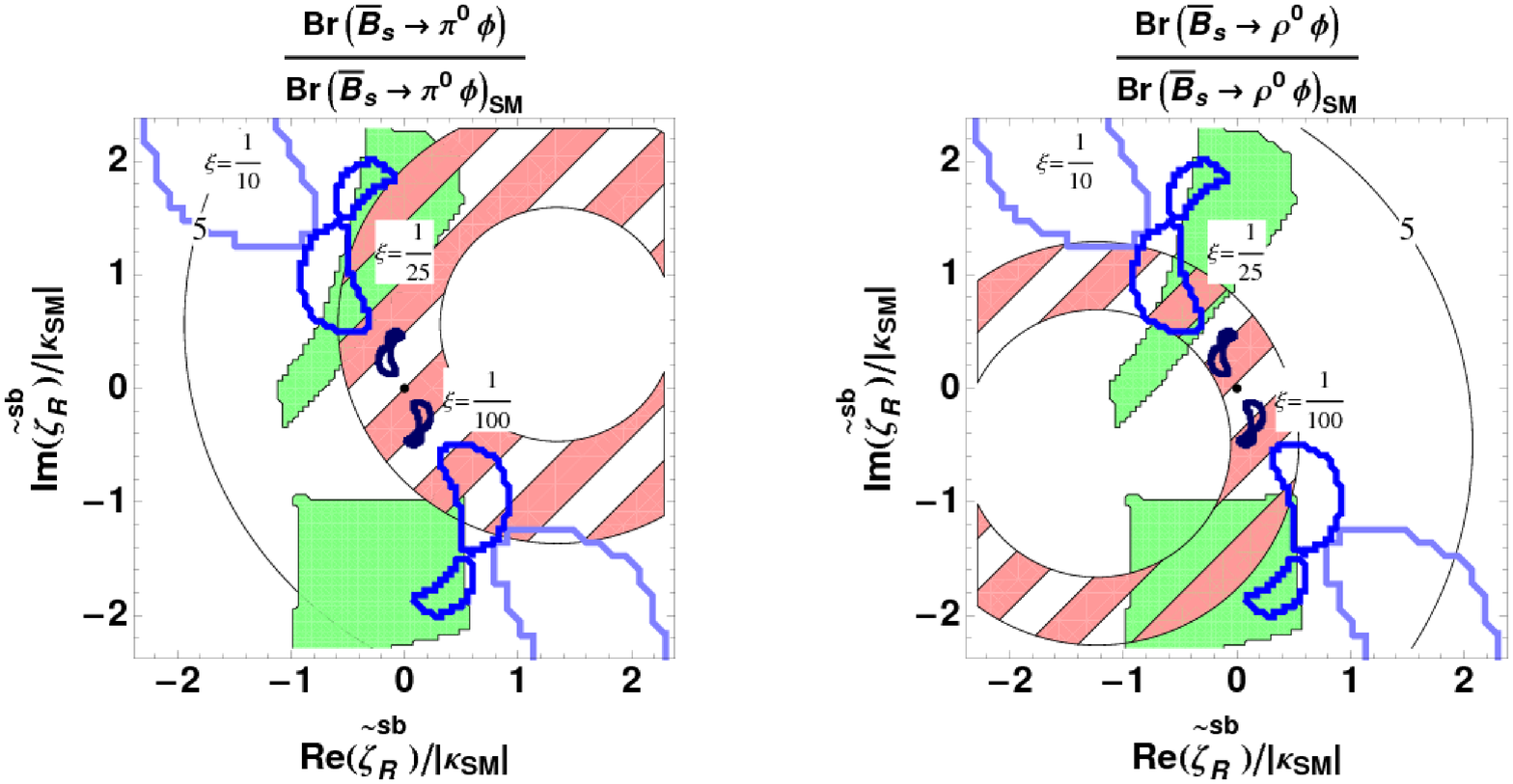}
\end{center}
 \caption{Enhancement factors of $\Br(\bar B_s \rightarrow \phi\pi^0)$ and $\Br(\bar B_s \rightarrow \phi\rho^0)$ for $\tilde \zeta_R^{sb}\neq 0$. The meaning of the coloured regions is the same as in fig.~\ref{figZPL}.}
 \label{figZPR}
\end{figure}

Considering eq.~(\ref{c21b}), one easily sees that the
three scenarios $\tilde\zeta^{sb}_L\neq 0$, $\tilde\zeta^{sb}_R\neq 0$ and
$\tilde\zeta^{sb}_L=\tilde\zeta^{sb}_R\neq 0$ exactly correspond  to the scenarios
$q_7\neq 0$, $q_9^{\prime}\neq 0$ and $q_7=q_9^{\prime}\neq 0$ in
our model-independent analysis, except for a normalisation factor.
In this way the exclusion regions from the $2\,\sigma$ constraints
and the confidence levels from the fit can be immediately read off
from figs.~\ref{figq7q9} and \ref{figq7q9P}, provided one rescales
the axes by an appropriate normalisation factor and rotates the
pictures by $180^{\circ}$ to take into account the minus signs
in eq.~(\ref{c21}).\medskip
\begin{table}[t]
 \centering
 \begin{tabular}{|c|c|r|r|r|} \hline
 \multicolumn{2}{|c|}{Scenario} & $\frac{\Br(\bar B_s \rightarrow
\phi\pi^0)}{\Br^{\rm SM}(\bar B_s \rightarrow \phi\pi^0)}$
                                & $\frac{\Br_L(\bar B_s \rightarrow
\phi\rho^0)}{\Br_L^{\rm SM}(\bar B_s \rightarrow \phi\rho^0)}$
                                & $\frac{\Br(\bar B_s \rightarrow
\phi\rho^0)}{\Br^{\rm SM}(\bar B_s \rightarrow \phi\rho^0)}$ \\ \hline
                   & $\tilde\zeta_L^{sb}\neq 0$                 &  2.5 (4.9) & 3.6 (5.6) & 3.3 (4.9) \\
$\xi=\frac{1}{25}$ & $\tilde\zeta_R^{sb}\neq 0$                 &  3.6 (5.7) & 3.7 (5.7) & 3.4 (5.1) \\
                   & $\tilde\zeta_L^{sb}=\tilde\zeta_R^{sb} $  &  1.0 (1.7) & 1.0 (1.7 )& 1.1 (1.8) \\ \hline
                   & $\tilde\zeta_L^{sb}\neq 0$                 &  1.9 (3.7) & 2.6 (4.0) & 2.4 (3.6) \\
$\xi=\frac{1}{50}$ & $\tilde\zeta_R^{sb}\neq 0$                 &  2.6 (4.1) & 2.6 (4.1) & 2.4 (3.7) \\
                   & $\tilde\zeta_L^{sb}=\tilde\zeta_R^{sb} $  &  1.0 (1.7) & 1.0 (1.7) & 1.1 (1.8) \\ \hline
 \end{tabular}
\caption{Maximal possible enhancement of the $B_s$ branching ratios compatible
with the constraints from all other hadronic decays and $B_s$-$\bar{B}_s$
mixing for the $Z^\prime$ model.}
\label{tab6}
\end{table}

In figs.~\ref{figZPL} and \ref{figZPR} we present our results for the
 $\zeta_L^{sb}$ and the $\zeta_R^{sb}$ scenarios with the
meanings of the green region and the red-hatched ring
being the same as in the preceeding sections. In addition the $2\sigma$
region for $\Delta_s$ is shown for different values of $\xi$. We recognise
that there is very little overlap of the region allowed by hadronic constraints with the region preferred by $\Delta_s$ in the
$\zeta_R^{sb}$ case. The same holds for the $\zeta_L^{sb}=\zeta_R^{sb}$
scenario not shown. This behaviour is easily understood: The observables
$\Delta A_{\textrm{CP}}$ and $\Delta_s$ both call for NP with a large
imaginary part. The branching ratios of hadronic $B$ decays depend
linearly on the real part of $\zeta_{L,R}^{sb}$  at leading order,
thus they pull the $\zeta_{L,R}^{sb}$ values towards the imaginary
axis. The observable $\Delta_s$, on the other hand, depends
quadratically on $\zeta_{L,R}^{sb}$ and favours values on the diagonal $\RE(\tilde \zeta^{sb}_{L,R})=-\IM(\tilde \zeta^{sb}_{L,R})$.
For the $\zeta_L^{sb}$ setting this situation is relaxed due to the weak
constraints from $B\to \pi K$ such that one can solve the two
experimental discrepancies in $\Delta A_{\textrm{CP}}$ and in $\Delta_s$
at the same time.\medskip

From the diagrams we see further that the $B_s$-$\bar{B}_s$ mixing
constraint is very tight. It prohibits large effects in
$\bar B_s\to\phi\pi^0,\phi\rho^0$ for realistic values of the parameter
$\xi\lesssim 1/25$. For $\xi=1/25$, which would correspond for
example to $g_{U(1)'}\sim g$ and $M_{Z'}\sim 400\gev$, and for $\xi=1/50$ we present
the maximum enhancement factors in tab.~\ref{tab6}. These numbers
are obtained abandoning the $1\,\sigma$ region of the $B\to \pi K$
fit and requiring only agreement with the $2\,\sigma$ constraints.
We find that enhancement of a factor $\sim 5$ is possible in the
$\zeta_L^{sb}$ and $\zeta_R^{sb}$ scenarios whereas no effect can
occur in the $\zeta_L^{sb}=\zeta_R^{sb}$ case because of
eq.~(\ref{eq:rtEWPhiNP}). For $\xi = 1/100$ the constraints from
$B_s$-$\bar{B}_s$ mixing become so strong that no effect in $\bar B_s\to\phi\pi^0,\phi\rho^0$ would be detectable.  A measurement
of a significant enhancement would therefore set a lower limit on
$\xi$, equivalent to an upper limit on the $Z'$ mass.\bigskip

\subsection{MSSM}\label{Susy}

Supersymmetric effects in B decays have been studied in an enormous
number of publications but most often hadronic decays have not been
considered in such studies because of their large theoretical
uncertainties. In the MSSM with conserved R-parity, all new flavour-changing
interactions can be related to the squark mass
matrices and enter all kinds of B decays via loops with virtual squarks
and gauginos/higgsinos. Therefore one can  expect supersymmetric contributions to
be of roughly the same size in hadronic modes as they are in leptonic,
semileptonic or radiative modes, so their effects will be most easily
found where the uncertainties are smallest. However, once hints for supersymmetry are found in clean decay channels,
one will also look for confirmations of these observations in other
modes. Therefore we find it interesting to study the possible size of
isospin-violation in the MSSM and whether large effects in the rare
decays $\bar B_s \rightarrow \phi \rho^0$ and $\bar B_s \rightarrow \phi
\pi^0$ can be expected or not, taking into account present experimental
constraints on supersymmetric flavour-violation. Besides, we investigate
whether the deviation of the $\Delta A_\text{CP}$ measurement from the
SM prediction can be explained in the MSSM, as it has been claimed
recently \cite{Huitu:2009st,Khalil:2009zf}.\medskip

Throughout this chapter, we use the MSSM conventions of the SUSY Les
Houches Accords (SLHA) \cite{Skands:2003cj,Allanach:2008qq} and diagonalise the
sparticle mass matrices exactly. We do not make use of the mass-insertion
approximation (MIA), which means that we are not limited to the case where
off-diagonal elements in the mass matrices are small with respect to the
diagonal elements.

\subsubsection{Flavour-violation in the down-squark sector}

First we consider the scenario where flavour-violation arises in the
down-squark sector, i.e.\ where the $6\times 6$ down-squark mass matrix contains
flavour-violating elements. In this case we expect the largest SUSY
effects in $b\rightarrow s$ transitions to stem from
gluino-(down-)squark loops since these loops come with the strong gauge
coupling $\alpha_s$. Neutral\-ino-squark loops arise from exactly the same
off-diagonal elements but are suppressed by weak couplings, so they can
be neglected to a first approximation.\medskip

The authors of ref.~\cite{Grossman:1999av} have analysed such a scenario and
have found that a significant amount of isospin-violation can only occur
via $b\rightarrow s\bar q q$ box diagrams with virtual gluinos and down
squarks ($q=u,d$) and not from photon- or Z-penguin diagrams. A
necessary condition therein is that the SUSY-breaking masses of the right-handed
first-generation squarks, $m_{\tilde u}^2$ and $m_{\tilde d}^2$, are very
different from each other. Recently, this idea has been seized in ref.~\cite{Imbeault:2008ge} and
studied in the light of new data, especially measurements of CP
asymmetries in $B\rightarrow \pi K$ decays and of $B_s$-$\bar{B}_s$
oscillations. It was found that the $B\rightarrow \pi K$ data can only
be reproduced in a tiny region of the parameter space of the model.\medskip

In contrast to these findings the authors of ref.~\cite{Huitu:2009st}
have found a large impact of the gluino-mediated photon penguin in a
mass-insertion calculation and state that sufficient isospin-violation
is generated to explain the  $\Delta A_\text{CP}$ data. However, we find
that this results from a missing factor $-\alpha/6\pi$ in eq.~(41) of
ref.~\cite{Huitu:2009st}. We have performed a scan over the MSSM
parameter space without using the mass-insertion approximation. Our full results for the $\Delta B=1$
Wilson coefficients are written
down in Appendix~\ref{susypeng}. We find the gluino-mediated
photon penguin to yield corrections below the 3\% level to the SM
coefficients of the EW penguin operators for all the points which
allow for a diagonalisation of the squark mass-matrices with
eigenvalues greater than $(100\, \text{GeV})^2$ and satisfy the
experimental constraints from $b\rightarrow s\gamma$. Such
corrections are negligible for the prediction of $\Delta A_\text{CP}$.\medskip

We conclude that we find no sizeable enhancement of EW penguins in the
MSSM with flavour-violation in the down-sector. Neither can we explain
the $\Delta A_\text{CP}$ discrepancy in this scenario nor can we expect
large NP effects in the decays $\bar B_s \rightarrow \phi \pi^0$ and $\bar
B_s \rightarrow \phi \rho^0$.\bigskip

\subsubsection{Flavour-violation in the up-squark sector}

Supersymmetric flavour-violation can also arise in the up-sector via
off-diagonal elements in the hermitian $6\times 6$ up-squark mass matrix. In this
scenario, penguin and box diagrams with virtual charginos and up-type
squarks can provide sizeable contributions to B decays. We have calculated
all of these diagrams, the results are given
in Appendix~\ref{susypeng}.\medskip

The small Yukawa couplings occuring in the quark-squark-chargino couplings and in the squark mass matrix strongly suppress the effect of certain off-diagonal elements of this matrix. The only relevant flavour-violating entries for $b\rightarrow s$ transitions are therefore those corresponding to $\tilde c_L-\tilde t_L$ and $\tilde c_L-\tilde t_R$ mixing. We define them via
\be
\hat m_{\tilde Q}^2 \equiv  m_{\tilde q}^2 \begin{pmatrix}  1 & 0 & 0\\ 0& 1 & \delta_{32}^{uLL*} \\ 0&   \delta_{32}^{uLL}  & 1   \end{pmatrix}
\ee
and
\be
(\mathcal M_{\tilde u}^2)_{62} = \frac{v_u}{\sqrt{2}} (\hat T_U)_{32} \equiv \delta_{32}^{uRL} m_{\tilde q}^2 .
\ee
with a generic squark mass $m_{\tilde q}$. The remaining flavour-conserving elements are specified by $\hat m_{\tilde u}^2= \text{Diag}(m_{\tilde q}^2,m_{\tilde q}^2,m_{\tilde t_R}^2)$, $(\hat T_U)_{ii}=(\hat T_U)_{33} \delta_{i3}$ and $\tan\beta = 10$.\medskip

\begin{figure}
\begin{center}
 \includegraphics[width=5.5cm]{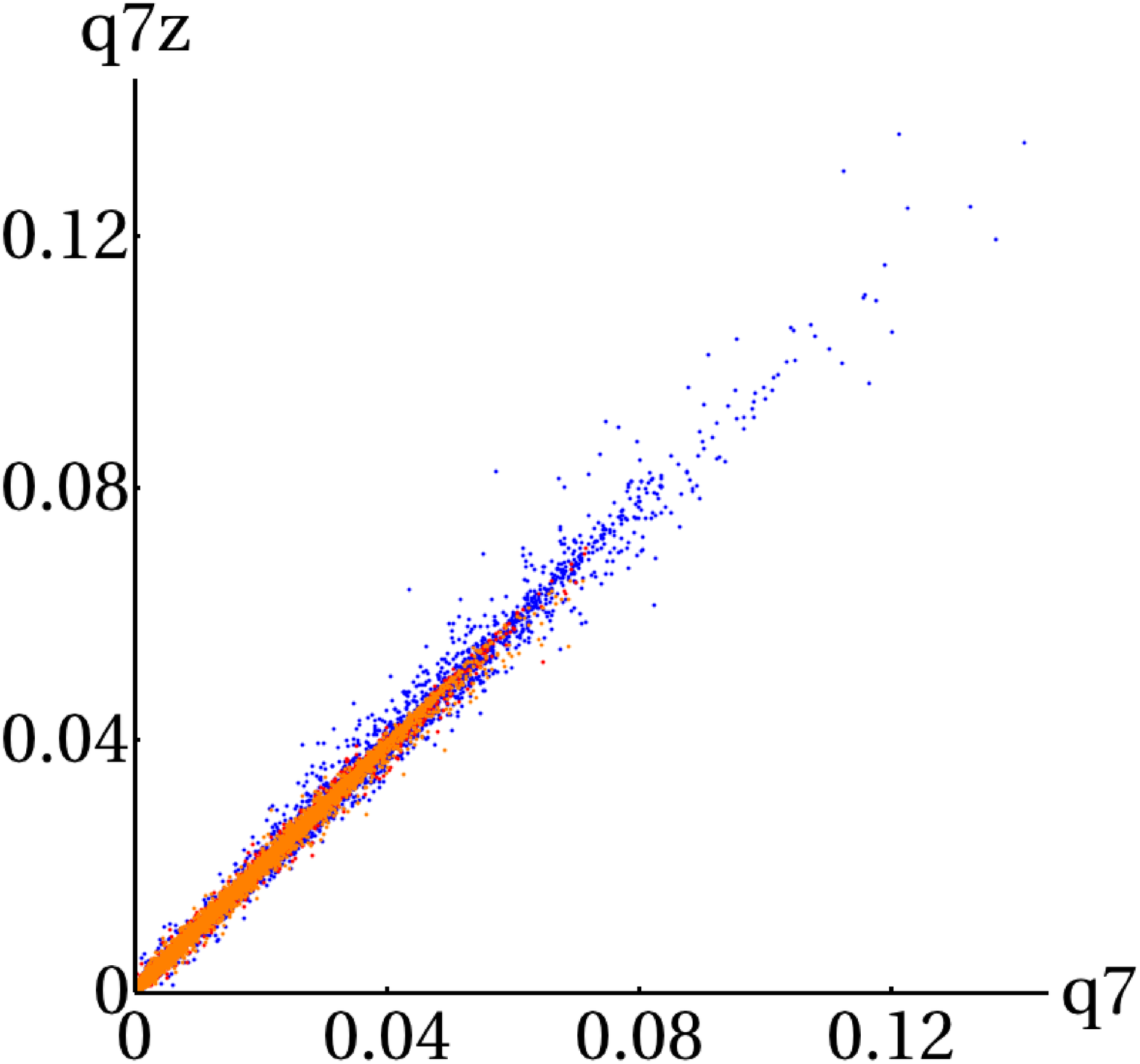} \hspace{2cm}
\includegraphics[width=5.5cm]{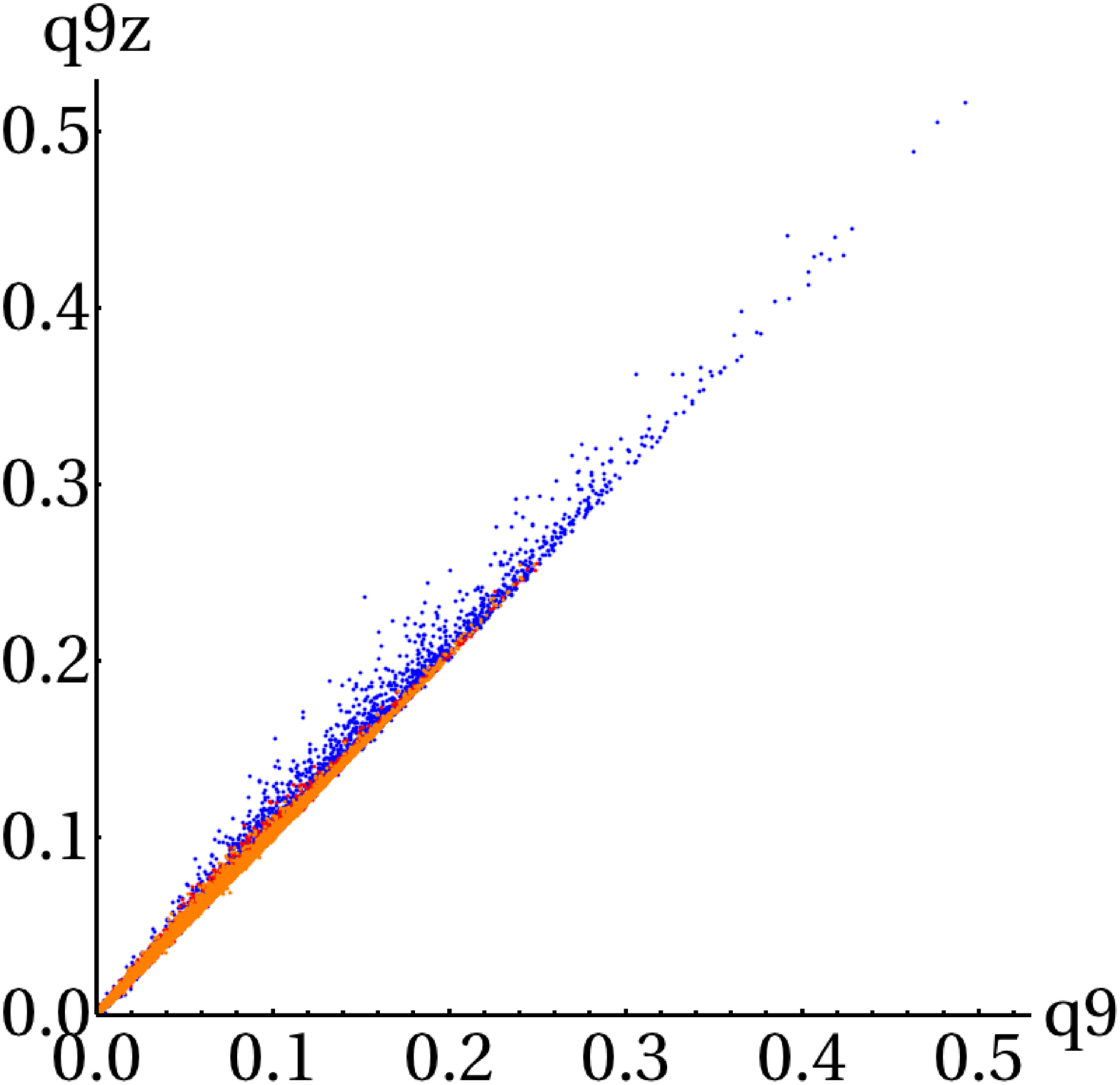}
 \caption{\small{Left: (Right:) Z-penguin contribution to $|q_7|$ ($|q_9|$) over the full $|q_7|$ ($|q_9|$) for 30000 points in the MSSM parameter space. Blue: points excluded by the bound on $|C_{7\gamma}|$. Red: points excluded by $\Br(\bar B\to X_s\gamma)$. Orange: points passing all constraints. Not displayed: points excluded by lower bounds on squark and chargino
masses.} \label{c9zoverc9}}
\end{center}
\end{figure}

\begin{figure}[t]
 \begin{center}
\begin{minipage}{6.5cm}
\psfrag{Rek}[cc][cc]{$\RE(\kappa_L^{sb})/|\kappa_\text{SM}|$}
\psfrag{Imk}[cc][cc]{$\IM(\kappa_L^{sb})/|\kappa_\text{SM}|$}
\includegraphics[width=6cm]{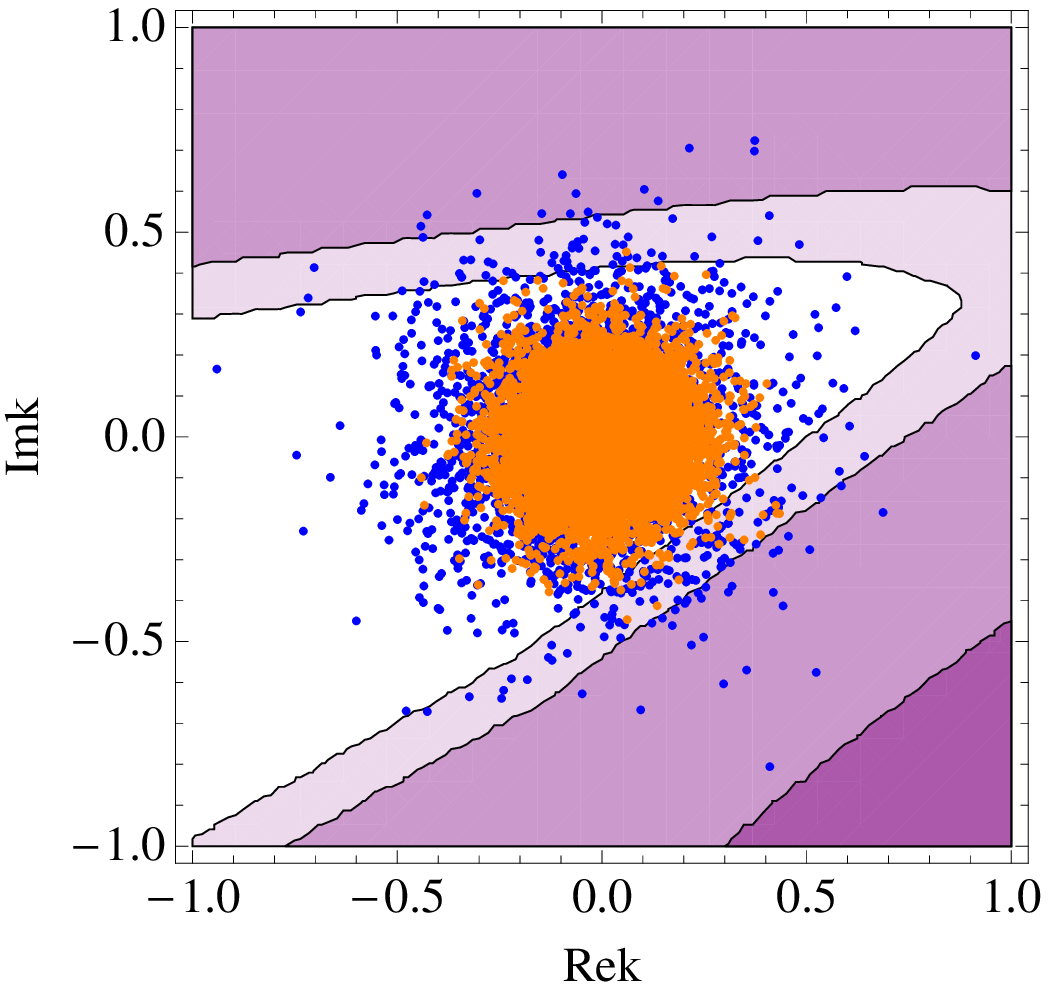}
\end{minipage} \qquad
\begin{minipage}{8cm}
 \begin{tabular}[b]{|c|c|c|} \hline
 $ m_{\tilde q}$, $ m_{\tilde t_R}$, $(\hat T_U)_{33}$    & 200 GeV  & 1000 GeV  \\ \hline
$ M_2$, $\mu$ & 100 GeV & 1000 GeV \\ \hline
$|\delta_{32}^{uLL}|, |\delta_{32}^{uRL}|$ & 0 & 1 \\  \hline
$\arg(\delta_{32}^{uLL}),\arg(\delta_{32}^{uRL})$ & 0 & $2\pi$ \\  \hline
 \end{tabular}
\end{minipage}
 \caption{Discrepancy between $\Delta A_\text{CP}$ in theory and experiment as a
function of $\kappa^{sb}_L/|\kappa_{\rm SM}|$ as defined in sec.~\ref{Z0P}. From light to dark the coloured
regions denote $2.2\sigma$, $2\sigma$ and $1\sigma$. On top we add the $\kappa_L^{sb}/|\kappa_{\rm SM}|$
values resulting from chargino-induced flavour-violating Z couplings in a parameter
scan as defined in the table. Blue (orange): Points (not) excluded by the bound on
$|C_{7\gamma}|$. Not displayed: Points excluded by the lower bounds on SUSY masses.\label{DACPsusy}}
\end{center}
\end{figure}

We have performed a scan over the free parameters in the value ranges defined in the table in fig.~\ref{DACPsusy} to identify possible sources of large isospin-violation. In fig.~\ref{c9zoverc9} we plot the Z-penguin contribution to $|q_7|$ and $|q_9|$ ($q_i$ is defined in eq.~(\ref{g23})) over the sum of all contributions to $|q_7|$ and $|q_9|$ for 30000 random points. We see that considering the Z-penguin only is a very good approximation for all points yielding large isospin-violation. In this way
our scenario is essentially equivalent to the one with a left-handed flavour-changing Z coupling $\kappa_L^{sb}$ discussed in sec.~\ref{Z0P} and we will stay in this approximation in the following. Since a non-vanishing $\kappa_L^{sb}$ breaks electroweak symmetry \cite{Buchalla:2000sk} it must involve the vacuum expectation values $v_{u,d}$. Therefore it is almost exclusively sensitive to $\delta_{32}^{uRL}$ and not to $\delta_{32}^{uLL}$ and we will neglect the latter in the following.\medskip

\noindent Having calculated the MSSM
mass spectrum for a given parameter point we apply the following constraints:
\begin{itemize}
 \item physical squark and chargino masses $\geq 100$ GeV ,
 \item $\Br(\bar B\to X_s\gamma)$ compatible with data at the $2\sigma$ level,
 \item chargino contribution $|C^\chi_{7\gamma}|\leq |C^\text{SM}_{7\gamma}|\approx 0.22$.
\end{itemize}
The last constraint ensures that fine-tuned points passing the $\Br(\bar B\to X_s\gamma)$ constraint are not considered.\medskip

\begin{figure}[t]
 \begin{center}
 \includegraphics[width=12cm]{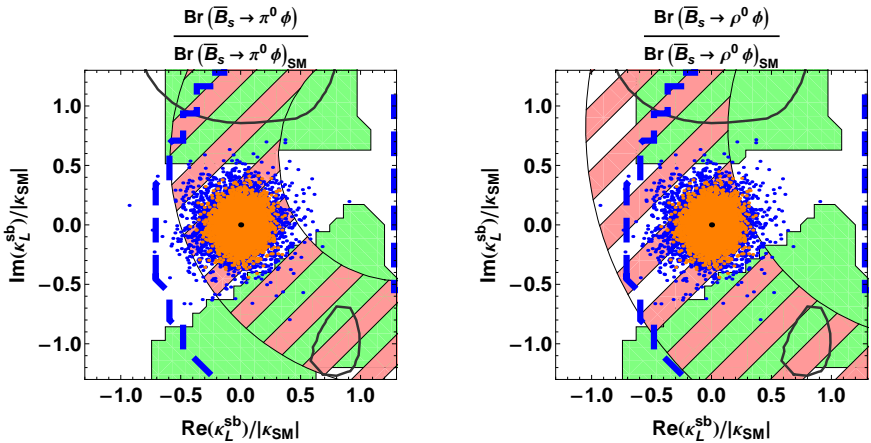}
 \caption{Zoom of the upper plots in fig.~\ref{figZ0}. On top we add the $\kappa_L^{sb}/|\kappa_\text{SM}|$
values resulting from chargino-induced flavour-violating Z couplings in a parameter
scan as defined in the table in fig.~\ref{DACPsusy} with the same colour coding as in
fig.~\ref{DACPsusy} \label{enhancementsusy}}
\end{center}
\end{figure}

It has been suggested \cite{Khalil:2009zf} that a non-vanishing $\delta_{32}^{uRL}$
can generate isospin-violating effects large enough to explain the
$\Delta A_\text{CP}$ discrepancy within the QCDF framework. This would
clearly be an interesting perspective for supersymmetric effects in
purely isospin-violating decays. However, we cannot confirm this statement
in our framework. In fig.~\ref{c9zoverc9} we find that $C_9$ can be enhanced by about 25\% and
$C_7$ by about 8\% with respect to the SM in the presence of a
non-vanishing $\delta_{32}^{uRL}$ \footnote{Including the naturalness constraint $|\delta_{32}^{uRL}|<0.59$ \cite{Crivellin:2008mq} the effects become even smaller.}. In section \ref{modind} we have seen that a 25\%
effect in $C_9$ or an 8\% effect in $C_7$ are not enough to generate a
large $\Delta A_\text{CP}$ and are also not sufficient to enhance the
branching fractions of $\bar B_s \rightarrow \phi \pi^0$ and $\bar B_s
\rightarrow \phi \rho^0$ sizeably. A NP effect of this size would
be hidden in the theoretical uncertainty and thus be unobservable.
These findings are illustrated in figs.~\ref{DACPsusy} and
\ref{enhancementsusy}, where we display on the one hand a figure similar
to fig.~\ref{fig:DeltaACP} in the complex $\kappa_L^{sb}$ plane and on the other
hand a zoomed version of the upper plots in fig.~\ref{figZ0}. To both figures
we add the $\kappa_L^{sb}$ values resulting from chargino-induced flavour-violating
Z couplings in a scan over 30000 points in the MSSM parameter space as defined
in the table in fig.~\ref{DACPsusy} to illustrate the statements of the last paragraph.\medskip

We note that there is an important difference between our calculation
and the one of \cite{Khalil:2009zf}, namely the treatment of strong
phases. We obtain all of these phases directly from QCDF,
where they are suppressed by either $\alpha_s(m_b)$ or
$\Lambda_\text{QCD}/m_B$. Thus we typically find small strong phases
even though their prediction comes with a large uncertainty. In the
framework of ref.~\cite{Khalil:2009zf} only absolute values
of penguin-to-tree ratios are predicted from QCDF whereas the
corresponding strong phases can assume arbitrary values between $0$
and $\pi$ \cite{khalilprivate}. In this way, large CP asymmetries can
be generated even without a large NP contribution and also the fact that
$A_{\rm CP}(B^-\to \pi^0 K^- )$ and $A_{\rm CP}(\bar{B}^0\to \pi^+ K^- )$
have opposite sign is no longer puzzling because the phases of the various
tree and penguin topologies are uncorrelated. In contrast, our
calculation reduces the $2.5\sigma$ SM discrepancy in $\Delta A_\text{CP}$
only marginally. From fig.~\ref{DACPsusy} we can read off that the vast majority of
the allowed points are outside the $2.2\sigma$ region. Only a few rather fine-tuned points
are between $2.2\sigma$ and $2\sigma$.\bigskip

\section{Conclusion}\label{concl}

In this article we have studied the possibility of probing isospin-violating NP
in hadronic $B$ decays. We have proposed to test the EW penguin sector of the
effective weak Hamiltonian via the decays $\bar B_s\to\phi\pi^0$ and $\bar B_s\to\phi\rho^0$ and provided a detailed
phenomenological analysis of these two modes in correlation to other hadronic
$B$ decays.\medskip

Our analysis is motivated by discrepancies found in $B\to \pi K$ decays,
which are to date the best-measured hadronic $b\rightarrow s$ decays. In
particular, the $2.5\,\sigma$ discrepancy found in the observable
$\Delta A_{\rm CP}$ can be interpreted as a sign of NP in the EW penguin
sector of the theory. We have demonstrated in a model-independent analysis that
this discrepancy can easily be resolved by an additional NP contribution
to the EW penguin operators $Q_{7}^{(\prime)},...,Q_{10}^{(\prime)}$ if
it is of the same order of magnitude as the leading SM coefficient
$C_9^{\textrm{SM}}$. An exception are parity-symmetric scenarios where
the contributions to $PP$ decays cancel. Whereas the solution in the
case of NP in $C_9^{(\prime)}$ is, as expected, due to a new contribution
to the colour-allowed EW penguin amplitude, we have pointed out that, in the
case of NP in $C_7^{(\prime)}$, the solution mainly comes about via a weak
annihilation contribution in the QCDF framework which has a surprisingly
large imaginary part. In particular we have found for the case of equal
new contributions to $C_7$ and $C_9$ that, even though these contributions
tend to cancel in the EW penguin amplitude, the $\Delta A_{CP}$ discrepancy
can still be solved via the EW penguin annihilation amplitude, a fact that
had not been noticed before. For various scenarios we have performed frequentist
fits to $B\to \pi K$ data. We have found the fit to work well for NP in
$C_9^{(\prime)}$ while NP in $C_7^{(\prime)}$ is only poorly constrained
from $B\to \pi K$ alone. Especially in this case, the $PV$ counterparts
$B\to \rho K$ and $B\to \pi K^*$, which carry a different interplay of
chiralities, give valuable additional information. We have seen that present
experimental data already set strong constraints: a new EW penguin
amplitude can be at most $\sim 5$ times larger than the SM one; on
the other hand a new amplitude of about the same size as the SM one
is required to solve the $\Delta A_{CP}$ discrepancy.\medskip

Whether the discrepancy in $\Delta A_{\rm CP}$ really is a manifestation
of NP, however, is not clear because the experimental data are still not
conclusive due to the large uncertainties in the theory prediction and
the still somewhat low statistics of the measurements. The long-standing
problem of large theoretical hadronic uncertainties in hadronic decays
can, however, be partially addressed exploiting the large variety of
hadronic decay channels. In the case at hand, the best test is provided
by the $\bar B_s\to\phi\pi^0$ and $\bar B_s\to\phi\rho^0$ modes, which are
purely isospin-violating decays and are dominated by the EW
penguin amplitude. Since these decays are not related to other hadronic
$B$ decays via flavour symmetries, their analysis requires a determination
of the hadronic matrix elements from first principles and we used the QCDF
approach to this end. From the full QCDF results we have derived simple approximate
expressions (eqs.~(\ref{eq:NumPEW})-(\ref{eq:Approx}), (\ref{eq:rtEWPhiNP}))
which reproduce the $\bar B_s\to\phi\pi^0,\phi\rho^0$ amplitudes with high accuracy
for arbitrary scenarios with NP in $C_{7}^{(\prime)},...,C_{10}^{(\prime)}$.
By quoting these formulae we facilitate the study of these decays without
an extensive implementation of the QCDF framework. \medskip

In this work we have performed the first analysis of the impact of NP in EW
penguins on $\bar B_s\to\phi\pi^0,\phi\rho^0$. A new EW penguin amplitude of
the same size as the SM one can easily enhance the $\bar B_s\to\phi\pi^0$
and $\bar B_s\to\phi\rho^0$ branching ratios by an order of magnitude. We have
performed a quantitative analysis parameterising NP in EW penguins in a
model-independent way, at the level of Wilson coefficients, and studied the maximum enhancement of the
$\bar B_s\to\phi\pi^0,\phi\rho^0$ branching ratios for various
scenarios, with respect to the result of our $B\to K\pi$ fits and with respect to constraints from other hadronic
$B$ decays. The results displayed in tab.~\ref{tab3} show that in many
cases a large enhancement is possible. Particular exceptions are
parity-symmetric models which have no impact on the $VV$ decay
$\bar B_s\to\phi\rho^0$ and scenarios with (approximately) equal contributions
to $C_7^{(\prime)}$ and $C_9^{(\prime)}$ which cancel in
$\bar B_s\to\phi\pi^0$.\medskip

A survey of concrete NP models has been performed in section \ref{moddep},
where we have considered a modified $Z^0$ penguin, a model with an
additional $U(1)^{\prime}$ gauge symmetry and the MSSM. In such
models additional constraints arise from the semileptonic decays
$\bar{B}\to X_s e^+e^-$ and $\bar{B}\to K^* l^+l^-$ and from
$B_s$-$\bar{B}_s$ mixing. In case of the modified-$Z^0$-penguin
scenario we have found that the semileptonic constraints still allow
for NP to an extent which is sufficient to resolve the
$\Delta A_{\textrm{CP}}$ discrepancy. On the other hand, they
prevent the $\bar B_s\to\phi\pi^0,\phi\rho^0$ decays from developing an
enhancement which beats the hadronic uncertainties of the SM
prediction. Therefore, a large effect measured in these decays
would rule out the modified $Z$ coupling. The semileptonic
constraints can for example be avoided in a model with an
additional $Z'$ boson whose couplings to leptons can be
switched off. Our analysis has shown that in this scenario
constraints from hadronic $B$ decays and $B_s$-$\bar{B}_s$
mixing can be fulfilled simultaneously only at the $2\,\sigma$
level. The tight constraints from $B_s$-$\bar{B}_s$ mixing limit
a potential enhancement of $\bar B_s\to\phi\pi^0,\phi\rho^0$ to a factor
$\sim 5$. Furthermore the occurrence of such a measurable enhancement
requires a large $g_{U(1)'}$ coupling and/or a light $Z'$ boson.
Finally our conclusion for the MSSM is that it is impossible to obtain a
new contribution larger than about 25\% in $C_9$  and about 8\% in $C_7$, which
is clearly not enough to generate a large $\Delta A_\text{CP}$ or
a significant enhancement of the $B_s$ decays. The $\Delta A_\text{CP}$
discrepancy can be reduced only marginally in this way. Only a few rather fine-tuned parameter points reduce it from $2.5\sigma$ to $2\sigma$.\medskip

We stress again that the decays $\bar B_s\to\phi\pi^0,\phi\rho^0$ are highly
sensitive to isospin-violating NP. Their measurement would complement
the analysis of $B\to K\pi$ decays and  could shed light on the
``$\Delta A_{\textrm{CP}}$ puzzle''. For this reason we strongly
encourage experimental efforts to measure these decays at LHCb
and at future Super-B factories.\bigskip

\vspace{1em}

\noindent

\subsubsection*{Acknowledgements}

We would like to thank U. Nierste for initiating the project
and for a careful reading of the manuscript. We are grateful to him and to
M. Beneke for useful discussions and suggestions. Furthermore we thank A.~Crivellin 
for helpful discussions on chirally enhanced corrections in SUSY, and
S. Khalil for helpful communications
on \cite{Huitu:2009st} and \cite{Khalil:2009zf}. This work is
supported by the DFG Sonder\-forschungs\-bereich/Transregio~9
``Computergest\"utzte Theoretische Teilchenphysik'' and by the EU
contract No.~MRTN-CT-2006-035482, \lq\lq FLAVIAnet''. L.V. acknowledges the
TTP Karlsruhe and the CERN Theory group for hospitality, and the
Alexander-von-Humboldt Foundation for support. L.H.\ and D.S.\
acknowledge the support by Ev.~Studienwerk Villigst and by Cusanuswerk,
respectively, and by the DFG Graduate College
\emph{High Energy Physics and Particle Astrophysics}. L.H. is further supported by 
the Federal Ministry of Education and Research (BMBF) under constract No.~05H09WWE. \bigskip

\appendix
\appendixpage
\addappheadtotoc
\boldmath
\section{General framework for the calculation of hadronic \texorpdfstring{$B$}{B} decays}\label{hadr}
\unboldmath

Throughout this work, we study hadronic B decays into two light mesons using
the framework of QCD factorisation
\cite{Beneke:1999br,Beneke:2000ry,Beneke:2001ev,Beneke:2003zv,Beneke:2006hg}.
This framework is based on the well-known effective weak Hamiltonian for $\Delta
B = 1$ transitions given in (\ref{g1a}) and described in section \ref{heff}. Matrix
elements of this Hamiltonian are calculated in an expansion in $\Lambda_\text{QCD}/m_B$.
We give a few details concerning the generalisation of QCDF to a
Hamiltonian containing operators with flipped chiralities in section \ref{matel}
and specify our input parameters in section \ref{input}. In section \ref{Observables}
we collect expressions for obtaining various observables from the QCDF-calculated
amplitudes.\bigskip

\subsection{The effective weak Hamiltonian}\label{heff}

The starting point for the analysis of hadronic B decays is the
parameterisation of high-energy transitions in terms of effective four-quark
operators multiplied by short-distance coefficients. In the SM, this leads to
the effective weak Hamiltonian in eq.~(\ref{g1a}). The short-distance coefficients
$C_i$ are calculated in the $\overline{\text{MS}}$ renormalisation scheme at the scale
$\mu_{\rm EW}\sim M_W$. Their low-scale values at $\mu_{\rm EW}\sim m_b$ are
obtained from renormalisation-group equations (RGE). In the SM, only $C_1$ is
${\cal O}(1)$, while $C_2$ and the QCD penguin coefficients $C_{3,\ldots,6}$ arise
at order $\mathcal O(\as)$ and the electroweak
penguin coefficients arise at order $\mathcal O(\alpha)$, albeit partly enhanced by
factors of $x_{tW}=m^2_t/m^2_W$ and/or $1/\sin^2\theta_W$. The complete analytical
expressions for these coefficients are written down e.g.\ in
\cite{Buchalla:1995vs}.\medskip

Besides the SM operators, we also consider the possibility that New
Physics gives rise to the so-called ``mirror'' penguin operators
$Q^{\prime}_{i}$, obtained from $Q_{i}$ by a global exchange of left- and
right-handed chiralities of the quark fields, $(V-A)\leftrightarrow(V+A)$. We
thus replace in eq.~(\ref{g1a})
\be
C_i Q_i \longrightarrow C_i Q_i + C^{\prime}_i Q^{\prime}_i.
\ee

Given the enhancement of $C_7(\mu_{\rm EW})$ and $C_9(\mu_{\rm EW})$ by
$x_{tW}$ and/or $1/\sin^2\theta_W$, we follow the modified RGE scheme
presented in \cite{Beneke:2001ev} for the SM coefficients and consider the enhanced terms as
leading-order coefficients, i.e.
\be\label{locoeffs}
C_7^\text{LO}(M_W) = \frac{\alpha}{4\pi} \frac{x_{tW}}{3} \quad , \quad C_9^\text{LO}(M_W)=
\frac{\alpha}{4\pi}\left[ \frac{x_{tW}}{3} + \frac{2}{3\sin^2\theta_W}(10 B_0(x_{tW})-4C_0(x_{tW})) \right]
\ee
with the loop functions $B_0(x)$ and $C_0(x)$ from \cite{Buchalla:1995vs}.
To be consistent we neglect at the same time electromagnetic corrections to the
QCD penguin coefficients as well as any mixing of  $C_7,\ldots,C_{10}$ into
$C_1,...,C_6$. Since this treatment improves the RGE evolution for $C_7,...,C_{10}$
and since it is exactly these coefficients we are interested in, it is well suited
for our analysis. By contrast, for the NP contributions we use the standard
treatment for the leading-order RGE.\bigskip

\subsection{Operator matrix elements in QCDF}\label{matel}

From the effective Hamiltonian (\ref{g1a}) the decay amplitude for the process
$\bar B\rightarrow M_1 M_2$ can be calculated as
\be\label{b3}
\mathcal A^{(h)}_{\bar B \rightarrow M_1 M_2} \,=\,
\langle M_1^{(h)} M_2^{(h)}|{\cal{H}_{\rm eff}} |\bar{B}\rangle,
\ee
where $h$ refers to a helicity amplitude in case of a decay into two vector
mesons, $\bar B \to V_1^h V_2^h$ with $h=0,+,-$.\medskip

In QCDF the matrix elements of the effective Hamiltonian are
organised in terms of flavour amplitudes $\alpha_i$ which are directly related to
the topologies of the underlying weak transition, for example colour-allowed EW
penguin, colour-suppressed tree etc. Analogous amplitudes $\beta_i$ represent
the corresponding weak annihilation transitions. The topological amplitudes
$\alpha_i$ in turn can be decomposed into coefficients $a_i$
 which are in direct correspondence to the operators $Q_i$ in the effective
Hamiltonian. At NLO these coefficients $a_i$ contain the perturbative
non-factorisable vertex-, penguin- and spectator-scattering corrections governed
by the factorisation formula. For a complete description we refer to
 \cite{Beneke:2003zv}. The expressions given there can easily be generalised to
account for the mirror operators by
adding a topological amplitude $\alpha_i^{\prime}$ to each of the $\alpha_i$
\cite{Beneke:2009eb}. A similar  generalisation applies to the annihilation
amplitudes $\beta_i$. Here we only need to add the expressions for
$\alpha_{3\rm EW,4 EW}^{\prime}$, which read
\bea\label{b5}\nn
\alpha^{\prime\,p}_{\rm 3 EW}(M_1 M_2) &=& \left\{
    \begin{array}{cl}
    -a^{\prime\,p}_9(M_1 M_2) + a^{\prime\,p}_7(M_1 M_2),
      & \quad \mbox{if~} M_1 M_2=PP, \\
     a^{\prime\,p}_9(M_1 M_2) + a^{\prime\,p}_7(M_1 M_2),
      & \quad \mbox{if~} M_1 M_2=PV, \\
     a^{\prime\,p}_9(M_1 M_2) - a^{\prime\,p}_7(M_1 M_2),
      & \quad \mbox{if~} M_1 M_2=VP, \\
    -a^{\prime\,p}_9(M_1 M_2) - a^{\prime\,p}_7(M_1 M_2),
      & \quad \mbox{if~} M_1 M_2=V^0V^0, \\
    -f^{M_1}_\pm\left(a^{\prime\,p}_9(M_1 M_2) + a^{\prime\,p}_7(M_1
M_2)\right),
      & \quad \mbox{if~} M_1 M_2=V^{\pm}V^{\pm},
    \end{array}\right. \\
\alpha^{\prime\,p}_{\rm 4 EW}(M_1 M_2) &=& \left\{
    \begin{array}{cl}
    -a^{\prime\,p}_{10}(M_1 M_2) - r_{\chi}^{M_2}\,a^{\prime\,p}_8(M_1 M_2),
      & \quad \mbox{if~} M_1 M_2=PP, \\
     a^{\prime\,p}_{10}(M_1 M_2) + r_{\chi}^{M_2}\,a^{\prime\,p}_8(M_1 M_2),
      & \quad \mbox{if~} M_1 M_2=PV, \\
     a^{\prime\,p}_{10}(M_1 M_2) - r_{\chi}^{M_2}\,a^{\prime\,p}_8(M_1 M_2),
      & \quad \mbox{if~} M_1 M_2=VP, \\
    -a^{\prime\,p}_{10}(M_1 M_2) + r_{\chi}^{M_2}\,a^{\prime\,p}_8(M_1 M_2),
      & \quad \mbox{if~} M_1 M_2=V^0V^0, \\
    f^{M_1}_\pm\left(-a^{\prime\,p}_{10}(M_1 M_2) +
r_{\chi}^{M_2}\,a^{\prime\,p}_8(M_1 M_2)\right),
      & \quad \mbox{if~} M_1 M_2=V^{\pm}V^{\pm}
    \end{array}\right.
\eea
with $f^{M_1}_\pm=F_{\mp}^{B\rightarrow M_1}(0)/F_{\pm}^{B\rightarrow M_1}(0)$
being a ratio of form factors, such that $f^{M_1}_+\sim m_B/\Lambda_{\rm QCD}$ and
$f^{M_1}_-\sim\Lambda_{\rm QCD}/m_B$ in the heavy-quark limit
\cite{Beneke:2006hg}. The NLO expressions for the coefficients $a_i^{\prime}$
are equivalent to the ones for the coefficients $a_i$ from \cite{Beneke:2003zv}
up to the replacement $C_i \rightarrow C_i^\prime$. Additionally in the transverse
amplitudes of $B\to VV$ decays one has to flip the helicities, i.e. the expressions
for $a_i^+$ are needed for ${a_i'}^-$ and vice versa.\medskip

The pattern of signs appearing in eq.~(\ref{b5}) is a consequence of the
fact that matrix elements of the mirror operators are related to the SM ones by
parity
\begin{equation}
   \langle M_1M_2 |Q_i|\bar{B}\rangle \,=\, -\eta_{M_1M_2}
               \,\langle M_1M_2 |Q_i^{\prime}|\bar{B}\rangle,
\end{equation}
which implies that the amplitudes involve the coefficients $a_i^{(\prime)}$ only
in the combinations \cite{Kagan:2004ia}
\begin{equation}\label{b6}
    a^p_i(M_1,M_2) \,-\, \eta_{M_1M_2}\,
a^{\prime\,p}_i(M_1,M_2).
\end{equation}
Here $\eta_{M_1M_2}=\pm 1$ is the parity of the final state.
For $PP$ and longitudinal $VV$ final states, we have $\eta_{M_1M_2} = 1$ whereas
for $PV$ final states $\eta_{M_1M_2} = -1$. In this manner left-handed  and
right-handed NP give rise to different signatures and correlations
among $PP$, $PV$ or $VV$ decays. Exploiting this feature can be very important
in order to probe the chirality structure of a potential NP model.\bigskip

\boldmath
\subsection{
Input parameters and tree to penguin ratios in \texorpdfstring{$B\to\pi K$,
$B\to\pi\pi$}{B to Pi,K and B to Pi,Pi}}\label{input}
\unboldmath

\begin{table}[htbp]
  \centering
  \begin{tabular}{|c|c|c|c|c|}
    \hline
    \multicolumn{5}{|p{13.7cm}|}{QCD scale and running quark masses [GeV]} \\[0.1cm] \hline
    $\Lambda_{\overline{\rm MS}}^{(5)}$ & $m_b(m_b)$ & $m_c(m_b)$ & $m_s$ & $m_q/m_s$  \\
    \hline
    0.231 & 4.2 & 1.3$\pm$0.2 & 0.090$\pm$ 0.020 & 0.0413\\
    \hline
    \multicolumn{5}{|p{13.7cm}|}{CKM parameters} \\[0.1cm] \hline
    $\lambda$ & $|V_{cb}|$ & $|V_{ub}/V_{cb}|$ & $\gamma$ & $\sin(2 \beta)$ \\
    \hline
    $0.225$ & $0.0415\pm0.0010$ & $0.085^{+0.025}_{-0.015}$ & $(70\pm10)^{\circ}$ & $0.673\pm 0.23$ \\
    \hline
    \multicolumn{5}{|p{13.7cm}|}{B meson parameters} \\[0.1cm] \hline
                    &                 & $B^-$ & $\bar{B}^0$ & $\bar{B}_s^0$ \\
    \hline
     Lifetime       & $\tau$[ps]      & $1.638$ & $1.525$ & $1.472$ \\
    \hline
     Decay constant & $f_B$[MeV]      & \multicolumn{2}{|c|}{$210\pm20$} & $240\pm20$ \\
                    & $\lambda_B$[MeV] & \multicolumn{2}{|c|}{$200^{+250}_{-0}$} & $200^{+250}_{-0}$ \\
    \hline
  \end{tabular}
  \begin{tabular}{|c|c|c|}
    \multicolumn{3}{|p{13.7cm}|}{Pseudoscalar-meson decay constants and Gegenbauer moments} \\[0.1cm] \hline
                            &      $\pi$    & $K$ \\
    \hline
     $f$[MeV]               &      $131$  & $160$ \\
     $\alpha_1,\alpha_{1,\perp}$ &      $0$  & $0.06\pm0.06$ \\
     \hspace{1cm}$\alpha_2,\alpha_{2,\perp}$\hspace{1cm} &      \hspace{1cm}$0.20\pm0.15$\hspace{1cm}  & $0.20\pm0.15$ \\
    \hline
  \end{tabular}
  \begin{tabular}{|c|c|c|c|}
    \multicolumn{4}{|p{13.7cm}|}{Vector-meson decay constants and Gegenbauer moments} \\[0.1cm] \hline
                            &      $\rho$ & $K^*$ & $\phi$ \\
    \hline
     $f$[MeV]               &      $209\pm1$  & $218\pm4$  & $221\pm3$ \\
     $f^{\perp}$[MeV]       &      $165\pm9$  & $185\pm10$ & $186\pm9$ \\
     $\alpha_1,\alpha_{1,\perp}$ &      $0$  & $0.06\pm0.06$ & $0$ \\
     \hspace{1cm}$\alpha_2,\alpha_{2,\perp}$\hspace{1cm} &       \hspace{0.5cm}$0.1\pm0.2$\hspace{0.5cm}  & \hspace{0.5cm}$0.1\pm0.2$\hspace{0.5cm} & $0\pm0.3$ \\
    \hline
    \multicolumn{4}{|p{13.7cm}|}{Pseudoscalar-meson form factor at $q^2=0$} \\[0.1cm] \hline
                            &      $B\to\pi$ & $B\to K$ & $B_s\to\bar{K}$\\
    \hline
     $F_0$             &      $0.25\pm0.05$  & $0.34\pm0.05$ & $0.31\pm0.05$ \\
    \hline
    \multicolumn{4}{|p{13.7cm}|}{Vector-meson form factor at $q^2=0$} \\[0.1cm] \hline
                            &      $B\to\rho$ & $B\to K^*$ & $B_s\to\phi$ \\
    \hline
     $A_0$             &      $0.30^{+0.07}_{-0.03}$ & $0.39\pm0.06$ & $0.38^{+0.10}_{-0.02}$ \\
     $F_+$             &      $0.00\pm0.06$          & $0.00\pm0.06$ & $0.00\pm0.06$ \\
     $F_-$             &      $0.55\pm0.06$          & $0.68\pm0.07$ & $0.65^{+0.14}_{-0.00}$ \\
    \hline
  \end{tabular}
  \caption{Summary of the theoretical input parameters for hadronic $B$ decays
into two light mesons. All scale-dependent quantities refer to $\mu=2\,\gev$
unless otherwise stated. The values represent the most up-to-date values taken from
\cite{Beneke:2003zv,Beneke:2006hg,Beneke:2005vv,Beneke:2006mk,Bartsch:2008ps,Beneke:2009ek,Amsler:2008zzb}.}
\label{tab0}
\end{table}

In the framework of QCDF the decay amplitudes depend on quite a few input
parameters such as form factors, Gegenbauer moments of light-cone
distribution amplitudes, etc. In tab.~\ref{tab0} we provide a list of up-to-date
values. We use the updated value of $\Lambda_{\overline{\rm MS}}^{(5)}$ as
obtained in \cite{Bethke:2009jm}; as of  $\lambda_B = \lambda_{B_s}$
we assume the lower value 200 MeV as suggested by exclusive hadronic decays,
see \cite{Beneke:2006hg,Beneke:2006mk,Beneke:2009ek,Bell:2009fm}. Particular
attention deserves the choice of $X_{H,A}$, i.e.\ the numbers which parameterise
our ignorance of non-perturbative physics occurring in spectator scattering
and weak annihilation due to the exchange of soft gluons. For $X_H$ we use
the definition of \cite{Beneke:2003zv},
\be
X_H = (1 + \rho_H e^{i \phi_H})\ln \frac{m_B}{\Lambda_h},
\ee
with a non-perturbative scale $\Lambda_h=500$ MeV. The default value is $\rho_H=0$
and we estimate the uncertainty setting $\rho_H=1$ and freely varying $\phi_H$ between
$0$ and $2\pi$.  In the light of experimental data, $X_{A}$ requires more attention.
The reasoning goes as follows.\medskip

\begin{figure}[ht]
\begin{center}
  \includegraphics[width=0.9\textwidth]{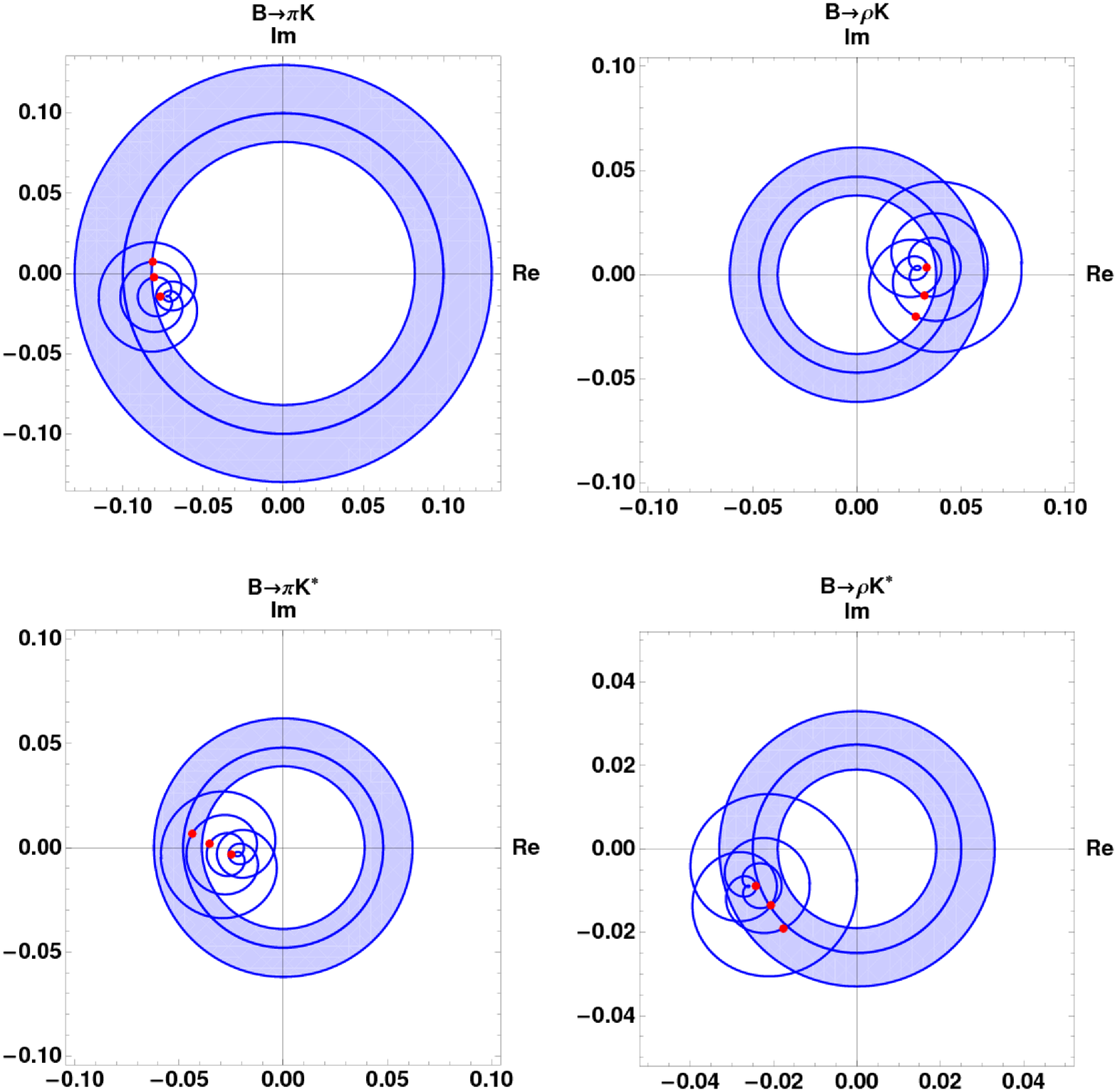}
\end{center}
  \caption{Experimental and theoretical values of the penguin-to-tree ratio defined
           in eq.~(\ref{z1}). For explanations see text.}
  \label{penguintotree}
\end{figure}

A good test of the QCDF hypothesis is to look at the
$\pi K$-penguin to $\pi\pi$-tree ratio \cite{Beneke:2003zv}, which
can be directly related to experimental observables:
\be\label{z1}
\left|\frac{\hat{\alpha}_4^c(\pi \bar{K})}{\alpha_1(\pi\pi)+\alpha_2(\pi\pi)}
\right| \simeq
\left|\frac{V_{ub}}{V_{cb}}\right|\frac{f_{\pi}}{f_K}
\left[\frac{\Gamma(B^-\to\pi^- \bar{K}^0)}{\Gamma(B^-\to\pi^-
\pi^0)}\right]^{1/2}
\stackrel{\text{exp.}}{=} 0.100^{+0.030}_{-0.018}.
\ee
To a good approximation this ratio relates a pure QCD-penguin decay to a
pure tree decay and allows to eliminate the uncertainty from the heavy-to-light
form factor. Since the tree decay $B^-\to\pi^- \pi^0$ suffers from small
uncertainties in QCDF and is expected to receive negligible contributions
from NP, eq.~(\ref{z1}) probes the accuracy of the QCDF prediction for the
penguin amplitude $\hat{\alpha}_4^c(\pi \bar{K})$. In absence of sizeable NP
contributions to QCD penguins it constrains the uncalculable weak-annihilation
contribution $\beta_3^c$, which is responsible for the lion's share of the
theoretical uncertainty in the penguin amplitude.
Though subleading in $\Lambda_{\rm QCD}/m_B$, $\beta_3^c$ is known to be
numerically enhanced.\medskip

\noindent
Using the parameters in tab.~\ref{tab0} and expressing $\beta_3^c$ via the
complex ${\cal O}(1)$ parameter $X_A$ as in ref.~\cite{Beneke:2003zv} with
\be
X_A = (1 + \rho_A e^{i \phi_A})\ln \frac{m_B}{\Lambda_h}
\ee
with the default value $\rho_A = 0$, we find
\be\label{z2}
\left|\frac{\hat{\alpha}_4^c(\pi \bar{K})}{\alpha_1(\pi\pi)+\alpha_2(\pi\pi)}
\right|
\stackrel{\text{SM}}{=} 0.078^{+0.025}_{-0.015},
\ee
which is slightly smaller than older results. Repeating the analysis of
\cite{Beneke:2003zv} for the $PV$ and $VV$ modes we obtain fig.~\ref{penguintotree}.
The plots show the experimental central values and uncertainties of the ratio in
eq.~(\ref{z1}) (circles around the origin) for $B\rightarrow\pi K$ and three related
decays. They are combined with ``lima\c{c}on'' curves representing the corresponding
theory predictions where the phase $\phi_A$ is freely varied between $0$ and $2\pi$
while we set $\rho_A=1$ $(1.5,2)$ for $PP$ and $PV$ modes and $\rho_A=0.6$ $(1.0,1.5)$
for $VV$ modes to obtain the blue (purple, yellow) curves. The red dots correspond to
$\rho_A=0,1,1.5$ ($\rho_A=0,0.6,1.0$ for $VV$ modes) with $\phi_A$ fixed as in the
scenario ``S4'' in \cite{Beneke:2003zv} for $PP$ and $PV$ modes and
$\phi_A=-40^\circ$ for $VV$.\medskip

These results lead us to the conclusion that, in the light of present data, we
prefer to change the widely used treatment with $\rho_A=0$ as default and
$\rho_A=1$ ($\rho_A=0.6$ for $VV$) for the variation of $\phi_A$ in order to
have a more conservative estimate of the theoretical uncertainty. Nevertheless
we confirm that $\rho_A\leq 2$ is a reasonable upper bound for weak-annihilation
contributions in QCDF. We adopt $\rho_A = 1.5$ ($\rho_A=1.0$ for $VV$ modes) as
our default value, keeping the default for $\phi_A$ as above, and estimate the
uncertainty with $\rho_A$ unchanged and $\phi_A\in [0, 2\pi)$.\bigskip

\boldmath
\subsection{Calculating observables in hadronic \texorpdfstring{$B$}{B} decays}\label{Observables}
\unboldmath

Starting from a  decay amplitude $\mathcal A(\bar B \to M_1 M_2)$  the corresponding
decay rate can be calculated as
\be\label{brformula}
\Gamma(\bar{B}\to M_1 M_2) = \frac{S}{16 \pi m_B} |\mathcal{A}(\bar{B}\to M_1 M_2)|^2,
\ee
with a symmetry factor $S$. We have $S=1/2$ if $M_1$ and $M_2$ are identical and $S=1$
otherwise. For decays into two vector mesons, where three different helicity amplitudes
exist, one replaces $|\mathcal{A}(\bar{B}\to M_1 M_2)|^2$ by a sum over the three
possible helicities of the final state, $\sum_{h=0,-,+}|\mathcal{A}^h(\bar{B}\to V_1 V_2)|^2$.
The branching ratios are easily calculated from $\Br(\bar{B}\to M_1 M_2) = \Gamma(\bar{B}\to M_1
M_2)/\Gamma_{\rm tot}$ with $\Gamma_{\rm tot}$ being the total decay width of the $\bar{B}$
meson. Our theoretical predictions always refer to CP-averaged branching ratios defined as
\be
\overline{\Br}(\bar{B}\to \bar{f}) = \frac{1}{2} \left(\Br(\bar{B}\to \bar{f}) +
\Br(B\to f)\right),
\ee
where $\bar{f},f$ stand for the final state $M_1 M_2$ and its CP-conjugated state.
The CP-conjugated decay rate is calculated by replacing $\mathcal A$ in
eq.~(\ref{brformula}) by the corresponding CP-conjugated amplitude. This amounts
to a change of sign for all the weak phases while strong phases are unchanged.
\medskip

The direct CP asymmetries read
\be
A_{\rm CP} = \frac{\Br(\bar{B}\to \bar{f}) - \Br(B\to f)}{\Br(\bar{B}\to
\bar{f}) + \Br(B\to f)}
\ee
and the longitudinal polarisation fraction $f_L(\bar{B}\to M_1 M_2)$ is defined as
\be
f_L(\bar{B}\to M_1 M_2) = \frac{|\mathcal{A}^0(\bar{B}\to V_1
V_2)|^2}{\sum_{h=0,-,+}|\mathcal{A}^h(\bar{B}\to V_1 V_2)|^2}.
\ee

\boldmath
\section{
Isospin-violating observables in \texorpdfstring{$B\to\pi K$}{B to Pi,K}}\label{IsoRatios}
\unboldmath

In this appendix we present the observables which are sensitive to isopin
violation in the $B\to\pi K$ decay modes and which we use to calculate our $2\,\sigma$
constraints.\medskip

First, one has ratios of any two different decay rates. Using the
parameterisation (\ref{g5}) and neglecting terms which are quadratic in the
$r_i$ as well as the annihilation contribution $\rEWA$ which has only a small real
part, one has 6 different ratios which read \cite{Yoshikawa:2003hb,Mishima:2004um}
\bea\label{y1}
   R_c^B&\equiv&2\,\frac{\overline{\Br} (B^-\to \pi^0K^-
)}{\overline{\Br}(B^-\to\pi^-\bar{K}^0)}
      \,\simeq\,1\,+\,2\,\RE(\rEW+\rEWC)-\,2\,\RE(\rT+\rC)\cos\gamma\,,\nn\\
   R_n^B&\equiv&\frac{1}{2}\,\frac{\overline{\Br} (\bar{B}^0\to\pi^+ K^-
)}{\overline{\Br}(\bar{B}^0\to\pi^0\bar{K}^0)}
      \,\simeq\,1\,+\,2\,\RE(\rEW+\rEWC)-\,2\,\RE(\rT+\rC)\cos\gamma\,,\nn\\
   R_c^K&\equiv&2\,\frac{\tau_0}{\tau_-}\,\frac{\overline{\Br} (B^-\to\pi^0 K^-
)}{\overline{\Br}(\bar{B}^0\to\pi^+ K^-)}
      \,\simeq\,1\,+\,2\,\RE(\rEW)-\,2\,\RE(\rC)\cos\gamma\,,\nn\\
   R_n^K&\equiv&\frac{1}{2}\,\frac{\tau_0}{\tau_-}\,\frac{\overline{\Br}
(B^-\to\pi^- \bar{K}^0)}{\overline{\Br}(\bar{B}^0\to\pi^0 \bar{K}^0)}
      \,\simeq\,1\,+\,2\,\RE(\rEW)-\,2\,\RE(\rC)\cos\gamma\,,\nn\\
   R_c^{\pi}&\equiv&\frac{\tau_0}{\tau_-}\frac{\overline{\Br} (B^-\to\pi^-
\bar{K}^0 )}{\overline{\Br}(\bar{B}^0\to\pi^+ K^-)}
      \,\simeq\,1\,+\,2\,\RE(\rT)\cos\gamma-\,2\,\RE(\rEWC),\nn\\
   R_n^{\pi}&\equiv&\frac{\tau_0}{\tau_-}\,\frac{\overline{\Br} (B^-\to\pi^0
K^-)}{\overline{\Br}(\bar{B}^0\to\pi^0 \bar{K}^0)}
\,\simeq\,1-\,2\,\RE(\rT+2\rC)\cos\gamma\,+\,2\,\RE(2\rEW+\rEWC)\,,\hspace{0.5cm}
\eea
where $\tau_0$ and $\tau_-$ are the lifetimes of the neutral and charged $B$
mesons, respectively. NP in EW penguins as in (\ref{g8})
enters the ratios $R_{c,n}^{B,K,\pi}$ through
\bea\label{y2}
 \RE(r_\text{EW})  &\rightarrow& \RE(r_\text{EW}) + \RE(\tilde{r}_\text{EW})
\cos\delta,\nn\\
 \RE(r_\text{EW}^{\textrm{C}})  &\rightarrow& \RE(r_\text{EW}^{\textrm{C}}) +
\RE(\tilde{r}_\text{EW}^{\textrm{C}}) \cos\delta.
\eea
Experimental data on the $R_{c,n}^{B,K,\pi}$ can be used to constrain the
NP contributions $\tilde{r}_\text{EW}$ and
$\tilde{r}_\text{EW}^{\textrm{C}}$. Note that the $R_{c,n}^{B,K,\pi}$ involve
different combinations of $\tilde{r}_\text{EW}$ and
$\tilde{r}_\text{EW}^{\textrm{C}}$ and thus they are sensitive to different
linear combinations of the electroweak penguin coefficients
$C_7^{(\prime)},...,C_{10}^{(\prime)}$. Therefore, it depends on the specific
NP scenario in consideration which of the $R_{c,n}^{B,K,\pi}$ give the
best constraints.\medskip

Beyond being responsible for the universal QCD penguin contribution, isospin
relations account for the approximate equation
\be\label{y3}
 \Gamma(B^-\to\pi^-\bar{K}^0)-2\,\Gamma(B^-\to\pi^0K^-)\,\approx\,
        2\,\Gamma(\bar{B}^0\to\pi^0\bar{K}^0)-\Gamma(\bar{B}^0\to\pi^+K^-)
\ee
known as Lipkin sum rule \cite{Lipkin:1998ie}. In the strict isospin limit both
sides of this equation vanish identically.  This is reflected in the fact that
$R^{B}_{c,n}$ in (\ref{y1}) are both equal to one, apart from isospin-violating
terms of order $\mathcal{O}(r_i)$. These linear terms are
generated by the interference of the isospin-violating parts of the amplitude
with the QCD penguin part. The special property of (\ref{y3}) is now that these
interference terms on the left- and righthand side of the approximate equation
cancel each other. For this reason (\ref{y3}) can be used to construct a purely
isospin-violating observable, namely
\be\label{y4}
R\,\equiv\,2\,\frac{\tau_-\,\overline{\Br}(\bar{B}^0\to\pi^0\bar{K}^0)\,+\,
\tau_0\,\overline{\Br}(B^-\to\pi^0K^-)}
{\tau_-\,\overline{\Br}(\bar{B}^0\to\pi^+K^-)\,+\,\tau_0\,\Gamma(B^-\to\pi^-\bar
{K}^0)}\,=\,1\,+\,\mathcal{O}(r_i^2)\,.
\ee\medskip

In a similar way it is possible to construct observables with a high sensitivity
to isospin violation out of the direct CP asymmetries. To this end we consider
the two differences
\bea\label{y5}
   \Delta A_{\textrm{CP}}^-&\equiv&A_{\textrm{CP}}(B^-\to\pi^0 K^-
)\,-\,A_{\textrm{CP}}(\bar{B}\to\pi^+ K^-)\,=\,\Delta A_{\textrm{CP}}\nn\\
   \Delta A_{\textrm{CP}}^0&\equiv&A_{\textrm{CP}}(B^-\to\pi^-
\bar{K}^0)\,-\,A_{\textrm{CP}}(\bar{B}\to\pi^0 \bar{K}^0).
\eea
In the parameterisation (\ref{g5}) and in the presence of NP in
electroweak penguins (\ref{g8}) one finds
$\Delta A_{\textrm{CP}}^0 = \Delta A_{\textrm{CP}}^-$ up to terms quadratic in
the $r_i$. The observable $\Delta A_{\textrm{CP}}^-$ is given in (\ref{g10}) and
represents the famous $\Delta A_{\textrm{CP}}$ showing a $2.5\sigma$
discrepancy with current data. A precise measurement of $\Delta A_{\textrm{CP}}^0$
could therfore shed light on the $\Delta A_{\textrm{CP}}$ discrepancy.\medskip

Since $\pi^0K_s$ is a CP-eigenstate into which both the $B^0$ and the $\bar{B}^0$
meson can decay, we have mixing-induced CP violation in this decay channel. The
corresponding observable $S_{\rm CP}$ is defined via the time-dependent CP
asymmetry as
\be\label{y7}
\frac{\Br(\bar{B}^0(t)\to f)-\Br(B^0(t)\to f)}{\Br(\bar{B}^0(t)\to
f)+\Br(B^0(t)\to f)}
\equiv S_{\rm CP}\sin(\Delta m_B t)-C_{\rm CP}\cos(\Delta m_B t), \\
\ee
where $C_{\rm CP}=-A_{\rm CP}$ is the direct CP asymmetry, up to a sign.
Although $S_{\pi K}$ is not sensitive to isospin-violation in particular, it
will be affected by a solution of the ``$\Delta A_{\textrm{CP}}$-puzzle''
via a NP contribution $\tilde{r}_{\textrm{EW}}$. The reason is that
$\tilde{r}_{\textrm{EW}}$ has to come with a large new weak phase $\delta$
in order to have substantial impact on $\Delta A_{\textrm{CP}}$.
Including the new electroweak contributions and
neglecting $\RE(\rEWA)$ we find
\be\label{y8}
S_{\rm CP}(\bar{B}^0\to \pi^0 \bar{K}^0) \simeq  \sin 2 \beta
+ 2 \RE \left( r_{\rm C}\right) \cos 2 \beta \sin\gamma
-2 \RE(\tilde r_{\rm EW}+\tilde r_{\rm EW}^C) \cos 2\beta \sin\delta.
\ee
We collect the theoretical and experimental results for the observables defined
here, as well as the CP-averaged branching ratios and direct CP asymmetries in
tab.~\ref{tab1}.\medskip

Finally we give the QCDF expressions of the ratios $r_i$ used above:
\bea\nn
r_\text{T} &=& -\left| \frac{\lambda_u^{(s)}}{\lambda_c^{(s)}} \right|
\frac{\alpha_1(\pi K)}{\hat\alpha_4^c(\pi K)}, \hspace{1.2cm}
r_\text{C} = -\left|\frac{\lambda_u^{(s)}}{\lambda_c^{(s)}} \right| \frac{A_{K
\pi}}{A_{\pi K}} \frac{\alpha_2( K \pi)}{\hat\alpha_4^c(\pi K)}, \\
r_\text{EW} &=& \frac 3 2 \frac{A_{K \pi}}{A_{\pi K}}
\frac{\alpha^c_\text{3,EW}(K \pi)}{\hat\alpha_4^c(\pi K)},  \hspace{0.6cm}
r^C_\text{EW} = \frac 3 2 \frac{\alpha^c_\text{4,EW}(\pi K)}{\hat\alpha_4^c(\pi
K)}, \hspace{0.8cm}
r^A_\text{EW} = \frac 3 2 \frac{\beta^c_\text{3,EW}(\pi K)}{\hat\alpha_4^c(\pi
K)}.
\eea\bigskip

\section{The fit}\label{fit}

The basic idea of our fit is quite simple: we calculate within a NP
scenario the expected values for a set of observables as a function of
NP parameters $q_i$. We then compare these values at each point of a grid in the
$\{q_i\}$ parameter space to experimental data. The points for which the
experimental and the theoretical results are closest are most likely to be
realised, i.e.\ they represent the $q_i$-values for which the theoretical
prediction describes best the experimental measurements. Technically, this
comparison is performed by evaluating at each grid point the $\chi^2$ function
\be\label{e51}
\chi^2(\{q_i\})=\sum_{j}\frac{(x_{j\,\rm theo}(\{q_i\})-x_{j\,\rm
exp})^2}{\sigma_{j\,\rm exp}^2},
\ee
where the sum runs over different observables $x_j$. In this notation $x_{j\,\rm
theo}$ represents the theoretical prediction of the observable and $x_{j\,\rm
exp}$ is the corresponding experimental mean value. $\sigma_{j\,\rm exp}$ stands
for the $1\sigma$ experimental uncertainty (symmetric around the mean).\medskip

The non-trivial part of the analysis is the implementation of the theoretical
error. Here we follow the \textit{R}fit scheme \cite{Hocker:2001xe}. Our basic
assumption is that experimental data approximatively yield a Gaussian
distribution of an observable but a theoretical calculation does not. The latter
depends on a set of input parameters like form factors, decay constants etc. for
which no probability distribution is known. The \textit{R}fit scheme corresponds
to a frequentist approach and it assumes no particular distribution for the theory
parameters, only that they are constrained to certain allowed ranges. The
theoretical and experimental uncertainties are then combined in the following
$\chi^2$ function:
\be\label{e52}
\chi^2 =\sum_j \left\{\begin{array}{cl}
           \frac{(|x_{j\,\rm exp}-x_{j\,\rm theo}|-\sigma_{j\,\rm
theo})^2}{\sigma_{j\,\rm exp}^2}
           & \mbox{  if  }|x_{j\,\rm exp}-x_{j\,\rm theo}|>\sigma_{j\,\rm theo},
\\
           0 & \mbox{  otherwise}.\\
         \end{array} \right.
\ee
Here we suppress the dependence on the parameters $\{q_i\}$. Since we often
encounter asymmetric theory intervals, notated as $(x_{i\,\rm
theo})^{+\sigma_{i\,\rm theo,\,sup}}_{-\sigma_{i\,\rm theo,\,inf}}$, we
generalise eq.~(\ref{e52}) to
\be\label{e53}
\chi^2 =\sum_j \left\{\begin{array}{cl}
           \frac{(x_{j\,\rm theo}-\sigma_{j\,\rm theo,\,inf}-x_{j\,\rm
exp})^2}{\sigma_{j\,\rm exp}^2}
           & \mbox{  if  } x_{j\,\rm exp} < (x_{j\,\rm theo}-\sigma_{j\,\rm
theo,\,inf}), \\
           \frac{(x_{j\,\rm exp}-(x_{j\,\rm theo}+\sigma_{j\,\rm
theo,\,sup}))^2}{\sigma_{j\,\rm exp}^2}
           & \mbox{  if  } x_{j\,\rm exp} > (x_{j\,\rm theo}+\sigma_{j\,\rm
theo,\,sup}), \\
           0 & \mbox{  otherwise}.\\
         \end{array} \right.
\ee
We calculate $\sigma_{j\,\rm theo,\,sup}$ and $\sigma_{j\,\rm theo,\,inf}$ using
the theory input given in tab.~\ref{tab0} and Appendix~\ref{input}, adding the
resulting uncertainties in quadrature.\medskip

Using the $\chi^2$ function in eq.~(\ref{e53}) it is possible to define confidence levels
(CLs) by means of the function
\be\label{e54}
{\rm CL}(\{q_i\}) = \frac{1}{\sqrt{2^{N_{\rm dof}}} \Gamma(N_{\rm dof}/2)}
\int_{\Delta \chi^2(\{q_i\})}^{\infty} e^{-t/2}t^{N_{\rm dof}/2 -1}dt,
\ee
where $\Delta \chi^2$ is the $\chi^2$ function after subtraction of its minimum:
$\Delta \chi^2 = \chi^2 -{\rm min}(\chi^2)$. $N_{\rm dof}$ is the number of
degrees of freedom of free model parameters.
Setting ${\rm CL} = 1 - 68.27/100$, ${\rm CL} = 1 - 95.45/100$, ${\rm CL} = 1 -
99.73/100$ and ${\rm CL} = 1 - 99.99/100$ we find the $1\,\sigma$, $2\,\sigma$,
$3\,\sigma$ and $5\,\sigma$ confidence levels respectively.\medskip

In our fits we only include quantities which are derived from \textit{independently}
measured observables, i.e.\  $\Delta A_{\rm CP}^0$, $\Delta A_{\rm CP}^-$,
$S_{\rm CP}(\bar{B}^0\to\pi^0 \bar{K}^0)$ and one out of the three categories of
ratios $R^{B,K,\pi}$ (see Appendix~\ref{IsoRatios}). Thereby we select the
category which is most sensitive to the NP scenario under consideration. In this way we
can on the one hand avoid to overweight a particular observable in the fit and
on the other hand avoid that the fit is pulled to large $q_i$-values by
discrepancies of quantities carrying only a small sensitivity on NP. We note
here that, since the $R^{B,K,\pi}_{(c,n)}$ are ratios of branching fractions,
which we assume to be Gaussian in experiment, their probability distributions
derived from experimental data are not exact Gaussians as required by eqs.
(\ref{e52}-\ref{e54}). Comparing fits to ratios of branching fractions to fits
to the differences of the corresponding branching ratios (which follow a Gaussian
distribution), we have checked that the qualitative outcome of the fits in terms of
preferred regions in the complex $q_i$-planes is not tarnished but that the contour
lines are sharpened due to the reduction of theoretical uncertainties in the
ratios.\bigskip

\section{SUSY contributions to penguin coefficients}\label{susypeng}

In this section we quote our results for the initial conditions of the Wilson
coefficients $C_{3,..,10}$, $C_{7\gamma}$ and $C_{8g}$ and of their mirror
counterparts at the SUSY mass scale. We decompose the Wilson coefficients $C_{3,..,10}$ into
contributions $C_g$ from gluon penguins, $C_{\gamma}$ from photon penguins  and
$C_{B_{q}}^\text{LL1}$, $C_{B_{q}}^\text{LL2}$, $C_{B_{q}}^\text{LR1}$, $C_{B_{q}}^\text{LR1}$ ($q=u,d$) from box diagrams:
\begin{align}\nonumber
C_3 &= -\frac{1}{3}C_g   + \frac{1}{2} C_{B_u}^\text{LL1} + C_{B_d}^\text{LL1}, & C_4 &= C_g + \frac{1}{2} C_{B_{u}}^\text{LL2} + C_{B_{d}}^\text{LL2}, \\\nonumber
C_5 &= -\frac{1}{3}C_g  + \frac{1}{2} C_{B_{u}}^\text{LR1} + C_{B_{d}}^\text{LR1}, & C_6 &= C_g  + \frac{1}{2} C_{B_{u}}^\text{LR2} + C_{B_{d}}^\text{LR2}, \\\nonumber
C_7 &= C_\gamma  +  C_{B_{u}}^\text{LR1} - C_{B_{d}}^\text{LR1}, & C_8 &=   C_{B_{u}}^\text{LR2} - C_{B_{d}}^\text{LR2}, \\
C_9 &= C_\gamma  +  C_{B_{u}}^\text{LL1} - C_{B_{d}}^\text{LL1}, & C_{10} &=  C_{B_{u}}^\text{LL2} - C_{B_{d}}^\text{LL2},
\end{align}
The coefficients $C_i'$ are obtained from these expressions by the replacements $L\leftrightarrow R$ and $C_{g,\gamma}\rightarrow C'_{g,\gamma}$. In addition $Z$-penguins contribute to $C_i^{(')}$ according to eq.~(\ref{c11}). The individual contributions are obtained by calculating chargino-(up-)squark loops and gluino-(down-)squark loops. We do not consider neutralino-(down-)squark exchange since it is suppressed with respect to the gluino contributions involving the strong coupling. Our results are in agreement with similar calculations in the context of semileptonic $B$ decays \cite{Bertolini:1990if,Buras:2004qb,Bobeth:2004jz} and with the gluino boxes in ref.~\cite{Imbeault:2008ge}.\medskip

We neglect additional operators arising from $b\rightarrow s \bar b b$ transitions
because they contribute to $B$ decays only in higher orders. Moreover we neglect
box diagrams with more than one flavour-violating squark line. As for the squarks
of the first two generations we assume approximate degeneracy of the corresponding
elements of $\hat m_{\tilde Q}^2$, likewise for $\hat m_{\tilde u}^2$ and
$\hat m_{\tilde d}^2$. Yukawa couplings of the first two generations are set to zero.
In this way the box diagrams depend on common masses $m_{\tilde u_L}=m_{\tilde d_L}$
($m_{\tilde u_R}$ and $m_{\tilde d_R}$) for the left-handed
(right-handed) squarks of the first and second generation.\medskip

We use the conventions of the SUSY Les Houches Accords
\cite{Skands:2003cj,Allanach:2008qq} with only one exception: The CKM matrix is
denoted as $V$ whereas the chargino mixing matrices are named $\mathcal U$ and
$\mathcal V$ instead of $U$ and $V$.\medskip

\subsection{Chargino contributions}

We use the loop functions written down in \cite{Buras:2004qb,Bobeth:2004jz} and the mass ratio
\be
x_{u_r c_m} = \left(\frac{m_{\tilde u_r}}{m_{\tilde \chi_m^+}}\right)^2.
\ee
Generation indices $i,j=1,2,3$, squark indices $r,s=1,...,6$ and chargino indices $m,n=1,2$ are always summed over in the following. We abbreviate the chargino-quark-squark couplings by
\begin{eqnarray}
\Gamma^L_{rim} &=& (g \mathcal{R}^{u*}_{rk} \mathcal{V}_{m1}-\mathcal{R}^{u*}_{r,k+3} \mathcal{V}_{m2} Y_{u_k})\,(\delta_{kj}+\Delta U^{u*}_{L,kj})\,V^*_{ji},\nonumber\\ 
\Gamma^R_{rim} &=& \mathcal{U}^*_{m2}\,\mathcal{R}^{u*}_{ra}\,(\delta_{ak}+\Delta U_{L,ak}^{u*})\,V_{kl}^*\,
   (\delta_{lj}-\Delta U^{d*}_{L,lj})\,Y_{d_j}^{(0)}\,(\delta_{ji}+\Delta U^{d*}_{R,ji})\,. 
\end{eqnarray}
The matrices $\Delta U_L^d$, $\Delta U_R^d$ and $\Delta U_L^u$ account for chirally enhanced corrections to the tree-level quark-squark-chargino coupling. Explicit expressions for these matrices are given in Ref.~\cite{Crivellin:2008mq}. The quantity $Y_{d_j}^{(0)}$ represents the modified Yukawa coupling incorporating $\tan\beta$-enhanced corrections (resummed to all orders in perturbation theory). An explicit formula for $Y_{d_j}^{(0)}$ permitting complex SUSY mass parameters can be found in Ref.~\cite{Hofer:2009xb}.\medskip

Neglecting Yukawa couplings of the first two generations we find that $C'_{3,..,10}=0$. The coefficients $C_{3,..,10}$ are constructed from
\begin{align}\label{cgammacha}
C_\gamma = & \frac{\alpha}{24 \sqrt{2}  G_F \pi  V_{33} V^*_{32}   (m_{\tilde \chi^+_m})^2 }   \Gamma^L_{r2m} \Gamma^{L*}_{r3m} h_3^{(0)}(x_{u_r c_m}),  \displaybreak[1]  \\
C_g =  & \frac{\alpha_s}{ 32 \sqrt{2} G_F \pi  V_{33} V^*_{32} (m_{\tilde\chi^+_m})^2 }  \Gamma^L_{r2m} \Gamma^{L*}_{r3m} h_4^{(0)}(x_{u_r c_m}), \\
 \begin{split}
 C_{B_u}^{\text{LL1}} =  & \frac{\alpha}{24 s_W^2 \sqrt{2} G_F \pi
    V_{33} V^*_{32}}   m_{\tilde\chi^+_m} m_{\tilde\chi^+_n}     \mathcal{U}_{n1} \mathcal{U}^*_{m1} \Gamma^L_{r2n} \Gamma^{L*}_{r3m} D_0(m_{\tilde d_L}^2,m_{\tilde u_{r}}^2,m_{\tilde\chi^+_m}^2,m_{\tilde \chi^+_n}^2),
 \end{split}\\
\begin{split}
C_{B_d}^{\text{LL1}} =  & \frac{\alpha}{48 s_W^2 \sqrt{2} G_F \pi   V_{33} V^*_{32}}
      \mathcal{V}_{m1} \mathcal{V}^*_{n1}  \Gamma^L_{r2n} \Gamma^{L*}_{r3m}  D_2(m_{\tilde u_L}^2,m_{\tilde u_r}^2,m_{\tilde\chi^+_m}^2,m_{\tilde\chi^+_n}^2),
\end{split}\\
\begin{split}
\kappa_L^{sb} = & \frac{1}{32\pi^2}  \Gamma^L_{r2n} \Gamma^{L*}_{s3m}  \left( \delta _{mn} C_2(m_{\tilde\chi^+_m}^2,m_{\tilde u_s}^2,m_{\tilde u_r}^2)  \sum_{k=1}^3 \mathcal{R}^u_{rk} \mathcal{R}^{u*}_{sk}\, +  \right. \\
& \left. \delta_{sr}  \left \{ 2 C_0(m_{\tilde u_s}^2,m_{\tilde\chi^+_m}^2,m_{\tilde\chi^+_n}^2) m_{\tilde\chi^+_m} m_{\tilde\chi^+_n} \mathcal{U}_{n1} \mathcal{U}^*_{m1} - C_2(m_{\tilde u_s}^2,m_{\tilde\chi^+_m}^2,m_{\tilde\chi^+_n}^2)  \mathcal{V}_{m1} \mathcal{V}^*_{n1}\right\}  \vphantom{\sum_s } \right)  .
\end{split}
\end{align}

\noindent
The remaining box coefficients vanish. The magnetic and chromo-magnetic coefficients are given by
\begin{align}
C_{7\gamma} = & \frac{1}{8 \sqrt{2} G_F  V_{33} V^*_{32} (m_{\tilde\chi^+_m})^2}
   \Gamma^L_{r2m} \left(   \frac{m_{\tilde\chi^+_m}}{m_b}  \Gamma^{R*}_{r3m} h_2^{(0)}(x_{u_r c_m}) -  \Gamma^{L*}_{r3m} h_1^{(0)}(x_{u_r c_m})   \right), \\
C_{8g} = & \frac{1}{8 \sqrt{2} G_F  V_{33} V^*_{32} (m_{\tilde\chi^+_m})^2 }  \Gamma^L_{r2m}  \left(   \frac{m_{\tilde\chi^+_m}}{m_b} \Gamma^{R*}_{r3m} h_6^{(0)}(x_{u_r c_m}) - \Gamma^{L*}_{r3m} h_5^{(0)}(x_{u_r c_m})   \right).\label{c8gcha}
\end{align}\medskip

\medskip

\subsection{Gluino contributions}

Here we use the loop functions written down in \cite{Bertolini:1990if,Buras:2004qb} and
\begin{equation}
F_{10}(x) =  \frac{3-3x+(2+x)\log(x)}{12(x-1)^2}.
\end{equation}
These functions depend on the mass ratio
\be
x_{g d_s} = \left(\frac{m_{\tilde g}}{m_{\tilde d_s}}\right)^2.
\ee
Writing
\begin{equation}
   G^L_{si}\,=\,\mathcal{R}^{d}_{sj}\left(\delta_{ji}+\Delta U_{L,ji}^d\right),\hspace{2cm} 
   G^R_{si}\,=\,\mathcal{R}^{d}_{s,j+3}\left(\delta_{ji}+\Delta U_{R,ji}^d\right)
\end{equation}
we split the $6\times 6$ down-squark mixing matrix $\mathcal{R}$ into a lefthanded and a righthanded $6\times 3$ block $G^L$ and $G^R$. The matrices $\Delta U_L^d$ and $\Delta U_R^d$ account for chirally enhanced corrections to the tree-level quark-squark-gluino coupling \cite{Crivellin:2009ar}. With a sum over all indices understood, we find ($q=u,d$)\footnote{The sign of $D_2$ in $C_{B_q}^\text{LL2}$ differs from ref.~\cite{Grossman:1999av}.}
\begin{align}\label{cgammaglu}
 C_\gamma = & -\frac{4 \sqrt{2} \alpha \alpha_s}{27 G_F V_{33} V^*_{32} m_{\tilde d_s}^2  }  G^{L*}_{s2} G^L_{s3}   \, F_6(x_{gd_s}), \displaybreak[1] \\
C_g =  &  \frac{\alpha_s^2}{2\sqrt{2} G_F V_{33} V^*_{32} m_{\tilde d_s}^2}   G^{L*}_{s2} G^L_{s3} \left( C_F F_6(x_{gd_s}) - C_A F_{10}(x_{gd_s}) \right),\\
C_{B_q}^\text{LL1} = & \phantom{- }   \frac{ \alpha_s^2 \, G^L_{r3}  G^{L*}_{r2} }{108 \sqrt{2} G_F V_{33} V^*_{32}} \left( 20 m_{\tilde g}^2 D_0(m_{\tilde g}^2, m_{\tilde g}^2,  m_{\tilde d_r}^2,  m_{\tilde q_L}^2 ) +  D_2(m_{\tilde g}^2, m_{\tilde g}^2, m_{\tilde d_r}^2, m_{\tilde q_L}^2) \right) ,  \\
C_{B_q}^\text{LL2} = &  -   \frac{ \alpha_s^2 \, G^{L}_{r3}  G^{L*}_{r2} }{36 \sqrt{2} G_F V_{33} V^*_{32}} \left( 4 m_{\tilde g}^2 D_0(m_{\tilde g}^2, m_{\tilde g}^2,  m_{\tilde d_r}^2,  m_{\tilde q_L}^2 ) -7  D_2(m_{\tilde g}^2, m_{\tilde g}^2, m_{\tilde d_r}^2, m_{\tilde q_L}^2) \right),\\
C_{B_q}^\text{LR1} = & -   \frac{ \alpha_s^2 \, G^{L}_{r3}  G^{L*}_{r2} }{54 \sqrt{2} G_F V_{33} V^*_{32}} \left(  m_{\tilde g}^2 D_0(m_{\tilde g}^2, m_{\tilde g}^2,  m_{\tilde d_r}^2,  m_{\tilde q_R}^2 ) +5  D_2(m_{\tilde g}^2, m_{\tilde g}^2, m_{\tilde d_r}^2, m_{\tilde q_R}^2) \right),\\
C_{B_q}^\text{LR2} = & -   \frac{ \alpha_s^2 \, G^{L}_{r3}  G^{L*}_{r2} }{18 \sqrt{2} G_F V_{33} V^*_{32}} \left(7  m_{\tilde g}^2 D_0(m_{\tilde g}^2, m_{\tilde g}^2,  m_{\tilde d_r}^2,  m_{\tilde q_R}^2 ) -  D_2(m_{\tilde g}^2, m_{\tilde g}^2, m_{\tilde d_r}^2, m_{\tilde q_R}^2) \right),\\
\kappa_L^{sb} = & \frac{   \alpha_s}{3\pi }  \,  \sum_{i=1}^3 G^L_{r3} G^R_{si} G^{R*}_{ri} G^{L*}_{s2}   C_2(m_{\tilde g}^2,m_{\tilde d_r}^2,m_{\tilde d_s}^2).
\end{align}
The magnetic and chromo-magnetic coefficients are
\begin{align}
\begin{split}
C_{7\gamma} = & \frac{4\sqrt{2}\alpha_s\pi}{9 G_F V_{33} V^*_{32} m_{\tilde d_s}^2  }  \left( \frac{m_{\tilde g}}{m_b}   G^{L*}_{s2} G^{R}_{s3}  F_4(x_{gd_s}) -   G^{L*}_{s2} G^{L}_{s3} F_2(x_{gd_s})\right)
\end{split},\\
\begin{split}
C_{8g} = &    \frac{\sqrt{2} \alpha_s \pi}{2 G_F V_{33} V^*_{32} m_{\tilde d_s}^2 }  \left( \frac{m_{\tilde g}}{m_b} G^{L*}_{s2} G^{R}_{s3}  (C_A F_3(x_{gd_s}) - (2 C_F - C_A) F_4(x_{gd_s})) \right. \\
& \left. \phantom{\frac{m_{\tilde g}}{m_b}} -   G^{L*}_{s2} G^{L}_{s3} (C_A F_1(x_{gd_s}) - (2 C_F - C_A) F_2(x_{gd_s})) \right),\label{c8gglu}
\end{split}
\end{align}
with the group factors $C_F=\frac{4}{3}$ and $C_A =3$.

The corresponding expressions entering the Wilson coefficients $C_i'$ are obtained from eqs.~(\ref{cgammaglu})--(\ref{c8gglu}) by the simple replacement $L\leftrightarrow R$ with the exception of the $Z$-penguin which reads
\be
\kappa_R^{sb} =  -\frac{   \alpha_s}{3\pi }  \,  \sum_{i=1}^3 G^R_{r3} G^L_{si} G^{L*}_{ri} G^{R*}_{s2}   C_2(m_{\tilde g}^2,m_{\tilde d_r}^2,m_{\tilde d_s}^2).
\ee
\bigskip

\bibliography{Paper020111}
\bibliographystyle{JHEP-2}

\end{document}